\documentclass[twocolumn]{aastex62}

\usepackage{multirow}
\usepackage{graphicx}
\usepackage[caption=false]{subfig}
\usepackage{amsmath}
\usepackage{url}
\newcommand{\Sp}{{\it Spitzer\/}}
\newcommand{\eg}{{\it e.g.}}

\submitjournal{ApJ}

\shorttitle{Star Formation in the Outer Galaxy II.}
\shortauthors{Winston, Hora, \& Tolls}

\begin{document}

\title{ A Census of Star Formation in the Outer Galaxy II: The GLIMPSE360 Field}

\correspondingauthor{Elaine Winston}
\email{elaine.winston@cfa.harvard.edu}

\author[0000-0001-9065-6633]{Elaine Winston}
\affiliation{Center for Astrophysics $|$ Harvard \& Smithsonian,
60 Garden St.,
Cambridge MA 02138, USA}

\author[0000-0002-5599-4650]{Joseph L. Hora}
\affiliation{Center for Astrophysics $|$ Harvard \& Smithsonian,
60 Garden St.,
Cambridge MA 02138, USA}

\author[0000-0003-1841-2241]{Volker Tolls}
\affiliation{Center for Astrophysics $|$ Harvard \& Smithsonian,
60 Garden St.,
Cambridge MA 02138, USA}

\begin{abstract}

We have conducted a study of star formation in the outer Galaxy  from 65\degr$< l <$265\degr~in the region observed by the GLIMPSE360 program.  
This {\it Spitzer} warm mission program mapped the plane of the outer Milky Way with IRAC at 3.6 and 4.5~$\mu$m. 
We combine the IRAC, {\it WISE}, and 2MASS catalogs and our previous results from another outer Galaxy survey and identify a total of 47,338 Young Stellar Objects (YSOs) across the field spanning $>$180\degr\ in Galactic longitude. 
Using the $DBSCAN$ method on the combined catalog, we identify 618 clusters or aggregations of YSOs  having 5 or more members. 
We identify 10,476 Class I, 29,604 Class II, and 7,325 anemic Class II/Class III YSOs.  
The ratio of YSOs identified as members of clusters was 25,528/47,338, or 54\%.    
We found 100 of the clusters identified have previously measured distances in the {\it WISE} \ion{H}{2} survey. We used these distances in our spectral energy distribution (SED) fitting of the YSOs in these clusters, of which 96 had YSOs with $<3\sigma$ fits.
We used the derived masses from the SED model fits to estimate the initial mass function (IMF) in the inner and outer Galaxy clusters: dividing the clusters by Galactocentric distances, the slopes were  $\Gamma = 1.87 \pm 0.31$ 
above 3~M$_{\odot}$ for $R_{Gal} < 11.5$~kpc  and $\Gamma = 1.15 \pm 0.24$ above 3~M$_{\odot}$ for $R_{Gal} > 11.5$~kpc.
The slope of the combined IMF was found to be $\Gamma = 1.92 \pm 0.42$ above 3~M$_{\odot}$. 
These values are consistent with each other within the uncertainties, and with literature values in the inner Galaxy high-mass star formation regions. 
The slopes are likely also consistent with a universal Salpeter IMF.

\end{abstract}


\section{Introduction}

The study of star formation has been revolutionized with the launch of the {\it Spitzer} Space Telescope \citep[{\it Spitzer;} ][]{werner}.  
Much of the published work to date has focused on regions at Galactocentric radii less than the sun, and in nearby clouds.  The \Sp\ Legacy programs\footnote{\url{https://irsa.ipac.caltech.edu/data/SPITZER/docs/spitzermission/observingprograms/legacy/history/}}, executed early in the mission, provided a large dataset for studies of star formation. The original GLIMPSE survey \citep{Benjamin2003} and subsequent follow-on programs \citep{Churchwell2009} mapped the inner Galactic plane ($295^\circ < l < 65^\circ$) with IRAC at 3.6, 4.5, 5.8, and 8~\micron, and the MIPSGAL programs \citep{Carey2009} mapped the regions with MIPS at 24 and 70~\micron. The c2d program \citep{Evans2003} scanned large areas in nearby molecular clouds for low luminosity sources to obtain a sample of nearby solar-type stars for debris disk studies. The FEPS program \citep{Meyer2004} studied a large sample of young nearby solar-type stars to trace the evolution of circumstellar gas and dust from primordial planet-building stages in young circumstellar disks
through to older collisionally generated debris disks. Later Legacy and other large programs further contributed to our knowledge of star formation in nearby molecular clouds. The Gould's Belt program \citep{Allen2006} completed the observations of all prominent star-forming regions within 500~pc. \citet{Megeath2012,megeath} conducted studies of the Orion A and B clouds, identifying thousands of YSO candidates in this nearby massive star-forming region. Other programs mapped the massive star formation complexes such as Cygnus-X \citep{Hora2009} and Vela-Carina \citep{Majewski2007} at distances of $\sim$1 -- 2~kpc.  These and many other individual programs have produced a wealth of data on nearby star-forming regions which have been utilized in thousands of papers and will continue to be mined for years to come.

The outer Galaxy is a distinctly different environment to that of the inner Galaxy, with conditions seemingly less likely to efficiently form stars.(In this paper, we will use the term `outer Galaxy' to refer to clusters with a Galactocentric radius greater than $\sim$8~kpc and a Galactic longitude between roughly $65^\circ < l < 265^\circ$.) The efficiency with which a molecular cloud forms stars is thought to be dependent on its density, temperature, and chemical abundances \citep[e.g.][]{evans}.  
The metallicity of the Milky Way is believed to decline as a function of Galacto-centric radius \citep{rudolph}.   
Average temperatures in molecular clouds are found to be lower \citep{mead}, as is the cosmic ray flux \citep{bloemen}. 
Further, the volume density of molecular clouds in the outer Galaxy is lower, and so interaction rates and incidence of spiral arm crossings will be lower compared to inner Galaxy regions over the star forming lifetime of an individual cloud.

The {\it Spitzer} SMOG survey \citep{carey} was designed to help fill in our knowledge of star formation in the outer Galaxy by providing deep coverage of a field in the 
outer Galaxy in the IRAC and MIPS bands. Utilizing data from this survey,  we presented in our previous paper \citep[][Paper I]{winston2019} an initial study of the outer regions of our Galaxy, where environmental factors may impact on the star formation occurring there. In this paper, we present a census of star formation across the entire plane of our outer Milky Way galaxy using {\it Spitzer}'s GLIMPSE360 survey \citep{whitney08,whitney09}. This survey allows us to identify the young stellar populations of these clusters to better answer the question of whether the colder, less dense, and lower metallicity environment of the outer Galaxy affects the formation and evolution of young stars. With this work we will expand on our previous study by applying the techniques outlined in Paper I to the outer Galaxy covered by the GLIMPSE360 Galactic plane survey from $65^\circ < l < 265^\circ$, in a 3\degr~wide strip that follows the warp in the outer Galactic disk.
Here we identify young stellar objects (YSOs) across this $\sim$600 deg$^2$ strip from their excess infrared emission, locate clusters of YSOs indicating new regions of star formation,  determine the evolutionary class of the YSOs to analyze the protostellar ratio of 
the clusters, and make a preliminary assessment of the initial mass function (IMF) across clusters with known distances in the outer Galaxy.   

This paper is organized as follows. In Section~\ref{obs} we discuss the origins of the infrared catalogs.   
In Section~\ref{iding} we discuss contamination removal, and the identification of the YSOs and their evolutionary classification. 
We then discuss the spatial distribution of the young stars and the identification of stellar clusters in the outer Galactic 
fields in Section~\ref{spatial}. We discuss the fits to the SEDs of YSOs in clusters with known distances in Section~\ref{sedfitter}, and in Section~\ref{compare} compare our results to previous YSO catalogs constructed for the outer Galaxy.
Finally, a brief summary is presented in Section~\ref{summ}.

\section{Observations and Data Reduction}\label{obs}

\subsection{GLIMPSE360 Survey}

The GLIMPSE360 survey completes the coverage of the Galactic plane that began with the earlier GLIMPSE surveys.  
The observations were taken as part of a \Sp\ Warm Mission Exploration Science program\footnote{\url{https://irsa.ipac.caltech.edu/data/SPITZER/docs/spitzermission/observingprograms/es/}}, and were performed using the two short-wavelength IRAC bands at 3.6 and 4.5~$\mu$m \citep{fazio}.
The GLIMPSE360 point source archive covering Galactic longitudes from $65^\circ < l < 265^\circ$ was downloaded from the 
Infrared Science Archive (IRSA).  This catalog did not contain the SMOG field data which were previously reported in Paper I 
($102^\circ < l < 109^\circ$), and which were observed for the cryogenic \Sp\ Mapping of the Outer Galaxy (SMOG) program. 
It also excludes the central field surrounding the Cygnus star forming complex ($76^\circ < l < 82^\circ$), which forms part of the Cygnus-X survey, 
though the flanking fields in the Galactic plane are included in the archive.  
Two types of catalog are available: a highly reliable point source Catalog and a highly complete point source Archive. 
The Archive includes sources with spatial positions as close as 0\farcs5, while the Catalog excludes sources closer than 2$\arcsec$ in position.   
As in Paper I, the more complete Archive was used in this paper, to aid in the detection of fainter, more embedded YSOs. 

The IRAC observations at 3.6 and 4.5~$\mu$m were obtained in High Dynamic Range mode, which was comprised of three visits per mosaic position with 
0.6 and 12 second integrations, similar to the SMOG survey.  These data products were produced using the GLIMPSE team's pipeline.   
The GLIMPSE360 catalog contained a total of 49,378,042 sources. 
The mid-IR photometry was supplemented by $J$, $H$ and $K$-band photometry from the 2MASS point source catalog \citep{skr}.   
The photometric catalogues were merged using a 1\farcs6 matching radius.  
Documentation describing the GLIMPSE360 survey and the reduction process in detail is available on the IRSA 
website\footnote{\url{https://irsa.ipac.caltech.edu/data/SPITZER/GLIMPSE/doc/glimpse360_dataprod_v1.5.pdf}}.

\subsection{WISE Catalog}

The Wide-field Infrared Survey Explorer \citep[{\it WISE};][]{wright} provides mid-IR photometry  at 3.5, 4.6, 12, and 22~$\mu$m.  
The AllWISE catalog is an all-sky survey combining the cryogenic {\it WISE} All Sky survey and the NEOWISE post-cryogenic survey \citep{mainzer}.  
The catalog is available via the NASA/IPAC Infrared Science Archive (IRSA) archive.\footnote{\url{https://irsa.ipac.caltech.edu/Missions/wise.html}} 
Selecting all the sources in the GLIMPSE360 field was made, resulting in a regional catalog of 14,483,596 point sources.   
The {\it WISE} satellite has considerably lower spatial resolution when compared to \Sp, with a highest resolution of $\sim$6\farcs1 at 3.5~$\mu$m.   
However, the astrometric accuracy and matching to the 2MASS catalog are to within 1\farcs5, and so we performed catalog matching to the GLIMPSE360 
catalog at 1\farcs5.

\section{YSO Identification \& Classification}\label{iding}

Young stellar objects are most frequently identified by their excess emission at IR wavelengths. This emission arises from reprocessed 
stellar radiation in the dusty material of their natal envelopes or circumstellar disks. The infrared identification of YSOs is 
carried out by identifying sources that possess colors indicative of IR excess and distinguishing them from reddened and/or 
cool stars \citep{winston, all, gut1}.  

A full description of the criteria for identification and selection of YSO and non-YSO sources as applied to the datasets here is given in the Appendices.  
The following subsections outline the removal of background extragalactic objects, YSO selection methods for each dataset, the complete YSO catalog, and 
the evolutionary classification of the YSOs based on their excess IR emission.

\subsection{IRAC \& 2MASS}\label{irwsysos}  

\subsubsection{Contamination}\label{contam} 

The GLIMPSE360 field covers the majority of plane of the outer Galaxy, where the background sources suffer negligible extinction from the Galactic bar, and thus it is expected that 
many of the point sources detected will be active galactic nuclei (AGN) or star-forming galaxies (polycyclic aromatic hydrocarbon (PAH) galaxies).   Knots of emission in the 
structure of molecular clouds may also be mistaken for YSOs.  A further source of confusion in the YSO sample comes from sources with photometric contamination of the apertures by PAH emission.  

Such sources, which we will refer to as contaminants, were identified in the SMOG field as outlined in Paper I.  The same methods could not be applied to the GLIMPSE360 data, due to the lack of 5.8 and 8~$\mu$m photometry.
In order to constrain the photometric and color cuts to be applied to the GLIMPSE360 data to remove contaminants, the SMOG field data were utilized and cuts were determined from the 
characteristics of the contaminants in that field.  A description of the applied cuts is given in Appendix~\ref{irysos_appen}.  
Sources in the color/magnitude spaces where the contaminants were located were removed before the selection of YSOs was carried out. 

\begin{figure}
\epsscale{1.15}
\plotone{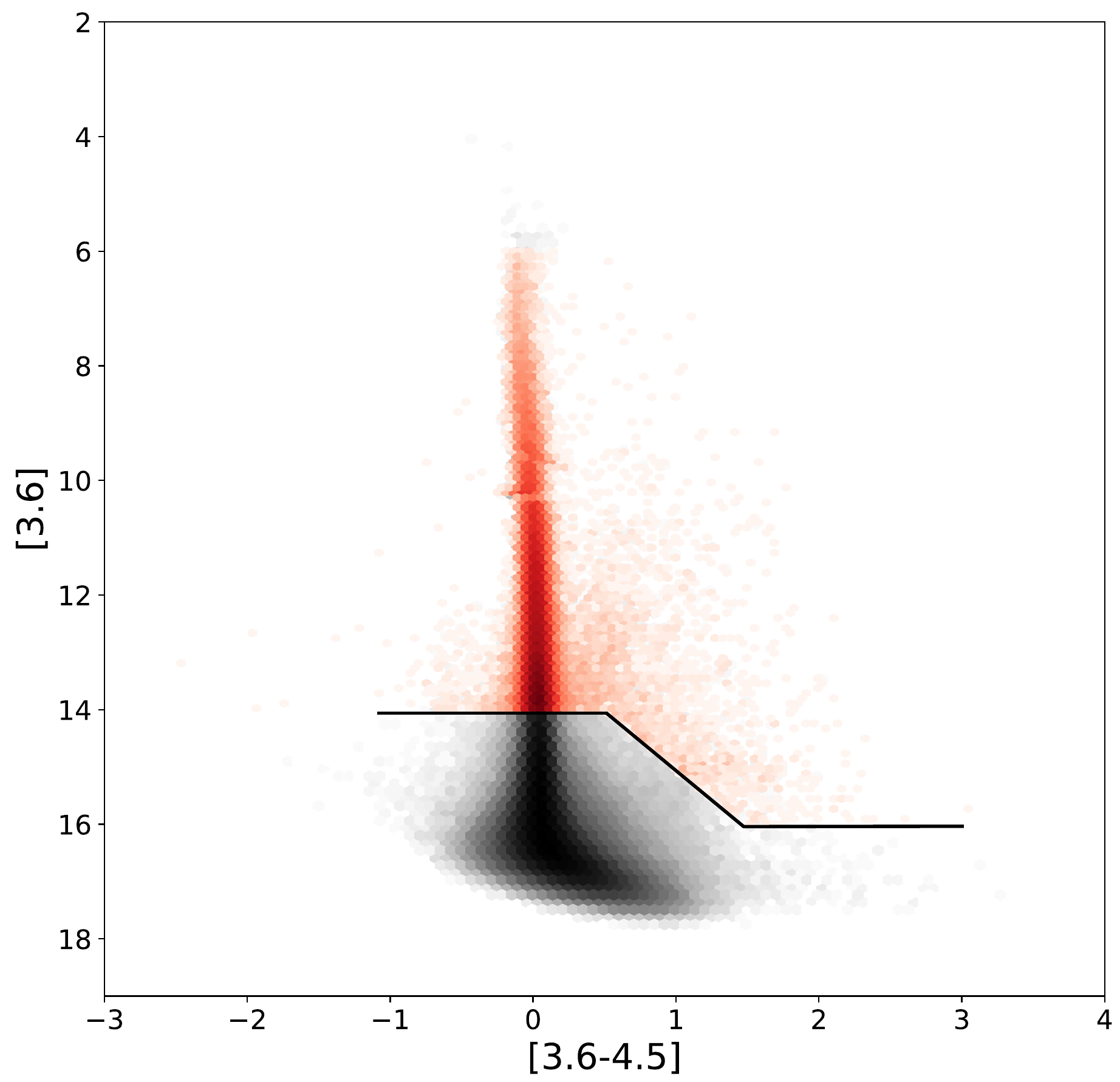}
\caption{ 
Example of color-magnitude of the IRAC [3.6] v [3.6 - 4.5] contaminants selection, 
in the 5\degr~section of the SFOG field in the range 180$^\circ < l< 185^\circ$. 
The red points represent the catalog after contaminants have been removed, the gray points are the selected contaminants.   
}
\protect\label{fig_contam}
\end{figure}

Figure~\ref{fig_contam} shows the criteria for the removal of contaminants in the GLIMPSE360 data for sources in a section of the field in the Galactic longitude range 180$^\circ < l< 185^\circ$. 
It was necessary to show only a subsection in order to enhance the clarity of the plot.

\subsubsection{IRAC YSO Selection}\label{iyso}

Young stellar objects were selected using a combination of color-color diagrams (CCDs) with IRAC and 2MASS+IRAC colors, as described in Appendix~\ref{irysos}.
Photometric uncertainties of $< 0.2$~mags and magnitudes fainter than the saturation limit were required in all the bands used for a {\it particular} color-color diagram to select YSOs.  
 
\begin{figure*}
\includegraphics[width=.335\linewidth]{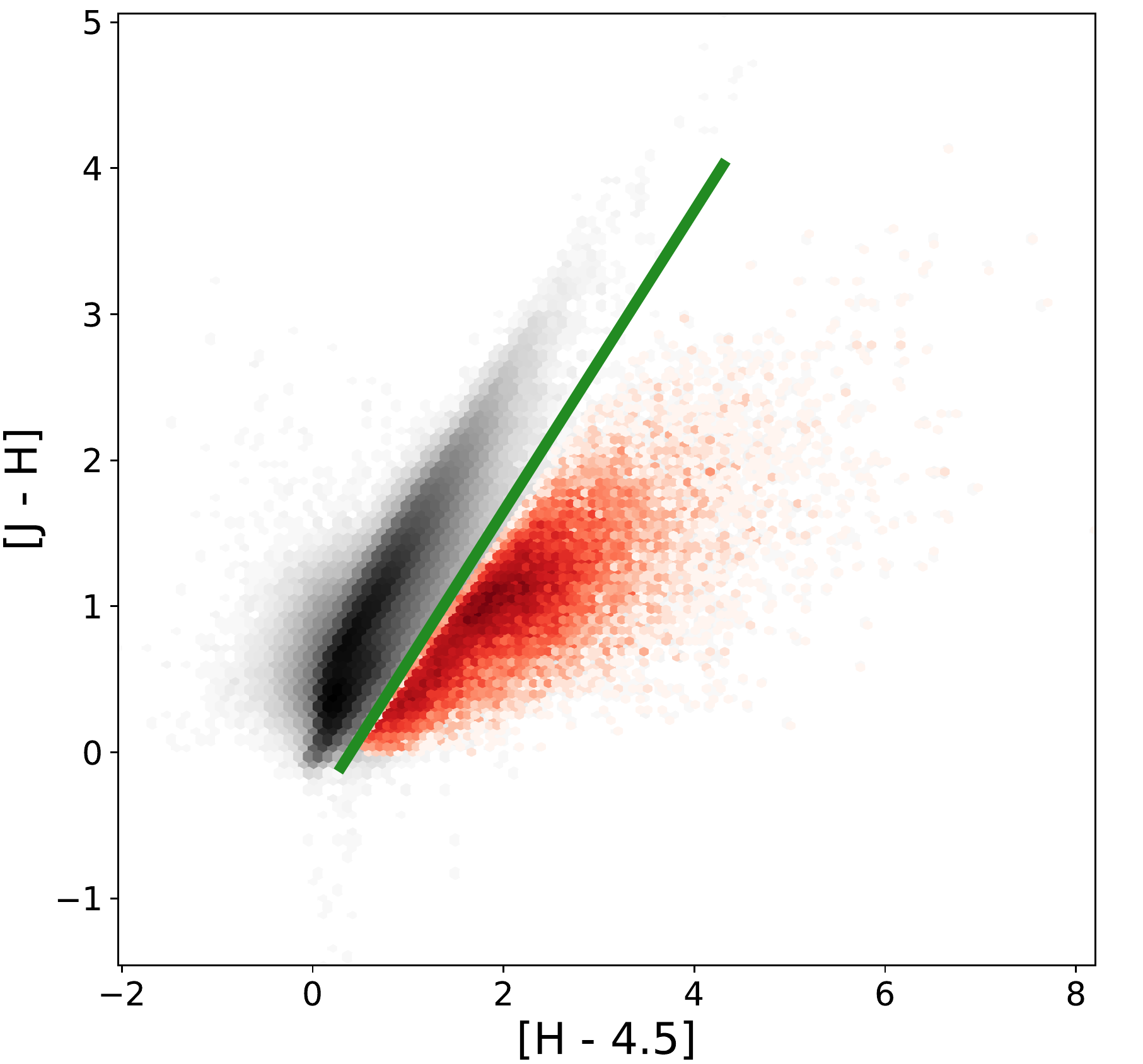}\includegraphics[width=.335\linewidth]{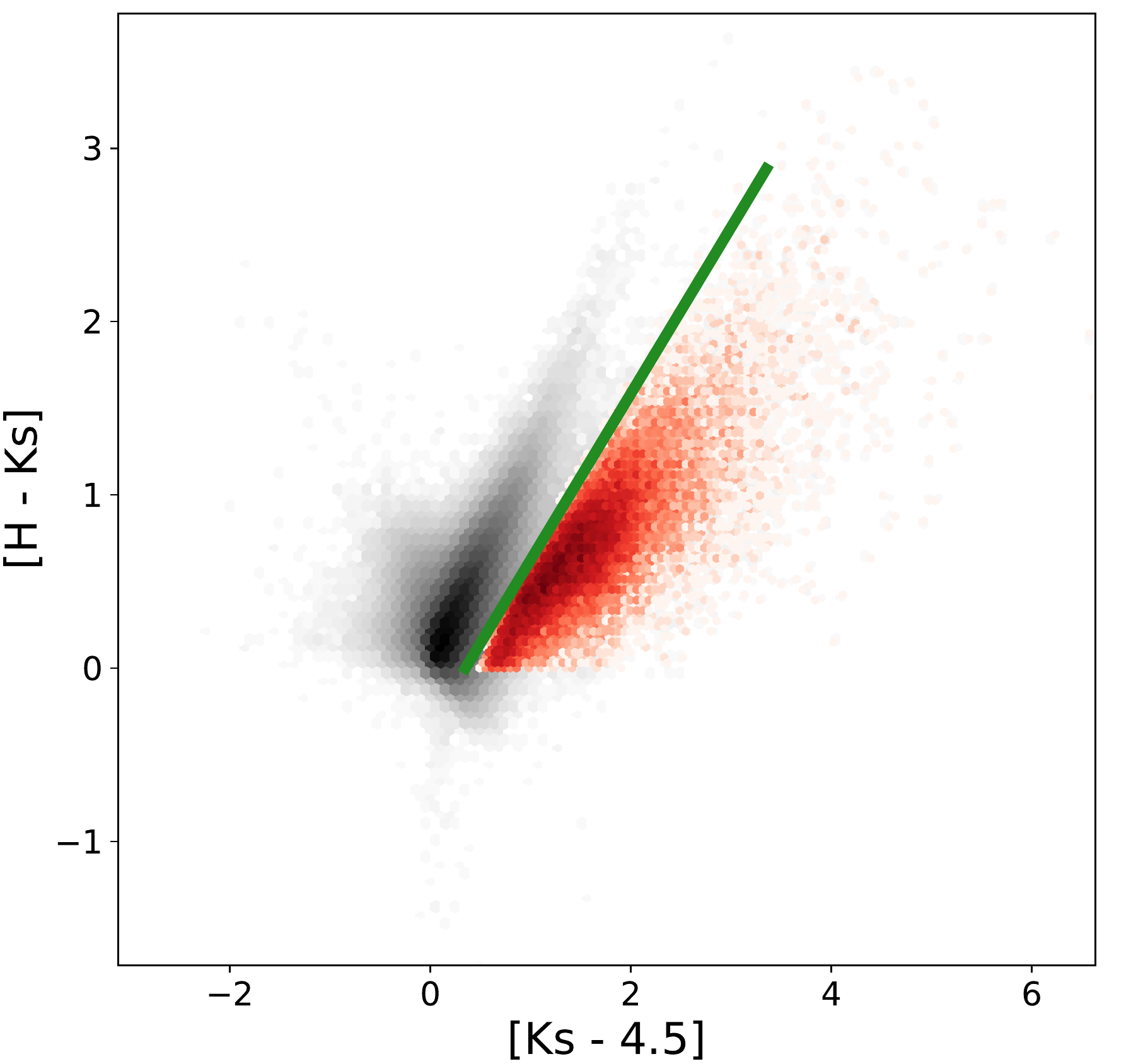}\includegraphics[width=.335\linewidth]{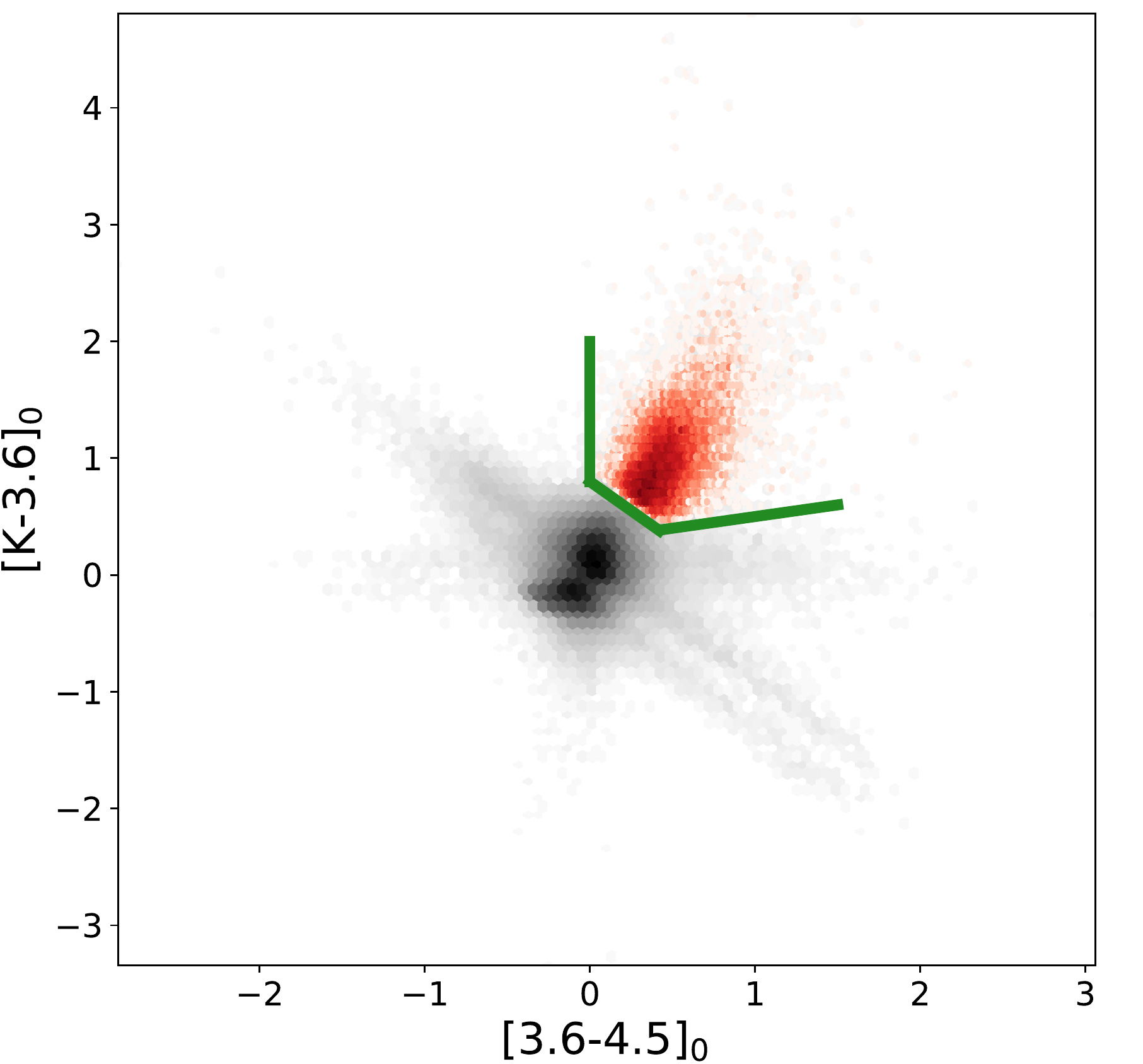}
\caption{ 
Selection of color-color diagrams of the near- and mid-IR 2MASS and IRAC YSO selection (red circles),   
overlaid on sources in the SFOG field (gray points) in the range 180$^\circ < l< 185^\circ$.
{\it Left:} 2MASS-IRAC [J-H] v [H - 4.5]   
{\it Center:} 2MASS-IRAC [H-K$_{s}$] v [K$_{s}$ - 4.5].  
{\it Right:} 2MASS-IRAC [K$_{s}$ -3.6] v [3.6 - 4.5].  
}
\protect\label{fig_irac}
\end{figure*}

The full GLIMPSE360 catalog contains 49,378,042 sources, of which 28,837 were identified as YSO candidates using the combined 2MASS and IRAC photometry. Hereafter we will refer to the YSO candidates as ``YSOs'', however a definitive classification would require a more detailed analysis of the spectra and other characteristics of each object.
Figure~\ref{fig_irac} shows the three source selection color-color diagrams used for the identification of YSOs for sources in the section of the field from 180$^\circ < l< 185^\circ$.

Remaining contaminants in the YSO sample that are not accounted for here include: galactic asymptotic giant branch (AGB) stars, and highly/unusually reddened field stars  
that may be confused for  anemic Class II (Class IIa) or Class III objects.  Such objects are expected to be scattered randomly over the field.  This issue is discussed further in Section~\ref{agbcontam} on AGB contamination.

\begin{figure*}
\plotone{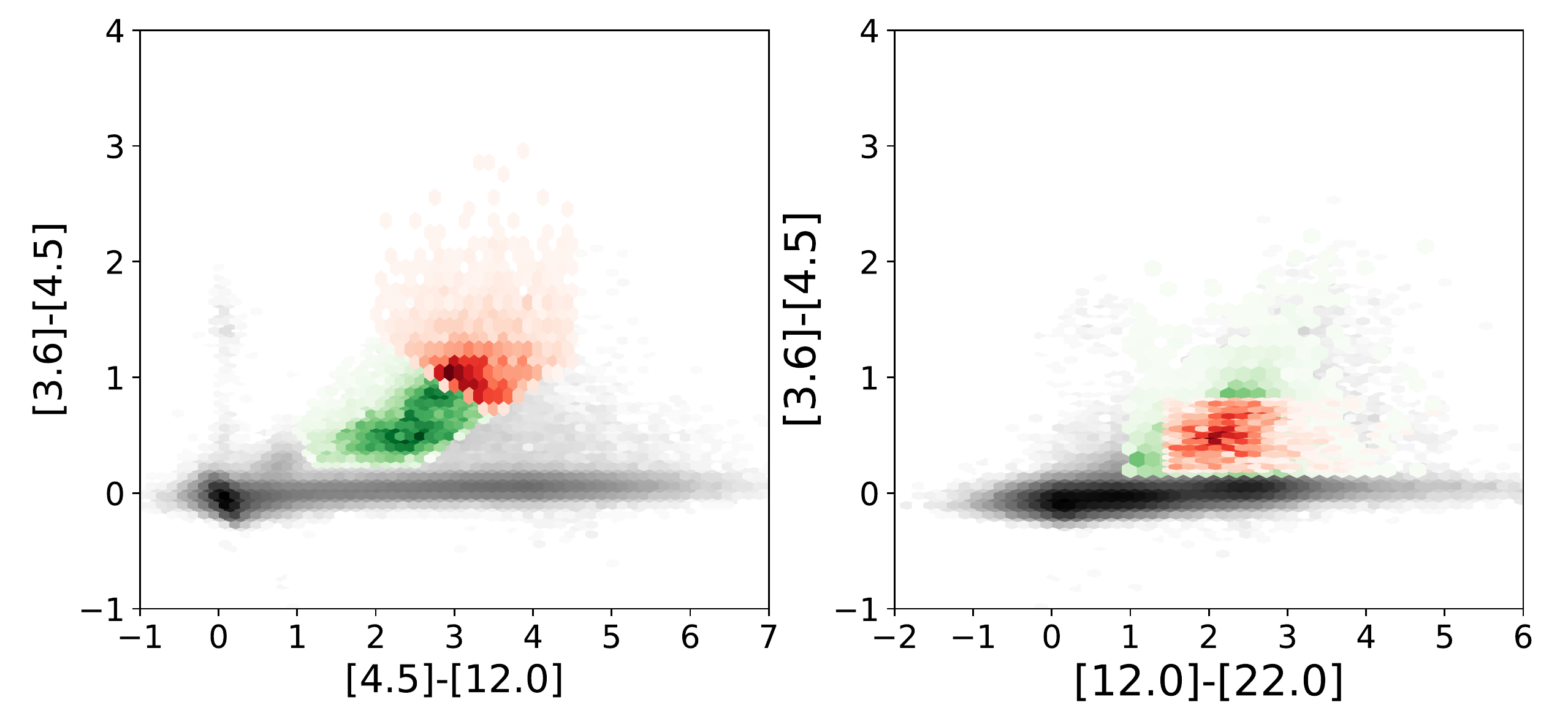}
\caption{ Selection of color-color diagrams of the IRAC+{\it WISE} YSO selection (red: excess, green: protostars) 
overlaid on sources in the SFOG field (gray points) in the range 180$^\circ < l< 185^\circ$. 
{\it Left:} IRAC+{\it WISE} [3.6 - 4.5] v [4.5 - 12].
{\it Right:} IRAC+{\it WISE} [3.6 - 4.5] v [12 - 22].
\label{fig_iracwise}}
\end{figure*}

\subsection{WISE Source Selection}\label{wise_compare}

The {\it WISE} catalog covering the GLIMPSE360 field comprised  $\sim$29\% the number of detections as the IRAC catalog. The long edges of the GLIMPSE360 field exhibit a sawtooth pattern due to the IRAC mapping procedure.   In our extraction of the {\it WISE} data from the all-sky catalog, we used a simple Galactic latitude cutoff that matches the largest extent of the IRAC data, so the edges of the {\it WISE} source distribution are smooth compared to the sawtooth edge of the GLIMPSE360 data.  
For this reason, a number of {\it WISE} sources fall into regions that are not included in the IRAC catalog and are identified solely by the {\it WISE} photometry on the edges of the field.  
Contaminant removal was undertaken using a procedure similar to the one used in Paper I and as outlined in Appendix~\ref{wise_ysos_appen}.  
The resulting catalog contained 543,457 point sources after the removal of spurious sources and extragalactic contaminants.   

A search for YSOs was performed using the four band {\it WISE} and 2MASS photometry; a detailed discussion is given in Appendix~\ref{wise_ysos_appen}.  
Of the 543,457 point sources, a total of 20,892 were identified as YSOs.

\subsection{Combined IRAC \& WISE Selection}\label{irws_compare}

Paper I compared the photometry in the IRAC bands 1 \& 2 and {\it WISE} bands 1 \& 2 for sources matched to within 1\farcs5. 
It was found that the photometric magnitudes matched well to within the uncertainties, which include measurement errors and possible variability of the sources themselves.  

In the GLIMPSE360 field, we used photometry from the two available IRAC bands to replace the appropriate {\it WISE} band fluxes where the sources matched to within  1\farcs5, and the {\it WISE} selection criteria were applied again to locate YSOs that may have reliable longer wavelength 
detections in {\it WISE}, but due to the shallower survey coverage or source confusion may not have been reliably detected at the shorter wavelengths. 
In these cases, the 2MASS selection criteria were not applied, since they replicate the selection made using the GLIMPSE360 data directly.      
The combined IRAC+{\it WISE} source list contained 318,588 objects.   A total of 11,196 YSOs were identified following our methodology, as outlined in Appendix~\ref{irws_ysos_appen}.   

Figure~\ref{fig_iracwise} shows the two source selection color-color diagrams used in the IRAC+{\it WISE} identification of YSOs for sources in the section of the field from $180^\circ < l < 185^\circ$.

\subsection{Combined SFOG YSO Catalog}\label{irysos}

The three sets of YSOs selections, 2MASS+IRAC (28,837 sources), IRAC+{\it WISE} (11,196 sources), and {\it WISE} (20,892 sources), were merged and common objects combined based on the source spatial position.  Sources in different sets that were within 1\farcs0 of each other were assumed to be the same source and combined.
The 2MASS+IRAC and IRAC+{\it WISE} catalogs also required a detection in the GLIMPSE360 catalog, which uniquely identifies each source.  
The unique list of YSOs in the GLIMPSE360 field contained 42,757 objects.  

The GLIMPSE360 YSOs were then combined with the SMOG field YSOs, to create the full catalog, which contained 47,405 candidate YSOs.  We call this the Star Formation in the Outer Galaxy, or SFOG catalog, and we refer to the total field covered by all of the component surveys as the SFOG field.
The SFOG catalog was then used for the following analysis.

\renewcommand{\arraystretch}{0.9}
\begin{deluxetable*}{ccl}
\tabletypesize{\scriptsize}
\tablecolumns{3}
\tablecaption{SFOG Field YSOs:  Photometry Table Description  }
\tablehead{   
\colhead{Column Number} & \colhead{Column ID}  & \colhead{Description}  
}
\startdata
0 &  designation &  GLIMPSE ID      \\  
1 &  2mass designation &   2MASS  ID     \\  
2 &  2mass cntr &  2MASS  counter     \\  
3 &  l   &  Longitude \\
4 &  b   &  Latitude \\
7 &  ra &  Right Ascension     \\  
8 &  dec  &   Declination     \\  
12 &  mag\_J & 2MASS J-band     \\  
13 &  dJ\_m &   2MASS J-band uncertainty     \\  
14 &  mag\_H &  2MASS H-band     \\  
15 &  dH\_m &  2MASS H-band uncertainty     \\  
16 &  mag\_K &  2MASS Ks-band     \\  
17 &  dKs\_m &  2MASS Ks-band uncertainty     \\   
18 &  mag3\_6 & IRAC band 1     \\  
19 &  d3\_6m & IRAC band 1 uncertainty     \\  
20 &  mag4\_5 & IRAC band 2     \\  
21 &  d4\_5m & IRAC band 2 uncertainty     \\  
22 &  mag5\_8 & IRAC band 3     \\  
23 &  d5\_8m & IRAC band 3 uncertainty     \\  
24 &  mag8\_0 & IRAC band 4     \\  
25 &  d8\_0m & IRAC band 4 uncertainty     \\  
78 &  designation\_1 &  WISE identifier     \\  
79 &  ra\_1 &  Right Ascension     \\  
80 &  dec\_1  &   Declination     \\  
93 &  w1mpro & WISE band 1     \\  
94 &  w1sigmpro & WISE band 1 uncertainty     \\  
97 &  w2mpro & WISE band 2     \\  
98 &  w2sigmpro & WISE band 2 uncertainty     \\  
101 &  w3mpro & WISE band 3     \\  
102 &  w3sigmpro & WISE band 3 uncertainty     \\  
105 &  w4mpro & WISE band 4     \\  
106 &  w4sigmpro & WISE band 4 uncertainty     \\  
150 &  tmass\_key &   WISE 2MASS  ID     \\  
154 &  j\_m\_2mass & WISE 2MASS J-band     \\  
155 &  j\_msig\_2mass &   WISE 2MASS J-band uncertainty     \\  
156 &  h\_m\_2mass &  WISE 2MASS H-band     \\  
157 &  h\_msig\_2mass &  WISE 2MASS H-band uncertainty     \\  
158 &  k\_m\_2mass &  WISE 2MASS Ks-band     \\  
159 &  k\_msig\_2mass &  WISE 2MASS Ks-band uncertainty     \\  
179 &  mag\_24 & MIPS band 1     \\  
180 &  d24\_m & MIPS band 1 uncertainty     \\ 
\enddata
\label{tab:IRphot} 
\tablecomments{Table \ref{tab:IRphot} is published 
in its entirety in the machine readable format.  A subset of the column identifiers are 
shown here for guidance regarding its form and primary content.}
\end{deluxetable*}

\begin{deluxetable*}{cchccccccc}
\tabletypesize{\scriptsize}
\tablecaption{SFOG Field YSOs:  Photometry Table Description \label{tab:photdescription} }
\tablehead{ 
\colhead{SF} & \colhead{Glimpse}  & \nocolhead{2MASS}  & \colhead{RA Dec}  & \colhead{IRAC}   & \colhead{WISE}  & \colhead{IRACWISE} & \colhead{SMOG}  & \colhead{Cluster}  & \colhead{Evolutionary}     \\[-0.3cm]
\colhead{ID}  & \colhead{ID}  & \nocolhead{ID} & \nocolhead{(J2000)}  & \colhead{YSO\tablenotemark{a}}  & \colhead{YSO}  & \colhead{YSO}  & \colhead{YSO}  &\colhead{Num\tablenotemark{b}}  &\colhead{Class\tablenotemark{c}} 
}
\startdata
SRC0 & SSTGLMA G064.5244+01.1116 & 19494624+2821339 & 19h49m46.2437s +28d21m33.8616s & 1 & 0 & 0 & 0 & -1 & 1 \\
SRC1 & SSTGLMA G064.5361+02.1303 & 19454732+2852593 & 19h45m47.3287s +28d52m59.394s & 1 & 0 & 0 & 0 & -1 & 1 \\
SRC2 & SSTGLMA G064.5721+01.1438 & 19494526+2825007 & 19h49m45.2604s +28d25m00.6924s & 1 & 0 & 0 & 0 & -1 & 1 \\
SRC3 & SSTGLMA G064.5808+01.1670 & 19494100+2826099 & 19h49m41.0153s +28d26m09.9204s & 1 & 0 & 0 & 0 & -1 & 1 \\
SRC4 & SSTGLMA G064.5925+02.9442 & 19424110+2920138 & 19h42m41.1101s +29d20m13.8336s & 1 & 0 & 0 & 0 & -1 & 1 \\
SRC5 & SSTGLMA G064.6126+01.2039 & 19493675+2828560 & 19h49m36.7589s +28d28m56.0388s & 1 & 0 & 0 & 0 & -1 & 1 \\
SRC6 & SSTGLMA G064.6262+01.0862 & 19500631+2826029 & 19h50m06.311s +28d26m02.9688s & 1 & 0 & 0 & 0 & -1 & 1 \\
SRC7 & SSTGLMA G064.6397+01.2037 & 19494054+2830196 & 19h49m40.5398s +28d30m19.8144s & 1 & 0 & 0 & 0 & -1 & 2 \\
SRC8 & SSTGLMA G064.6532+02.0310 & 19462692+2856045 & 19h46m26.9479s +28d56m04.6608s & 1 & 0 & 0 & 0 & -1 & 1 \\
SRC9 & SSTGLMA G064.6574+02.9303 & 19425328+2923114 & 19h42m53.2805s +29d23m11.598s & 1 & 0 & 0 & 0 & -1 & 2 \\
\enddata
\tablenotetext{a}{A ``1" in these YSO columns means the object was identified as a YSO based on this dataset (see Section~\ref{iding}).}
\tablenotetext{b}{The cluster number of the YSO is indicated in this column (see Table~\ref{tab:clusters}). A ``-1" indicates that there was no cluster affiliation identified.}
\tablenotetext{c}{The numerical YSO class, as described in Section~\ref{ysoevol}.}
\tablecomments{Table \ref{tab:photdescription} is published 
in its entirety in the machine readable format.  A subset of the column identifiers are 
shown here for guidance regarding its form and primary content.}
\end{deluxetable*}

Table~\ref{tab:IRphot} lists a selection of the column identifiers of the photometry table for the full list of identified YSOs.  
This table is a stacked table of subsets from the standard IPAC tables available on the IPAC website, for the GLIMPSE360, SMOG, and {\it WISE} datasets.  
It is available in its entirety in the online version of the paper.

\begin{figure*}
\epsscale{1.15}
\plotone{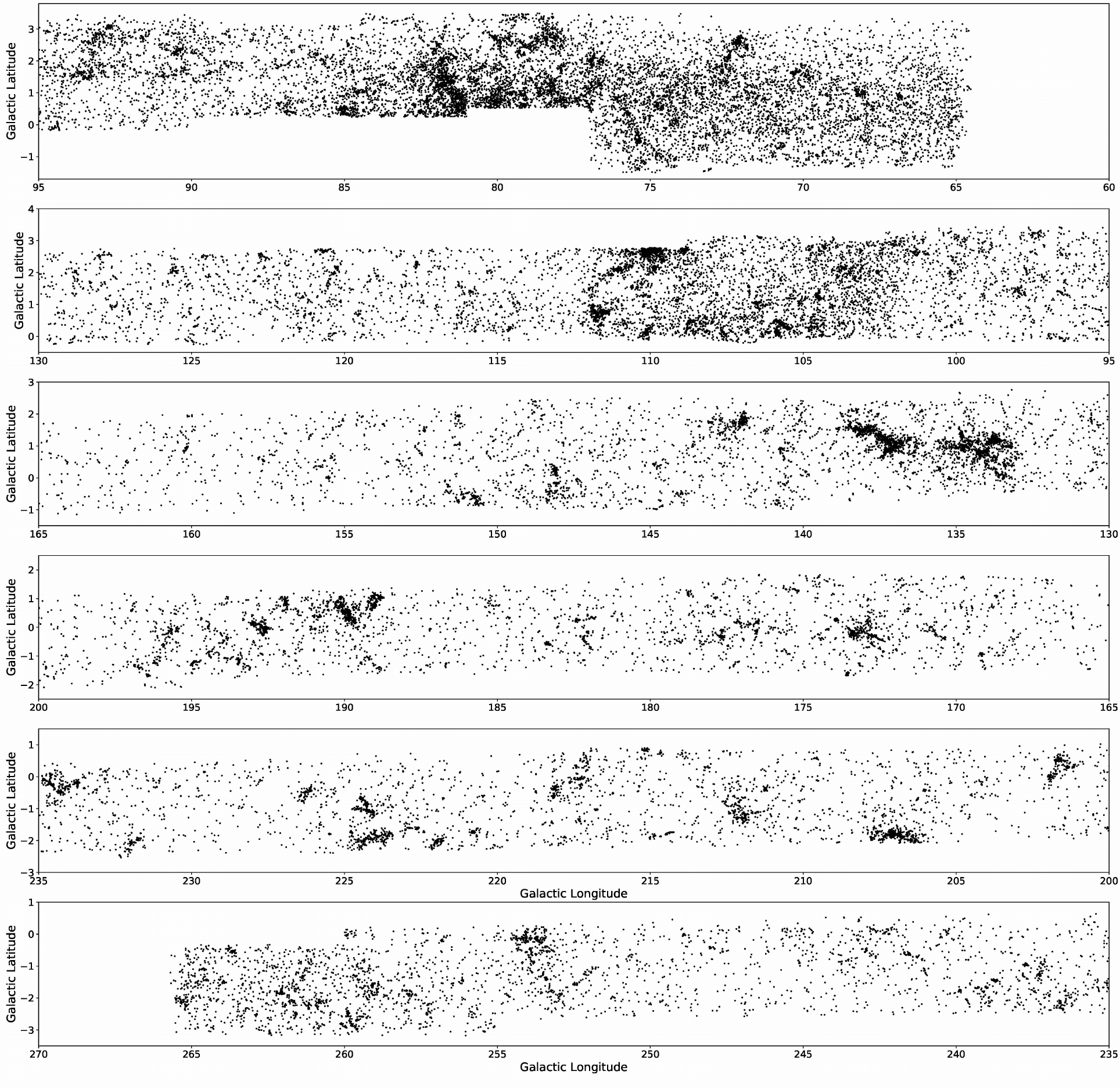}
\caption{ Spatial distribution of all identified YSOs in the SFOG catalog from the combined 2MASS-IRAC, IRAC+{\it WISE}, and {\it WISE} photometric 
selection criteria and also incorporating the SMOG YSOs.  
\label{fig_sda}}
\end{figure*}

The online SIMBAD catalog was searched for matches to the SFOG catalog within 2\arcsec, with 7,343 YSOs ($\sim$17\%) found to have a previous identification.  
Table~\ref{tab:photdescription} lists the object identifiers, positions, and by which selection method the YSO was identified.   Table~\ref{tab:Simbad} lists the YSOs that were found to have matches in the SIMBAD catalog along with their alternate identifications and object types. 
Figure~\ref{fig_sda} shows the spatial distribution of the identified YSOs over the outer plane of the Milky Way.

\begin{deluxetable*}{cccccchhh}
\tabletypesize{\scriptsize}
\tablecaption{SFOG Field Young Stellar Objects: SIMBAD matches within 2 arcsec  \label{tab:Simbad} }
\tablehead{ 
\colhead{SF} & \colhead{Glimpse}  & \colhead{RA}  & \colhead{Dec} & \colhead{SIMBAD}   & \colhead{OTYPE}  & \nocolhead{SIMBAD RA} & \nocolhead{SIMBAD Dec}  & \nocolhead{IDS}     \\[-0.3cm]
\colhead{ID}  & \colhead{ID}  & \colhead{(J2000)} & \colhead{(J2000)} & \colhead{ID}  & \colhead{}  & \nocolhead{(J2000)}  & \nocolhead{(J2000)}  &\nocolhead{} 
}
\startdata
SRC9 & SSTGLMA G064.6574+02.9303 & 295.722002 & 29.386555 & '2MASS J19425328+2923114' & AGN\_Candidate & 19 42 53.285 & +29 23 11.49 & '2MASS J19425328+2923114|1RXS J194254.4+292304|WISE J194253.28+292311.6' \\
SRC28 & SSTGLMA G064.8321+01.3951 & 297.34215 & 28.768332 & 'HBHA 2703-38' & Em* & 19 49 22.1216 & +28 46 05.975 & 'HBHA 2703-38|2MASS J19492211+2846059|[KW97] 40-1|Gaia DR2 2028700520637595776' \\
SRC31 & SSTGLMA G064.8407+02.9476 & 295.808845 & 29.554165 & 'V* V1279 Cyg' & Mira & 19 43 14.1288 & +29 33 14.814 & 'V* V1279 Cyg|CSV   8265|2MASS J19431412+2933146|SV* VV   217|Gaia DR2 2031986587236662784' \\
SRC35 & SSTGLMA G064.8492+00.3496 & 298.375033 & 28.249132 & '2MASS J19532999+2814568' & Candidate\_YSO & 19 53 29.997 & +28 14 56.82 & '2MASS J19532999+2814568|SSTGLMC G064.8492+00.3496|[RMB2008] G064.8492+00.3496' \\
SRC53 & SSTGLMA G064.9585+02.4134 & 296.408345 & 29.390237 & 'IRAS 19436+2916' & Star & 19 45 38.0039 & +29 23 24.681 & 'AKARI-IRC-V1 J1945380+292324|IRAS 19436+2916|2MASS J19453800+2923246|Gaia DR2 2031783796090088960' \\
SRC62 & SSTGLMA G064.9774+00.2589 & 298.538293 & 28.312292 & '2MASS J19540918+2818443' & Candidate\_YSO & 19 54 09.188 & +28 18 44.32 & '2MASS J19540918+2818443|SSTGLMC G064.9774+00.2589|[RMB2008] G064.9774+00.2589' \\
SRC67 & SSTGLMA G064.9890+00.2190 & 298.583948 & 28.301645 & 'IRAS 19523+2810'  & Star & 19 54 20.16 & +28 18 05.4 & 'AKARI-IRC-V1 J1954201+281805|IRAS 19523+2810' \\
SRC69 & SSTGLMA G064.9949+00.2704 & 298.537384 & 28.333228 & '2MASS J19540897+2819594' & Candidate\_YSO & 19 54 08.975 & +28 19 59.49 & '2MASS J19540897+2819594|SSTGLMC G064.9949+00.2703|[RMB2008] G064.9949+00.2703' \\
SRC73 & SSTGLMA G065.0172-00.0636 & 298.87522 & 28.179735 & 'OH  65.0   -0.1' & OH\/IR & 19 55 30.05 & +28 10 47.0 & 'OH  65.0   -0.1|AKARI-IRC-V1 J1955300+281046|IRAS 19534+2802|2MASS J19553005+2810469|MSX6C G065.0168-00.0636|SSTGLMC G065.0172-00.0637|WISE J195530.05+281047.0|[TVH89] 420' \\
SRC77 & SSTGLMA G065.0207+02.7296 & 296.129139 & 29.601918 & 'IRAS 19425+2928' & Star & 19 44 31.0007 & +29 36 06.694 & 'AKARI-IRC-V1 J1944310+293606|IRAS 19425+2928|2MASS J19443099+2936066|Gaia DR2 2031981192812098944' \\
\enddata
\tablecomments{Table \ref{tab:Simbad} is published 
in its entirety in the machine readable format.  A portion is
shown here for guidance regarding its form and content.}
\end{deluxetable*}

\subsubsection{Evolutionary Classification}\label{ysoevol}

Young stars evolve through a number of broad stages from the embedded core phase, through the protostellar phase where stellar accretion is still dominant, 
to the circumstellar disk phase where the envelope has dissipated and processing of disk material is ongoing, to the weak disk regime where the disk has 
dissipated and planets will have formed.  

A general evolutionary classification of YSOs in the SFOG field was carried out by measuring the slope, $\alpha$, of the spectral energy distribution (SED) across the mid-IR 
bandpasses \citep{lad84}.  In our SFOG catalog, many sources are detected with only the two shorter IRAC bands, while a smaller number also have the longer wavelength 
{\it WISE} data.  The SED slope was calculated based on the available photometric bands longward of the 2MASS K-band inclusive,  for each source by performing a least squares polynomial fit to the data. 

Protostellar objects (Class 0 and I) have a rising slope, $\alpha > 0$;   Class II sources are characterized by decreasing slopes between $-1.6 < \alpha < 0$, 
while Class IIa sources lack optically thick emission from a disk and possess decreasing slopes $\alpha < -1.6$, consistent with a weak emission above a stellar photosphere.  Truly diskless Class III objects show slopes between $-2.7 < \alpha < -2.0$  \citep{lad, lad06}. 
Many previous publications using \Sp\ data combine these two categories as Class III objects \citep[\eg][]{winston2019,saral,Hora2009}. 

As we do not have ancillary data, such as X-ray observations, to separate young  completely diskless Class III YSOs from field stars, we  cannot reliably differentiate between the weak disk bearing  Class IIa YSOs identified here and truly mid-IR diskless Class III YSOs \citep{win11, win18}. For this reason, we list the class of all objects with slope $\alpha < -1.6$ as being Class IIa/III in this paper. 

The identified YSOs were assigned classes as follows:  10,476 Class 0/I,  29,604 Class II sources, and 7,325 weak emission Class IIa/III stars.   
Further, some of the {\it WISE} identified sources were classified from the color-color diagram as candidate transition disk sources. This classification was not used in the subsequent analysis, with only the classification based on SED slope reported.    

\begin{figure*}
\epsscale{1.15}
\plotone{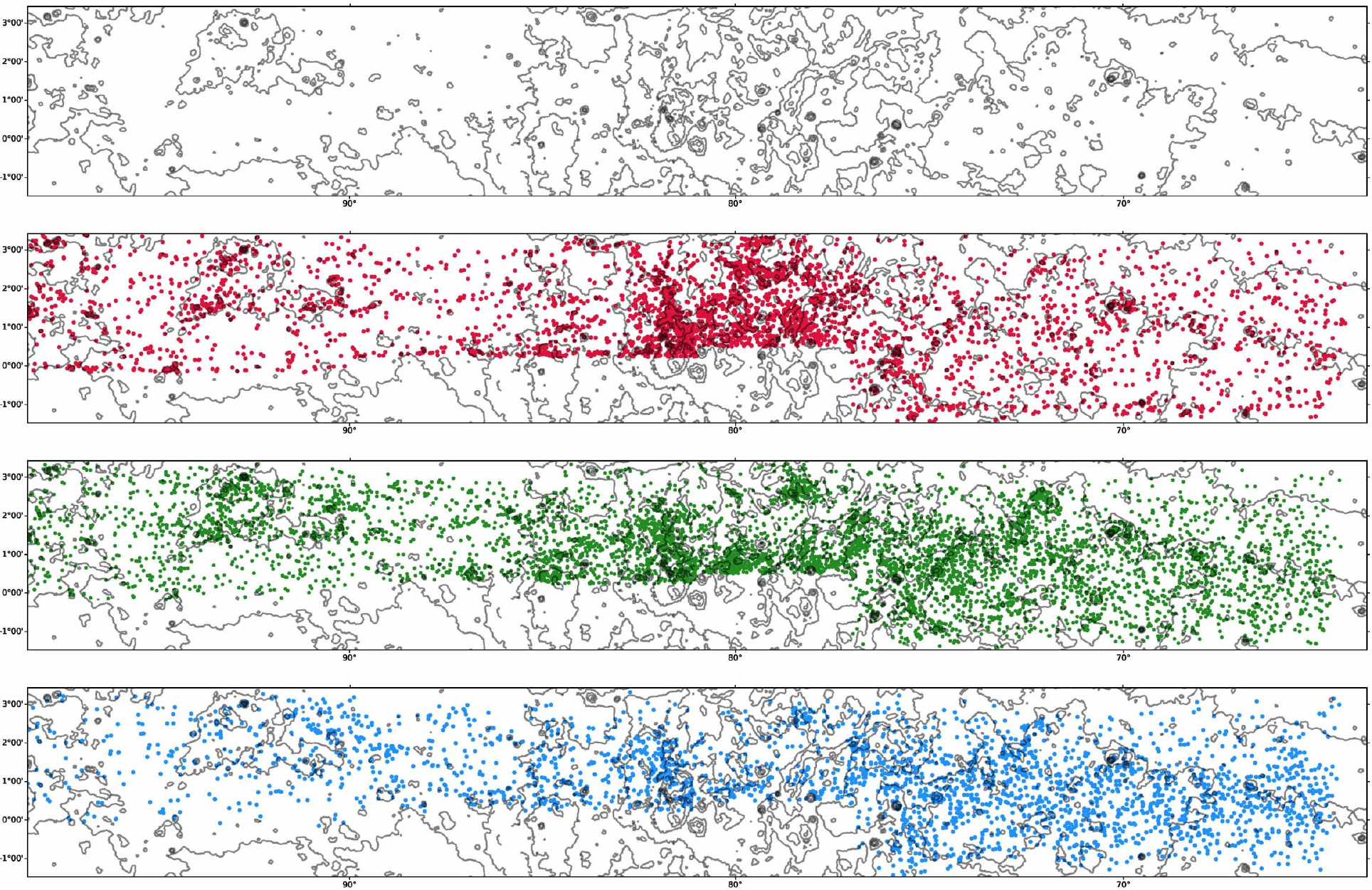}
\caption{ Spatial distribution by evolutionary classification of the identified YSOs in the SFOG catalog. Plotted in each panel are contours of the 100~\micron\ {\it IRAS} IRIS image \citep{miville2005} as an indication of the dust distribution along the outer Galactic plane.
Class I objects are plotted in red  in the second panel, Class II in green  in the third panel, and Class IIa/III in cyan points in the fourth panel.   
This figure shows only one-sixth of the entire field, the online version contains an electronic figure set containing all six panels covering the full SFOG field. 
\label{fig_sday}}
\end{figure*}

Figure~\ref{fig_sday} shows the spatial distribution of the identified YSOs by evolutionary classification over the plane.  
The classification of each YSO is listed in Table~\ref{tab:photdescription}.

\figsetstart
\figsetnum{5}
\figsettitle{SFOG YSO Spatial Distribution}

\figsetgrpstart
\figsetgrpnum{5.1}
\figsetgrptitle{60-95}
\figsetplot{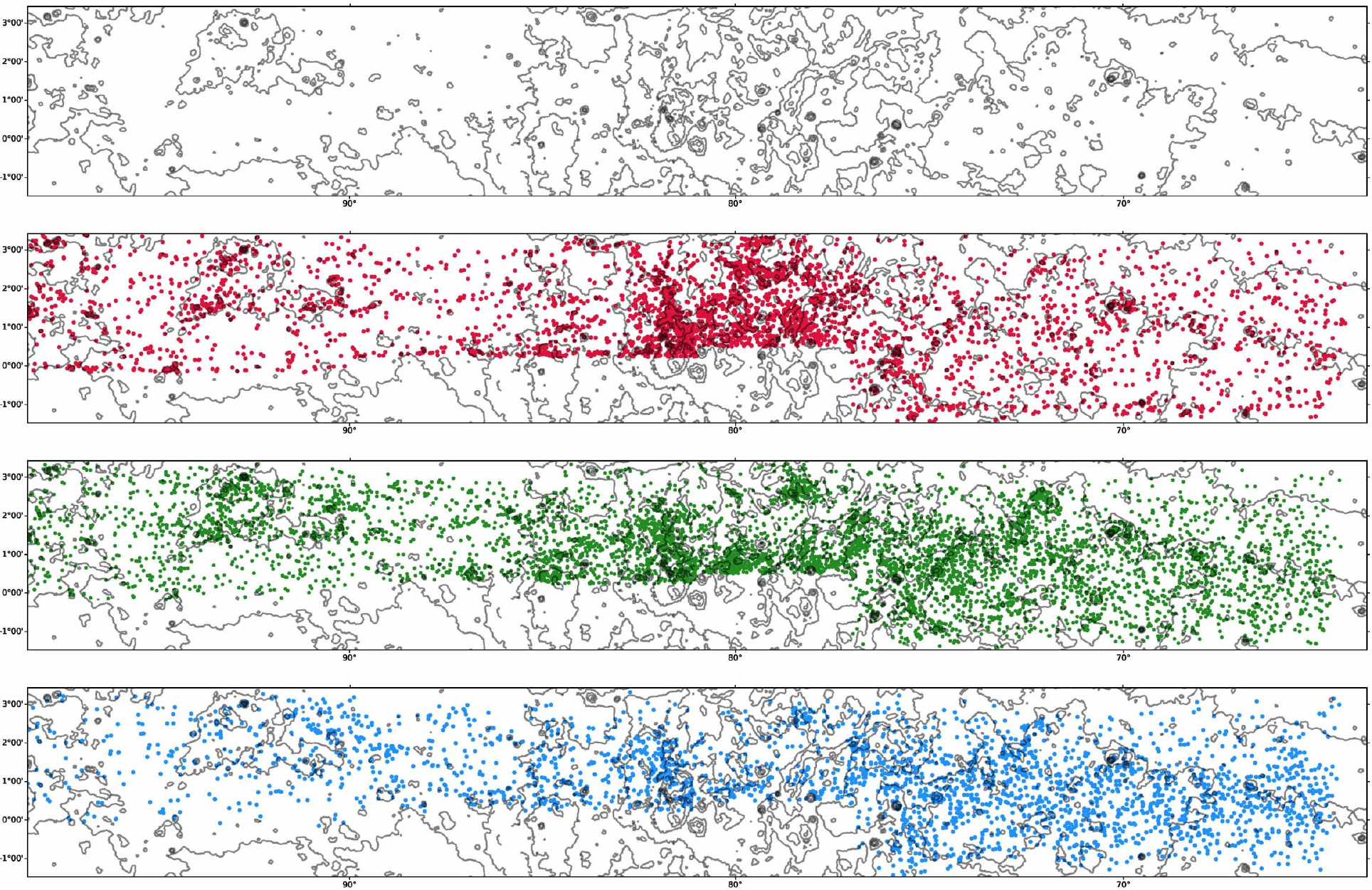}
\figsetgrpnote{Spatial distribution by evolutionary classification of the identified YSOs in the SFOG catalog. Plotted in each panel are contours of the 100~\micron\ {\it IRAS} IRIS image \citep{miville2005} as an indication of the dust distribution along the outer Galactic plane.
Class I objects are plotted in red in the second panel, Class II in green in the third panel, and Class IIa/III in cyan points in the fourth panel }
\figsetgrpend

\figsetend

\renewcommand{\arraystretch}{0.9}
\begin{deluxetable*}{rlrrrrl}
\tablecolumns{7}
\tabletypesize{\scriptsize}
\tablecaption{SFOG Young Stellar Object Cluster Descriptions }
\tablehead{ 
\colhead{Cluster} & \colhead{Cluster} & \colhead{\# of}  & \colhead{Central} & \colhead{Central}   & \colhead{Circular}  & \colhead{Associated}  \\[-0.3cm]
\colhead{Number} & \colhead{Name}  & \colhead{YSOs}  & \colhead{RA} & \colhead{Dec}   & \colhead{Radius}   & \colhead{Region}     \\[-0.3cm]
\colhead{}  & \colhead{} & \colhead{}  & \colhead{(J2000)} & \colhead{(J2000)}  & \colhead{deg}    & \colhead{} 
}
\startdata
104 &  G081.55+1.11  & 1177 & 309.049 & 42.535 & 1.200 & Cygnus-X\\
237 &  G133.95+1.00 & 730 & 36.725 & 61.818 & 0.808 & W3/W4\\
189 &  G109.96+2.63  & 704 & 343.771 & 62.535 & 0.826 & S 155\\
257 &  G137.20+1.07  & 632 & 42.959 & 60.566 & 1.035 & W5\\
357 &  G189.93+0.50 & 555 & 92.373 & 20.589 & 0.597 & S 252\\
527 &  G079.72+0.83  & 519 & 307.866 & 40.931 & 0.989 & Cygnus-X\\
389 &  G207.07$-$1.82  & 497 & 98.547 & 4.381 & 0.817 & Rosette; S 275, NGC 22\\
199 &  G111.67+0.73  & 444 & 348.712 & 61.478 & 0.688 & S 158,NGC 75\\
261 &  G138.08+1.53  & 422 & 44.976 & 60.561 & 0.987 & W5\\
523 &  G078.26+1.05  & 400 & 306.513 & 39.876 & 0.780 & Cygnus-X\\
375 &  G192.73$-$0.00 & 395 & 93.322 & 17.903 & 0.422 & S 254/258\\
360 &  G188.99+0.93  & 330 & 92.277 & 21.614 & 0.355 & S 247\\
572 &  G105.64+0.34  & 304 & 338.208 & 58.480 & 0.505 & S 138\\
417 &  G224.02$-$1.92  & 302 & 106.237 & -10.750 & 0.649 & Canis Majori\\
413 &  G224.31$-$0.93  & 288 & 107.243 & -10.548 & 0.429 & Canis Majori\\
429 &  G234.46$-$0.27  & 245 & 112.825 & -19.208 & 0.532 & NGC 2343\\
93 &  G077.04+0.98  & 237 & 305.682 & 38.838 & 0.773 & Cygnus-X\\
34 &  G072.08+2.57  & 217 & 300.588 & 35.577 & 0.334 & IRAS 20003+3524\\
96 &  G078.04+2.73  & 210 & 304.813 & 40.961 & 0.562 & Cygnus-X\\
585 &  G104.64+0.31  & 207 & 336.613 & 57.932 & 0.481 & IRAS 22246+5750\\
548 &  G108.73+0.32  & 205 & 343.558 & 59.915 & 0.483 & CO\tablenotemark{a}\\
316 &  G173.35$-$0.16  & 201 & 82.043 & 34.458 & 0.391 & S 234\\
453 &  G253.95$-$0.13  & 193 & 124.362 & -35.807 & 0.616 & GN 08.16.0\\
395 &  G212.06$-$1.14  & 192 & 101.444 & 0.260 & 0.419 & S 284\\
244 &  G134.79+0.99  & 185 & 38.357 & 61.500 & 0.781 & W4\\
194 &  G110.11+0.15  & 178 & 346.206 & 60.336 & 0.396 & IC 1470\\
123 &  G084.90+0.42  & 175 & 312.649 & 44.770 & 0.416 & Pelican\\
190 &  G108.89+2.68  & 170 & 341.671 & 62.100 & 0.322 & S 155\\
593 &  G104.52+1.27  & 169 & 335.447 & 58.689 & 0.303 & S 135\\
269 &  G141.97+1.77  & 165 & 51.812 & 58.761 & 0.281 & AFGL 490\\
408 &  G218.12$-$0.44  & 163 & 104.835 & -4.822 & 0.228 & S 287\\
94 &  G076.91+2.05  & 156 & 304.449 & 39.335 & 0.333 & Cygnus-X\\
566 &  G106.49+1.00  & 151 & 338.995 & 59.471 & 0.429 & IRAS 22344+5909\\
286 &  G148.10+0.20  & 134 & 58.939 & 53.815 & 0.254 & IRAS 03523+5343  \\
122 &  G084.51+1.03  & 130 & 311.629 & 44.851 & 0.383 & Pelican\\
66 &  G074.68+0.58  & 126 & 304.414 & 36.673 & 0.363 & S 104\\
191 &  G109.86+2.13  & 125 & 344.046 & 62.036 & 0.292 & S 155\\
387 &  G201.49+0.44  & 125 & 97.977 & 10.374 & 0.398 & IC 446\\
422 &  G226.28$-$0.54  & 123 & 108.550 & -12.103 & 0.338 & Canis Majori\\
77 &  G075.35$-$0.44  & 121 & 305.929 & 36.634 & 0.340 & DOBASHI 2314\\
420 &  G221.95$-$2.04  & 117 & 105.168 & -8.962 & 0.316 & DOBASHI 5043\\
536 &  G080.01+2.63  & 110 & 306.140 & 42.220 & 0.435 & Cygnus-X\\
163 &  G093.44+1.59  & 108 & 319.954 & 51.908 & 0.406 & NRAO 655\\
577 &  G103.70+2.15  & 107 & 333.188 & 58.967 & 0.315 & S 134\\
525 &  G079.00+2.47  & 105 & 305.550 & 41.299 & 0.405 & Cygnus-X\\
301 &  G150.68$-$0.70  & 103 & 61.189 & 51.446 & 0.421 & S 206\\
147 &  G090.47+2.30  & 101 & 315.908 & 50.231 & 0.312 & L 988\\
\enddata
\label{tab:clusters} 
\tablenotetext{a}{Molecular cloud with CO emission identified by \citet{2000ungerechts}}
\tablecomments{ The clusters are shown in order of decreasing number of YSO members. Only the clusters with more than 100 YSOs are shown here, Table \ref{tab:clusters} is published in the electronic version
in its entirety in the machine readable format. }
\end{deluxetable*}

\subsubsection{Sample Completeness}

The SFOG catalog is susceptible to the same difficulties in assessing incompleteness in the sample as the SMOG field, discussed in Paper I. 
In brief, the large spatial distribution and lack of data on the distances in the majority of the clusters means that estimates of the minimum 
mass YSOs detected are not possible.  
Incompleteness by evolutionary class is also difficult to quantify: because of the wavelength range of the 2MASS and IRAC data from 1.2 to 4.5 \micron, the survey is most sensitive to Class II pre-main sequence objects.  The {\it WISE} long wavelength bands at 12 and 22~\micron\ makes it more sensitive than GLIMPSE360 for the detection of embedded protostars, but it is hampered by lower resolution and sensitivity at the shorter wavelengths.

\section{Cluster Identification and Properties}\label{spatial}

\subsection{Identification with DBSCAN}
A cursory visual examination of the spatial distribution 
of the YSOs over the $\sim$600 deg$^2$ SFOG field shows evidence of clustering/clumping, as can be seen in Figure~\ref{fig_sda}.  
Following the method described in Paper I, we used the DBSCAN \citep{ester} density based algorithm to identify over-densities in the spatial distribution of the YSOs.
The values of the two free parameters, $\epsilon$ (the scaling size for clustering) and $MinPts$ (the minimum number of points required to define a dense region) were 
determined following the approach used by \citet{joncour} in the Taurus region.

\begin{figure*}
\plottwo{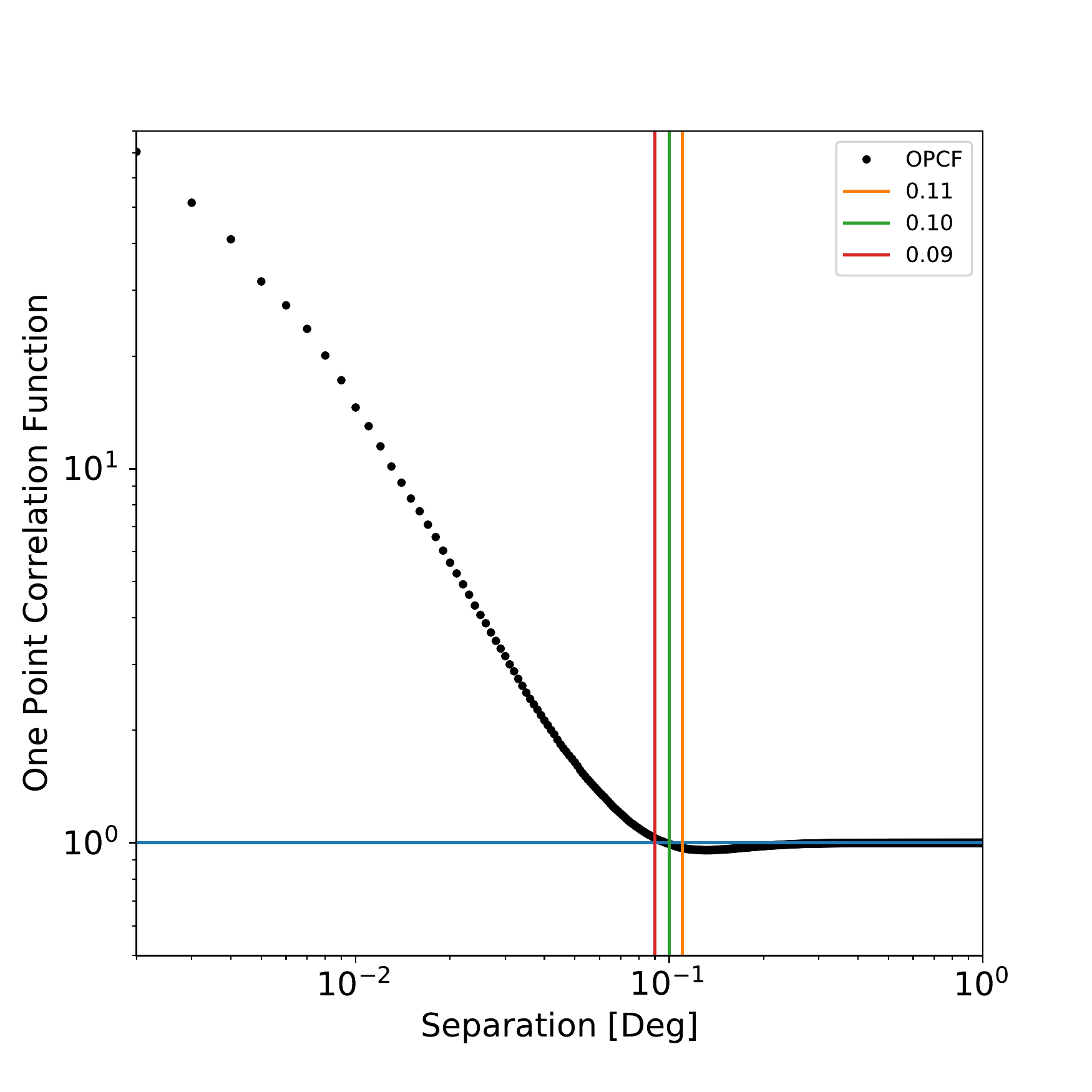}{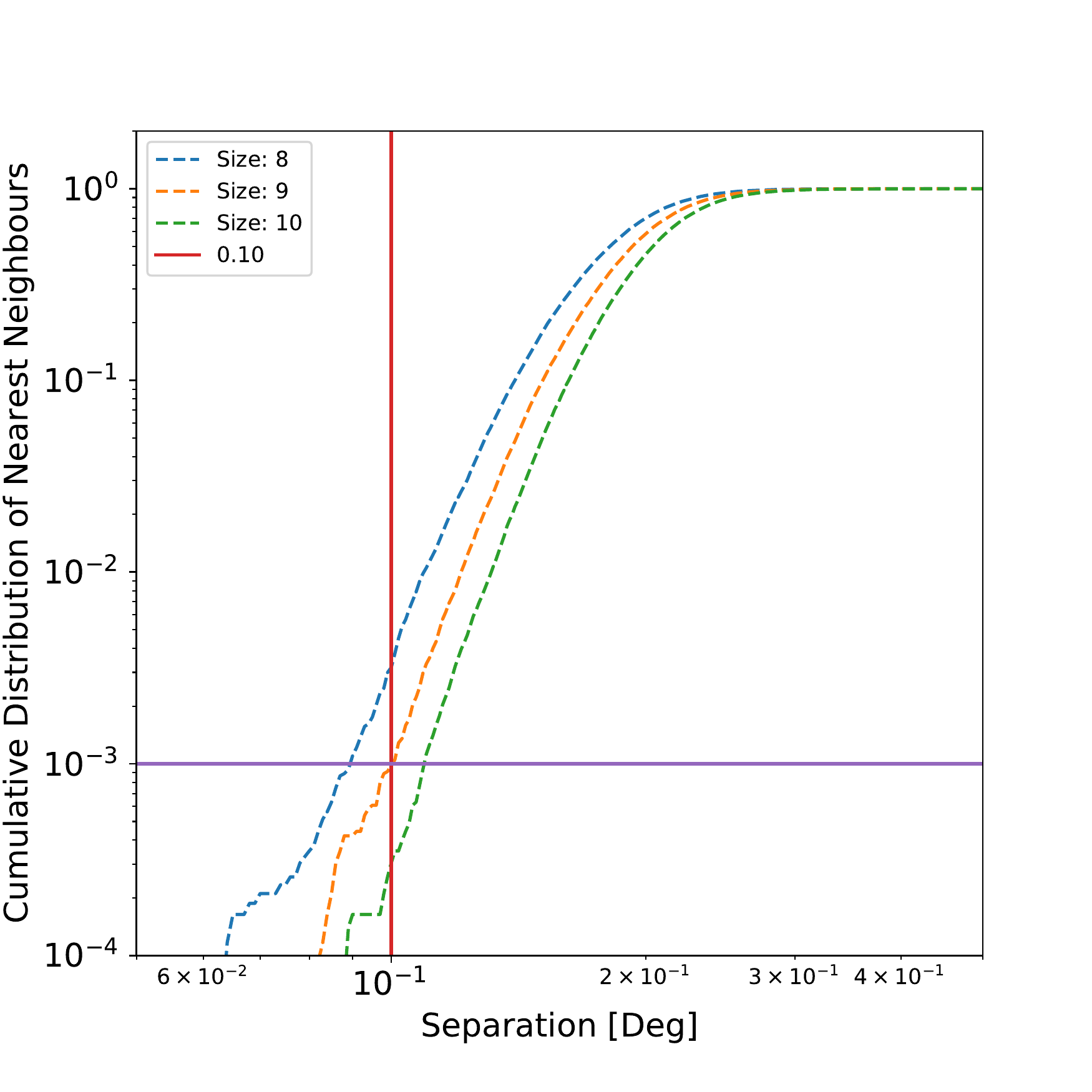}
\caption{  Selection of the criteria for the DBSCAN clustering algorithm.  
Left: One point correlation function showing the ratio of the YSO and random cumulative distributions, with a
crossing point at 0\fdg1 separation.
Right: Cumulative distribution of three nearest neighbour distributions at 8, 9, 10$^{th}$ nearest neighbours. 
The 10$^{-3}$ probability occurs at 0\fdg1 for a cluster density of 9 members.  }
\protect\label{fig_eps}
\end{figure*}

Figure~\ref{fig_eps}(a) shows the one point correlation function, which gives the ratio of the cumulative distributions of the identified YSOs and a random distribution over the same field. 
From this, the value of $\epsilon = $ 0\fdg1 was selected.  
Figure~\ref{fig_eps}(b) examines the cumulative distribution of three nearest neighbour distributions at 8, 9, 10$^{th}$ nearest neighbors for the random distribution showing that a probability 
of 0.001 occurs at 0\fdg1 for a minimum cluster size with 9 members.

\subsection{Cluster Properties}
\begin{figure*}
\epsscale{1.15}
\plotone{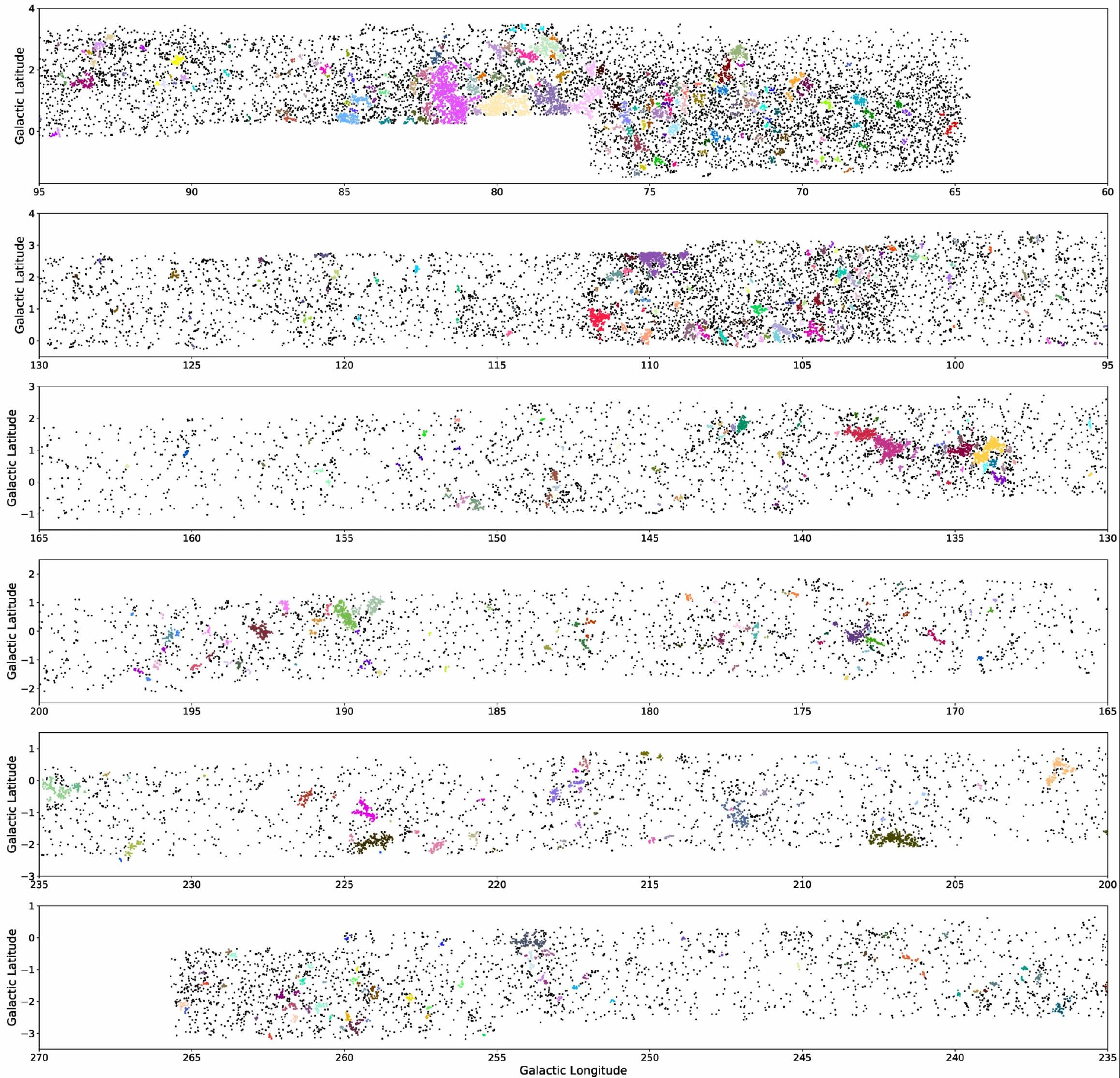}
\caption{ Spatial distribution of the identified YSOs showing the clusters identified by the DBScan method.  The sources in each cluster are color-coded.  
The black dots represent those YSOs not identified as belonging to a cluster.   }
\protect\label{fig_sdac}
\end{figure*}

With these values, 621 clusters were identified  in the SFOG field.  
Figure~\ref{fig_sdac} shows the spatial distribution of the identified clusters of YSOs over the plane.  The black YSOs are unclustered, with the clustered YSOs color-coded 
by identified cluster.  The clustered YSOs represented 54\% of the whole catalog, with 21,810 YSOs not clustered.  
The minimum cluster size was 5 members, the largest cluster identified contained 1,177 members.  The median cluster size is 17 members.
Of the 621 clusters, 133 have ten members or less, 25 clusters have 100-200 members, and 22 cluster have more than 200 identified members.

There is a possibility that some of the smaller clusters identified are not genuine stellar groupings, but chance over-densities in the field.  To determine the statistical likelihood that the identified groupings are true clusters, we repeated the DBSCAN analysis for 1000 iterations of randomly distributed points covering a similar area as the SFOG field.  Using the same values of $\epsilon = $ 0\fdg1 and $MinPts=9$, we varied the number of fake 'stars' as follows: 21810, 29604, 36000, 43000, 47405.  These are the numbers of unclustered YSOs, Class II YSOs, two intermediate values, and the total number of YSOs in the SFOG catalog.  This range was used to take into account the known clustering of objects, which reduces the overall density in the field, and thus reduces the likelihood of a 'fake' cluster being detected.  The average number of fake clusters detected for each run was:  0.04, 0.43, 1.9, 6.9, 13.8.  We can therefore assume that between 0-14 of the 621 clusters found by DBSCAN may be misidentified random alignments.  The 'fake' clusters in our random trials all have $\leq$10 members, and so if any clusters in our SFOG list in Table \ref{tab:clusters} are chance overdensities rather than true clusters, they would most likely be in the subset of 133 clusters with $\leq$10 YSOs.

There is  the further possibility that the unclustered YSOs are contaminants; foreground or background Galactic field dwarfs or AGB stars with either a high extinction or a small amount of dusty 
material surrounding them that leads to an excess of flux in the mid-IR or extragalactic sources.  Spectroscopic analysis would be necessary to secure the identification of 
all sources.

\begin{center}
\begin{figure*}
\includegraphics[width=.33\linewidth]{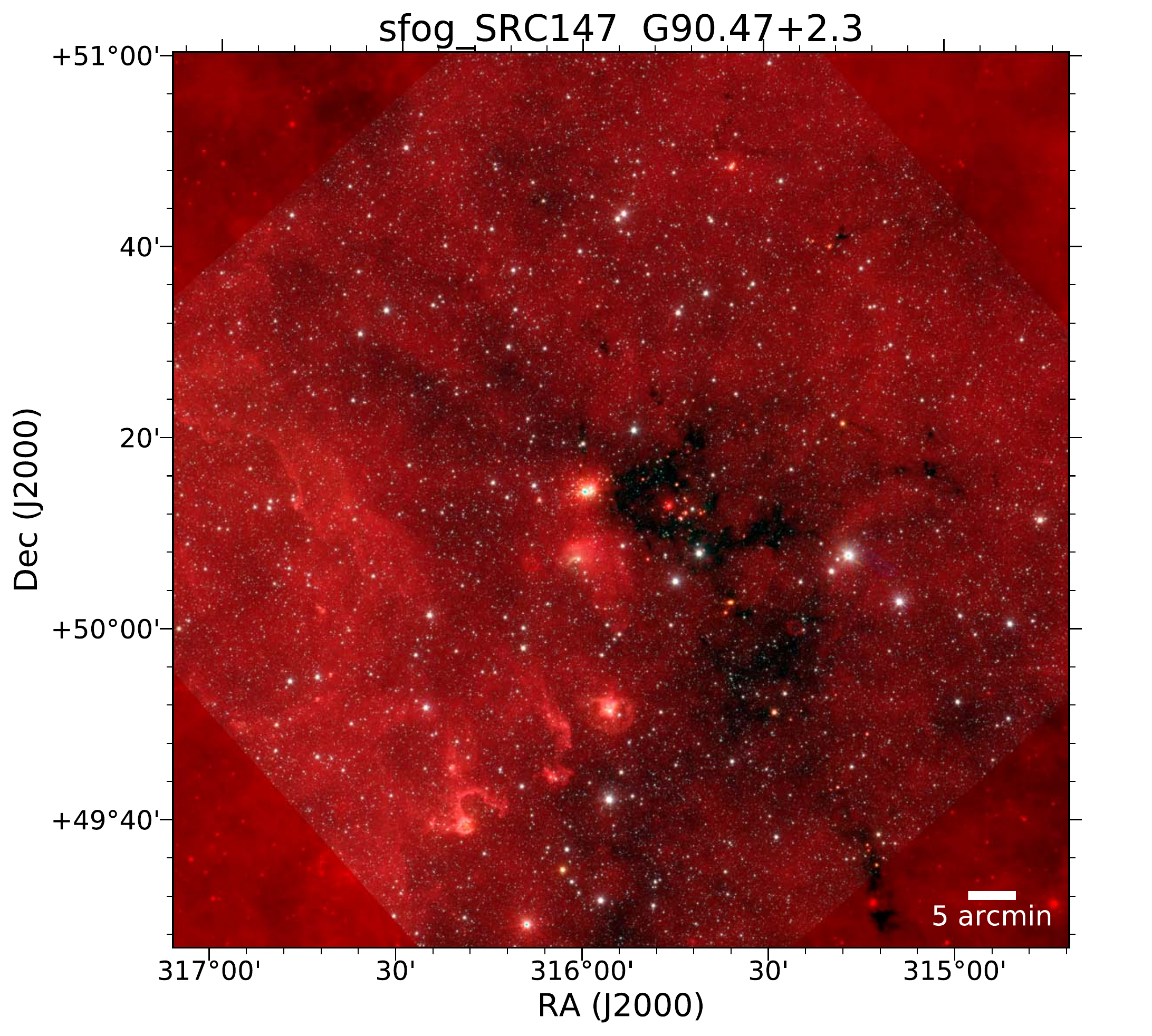}\includegraphics[width=.33\linewidth]{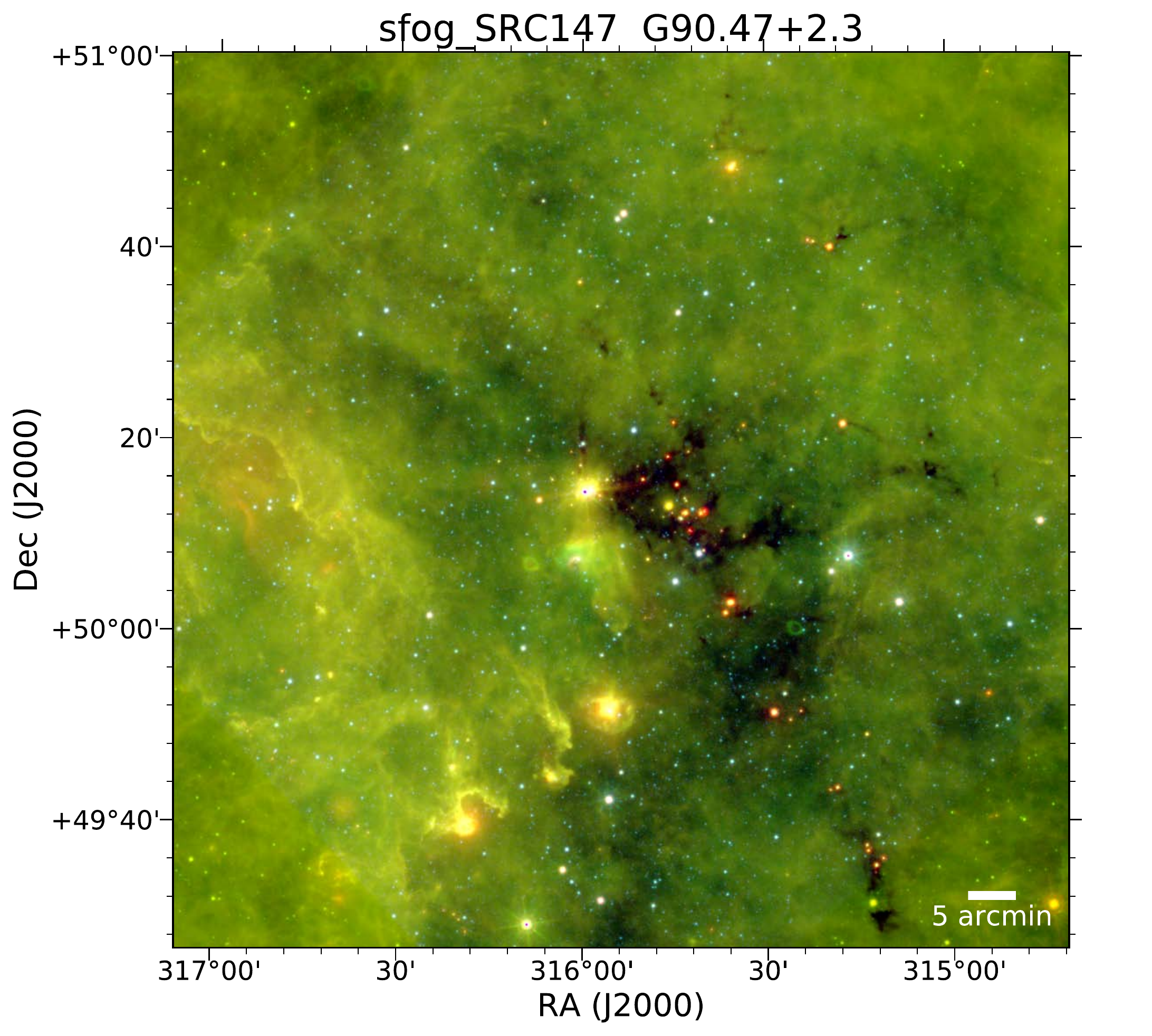}\includegraphics[width=.33\linewidth]{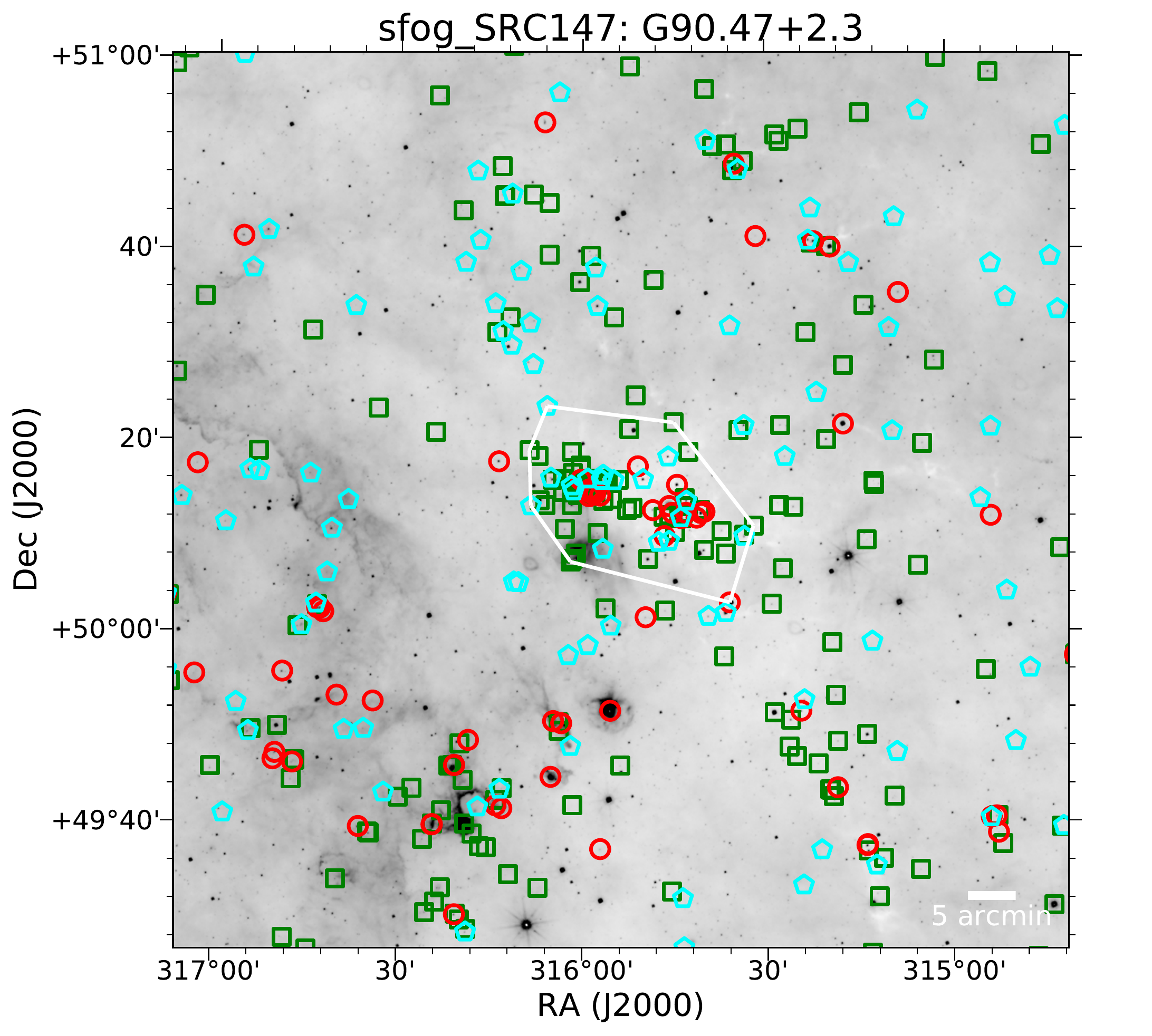}
\includegraphics[width=.33\linewidth]{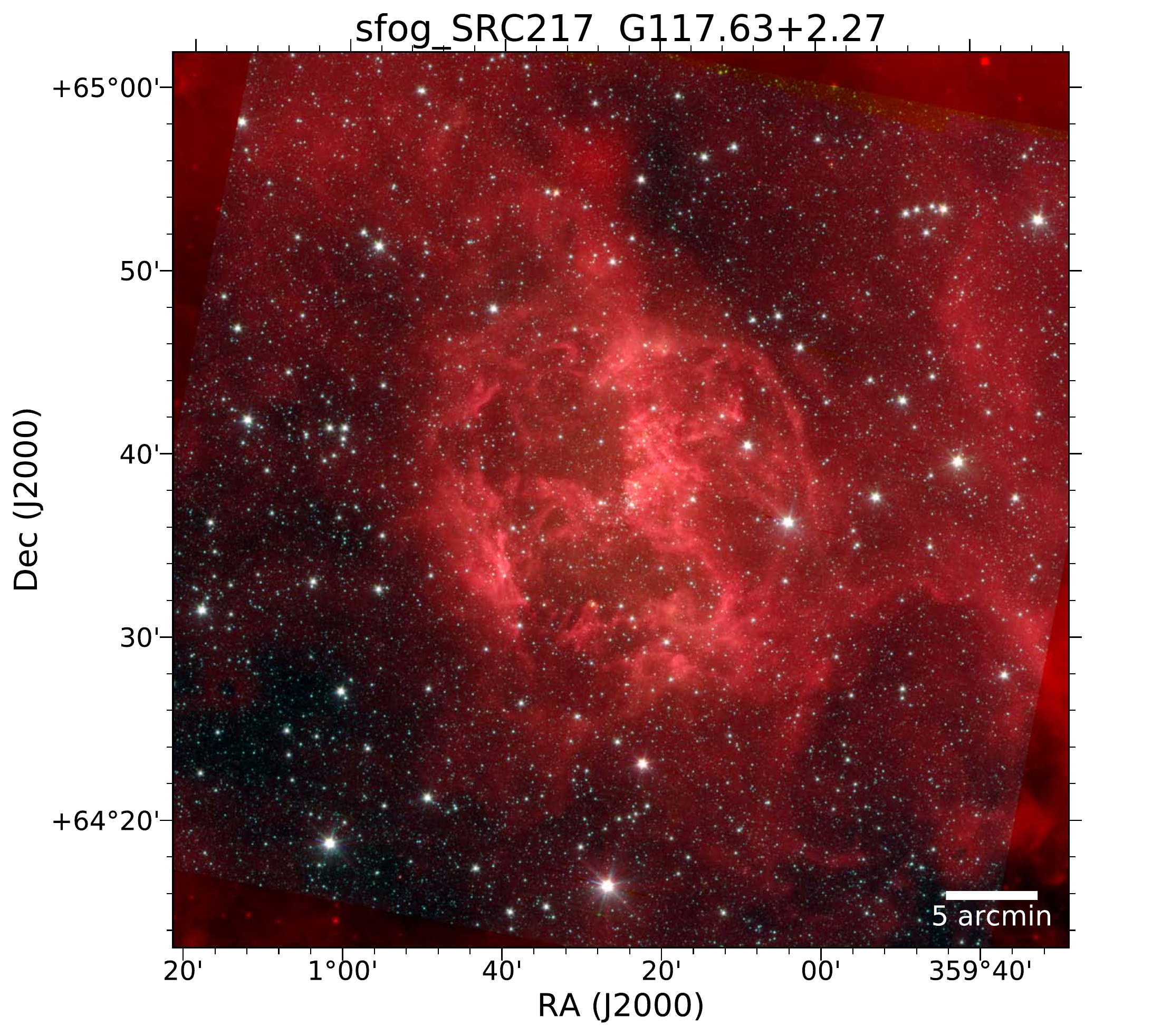}\includegraphics[width=.33\linewidth]{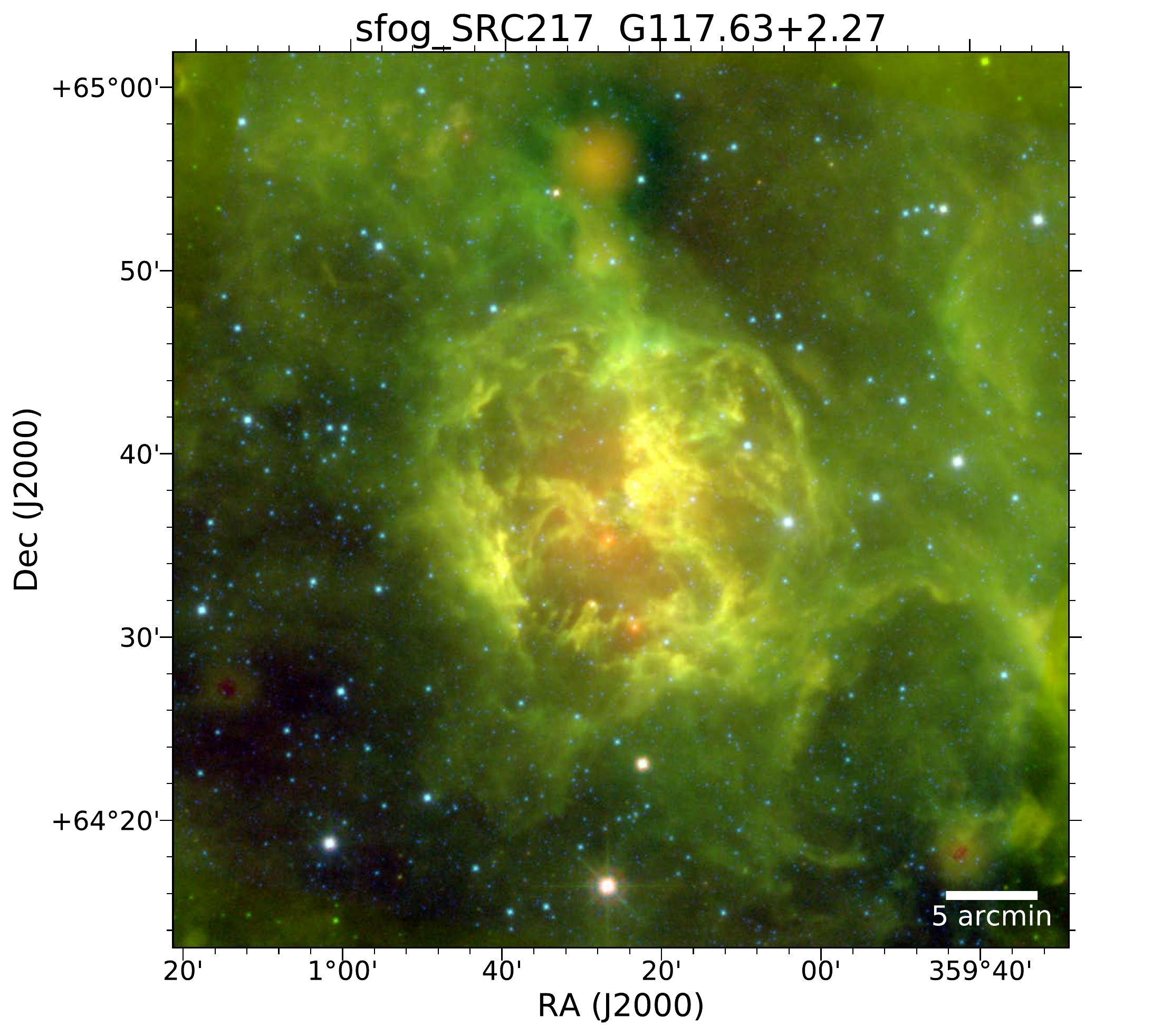}\includegraphics[width=.33\linewidth]{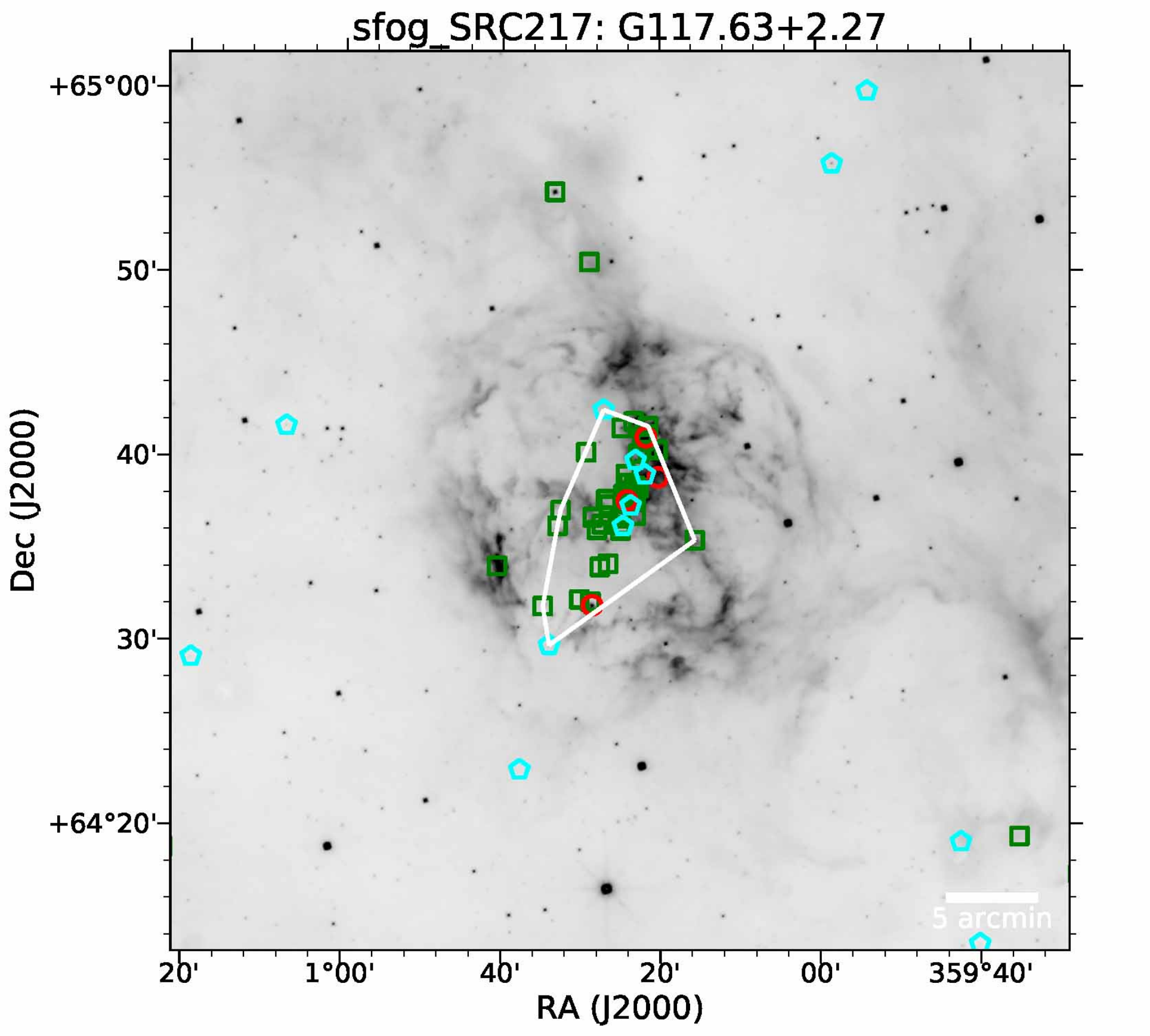}
\includegraphics[width=.33\linewidth]{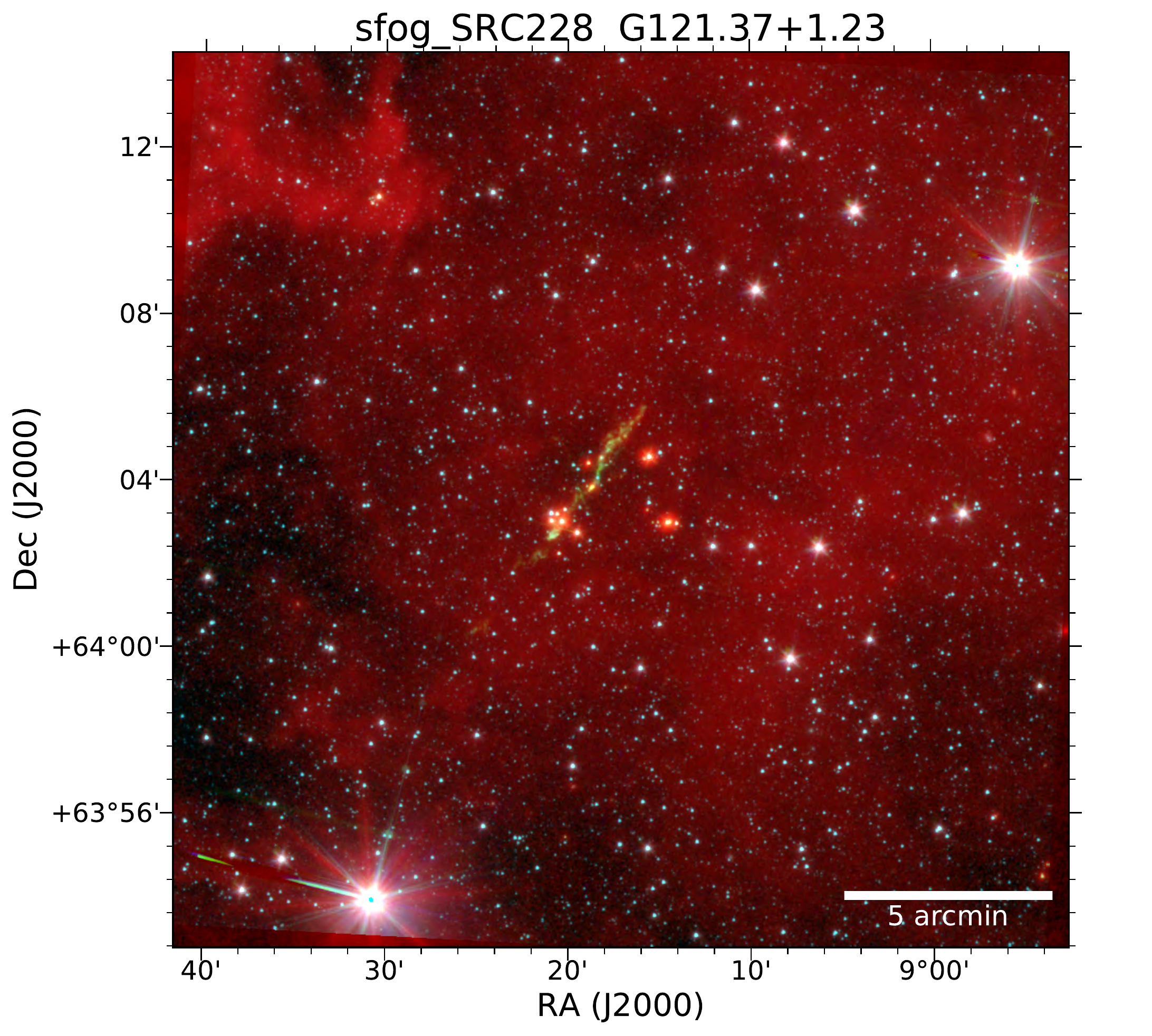}\includegraphics[width=.33\linewidth]{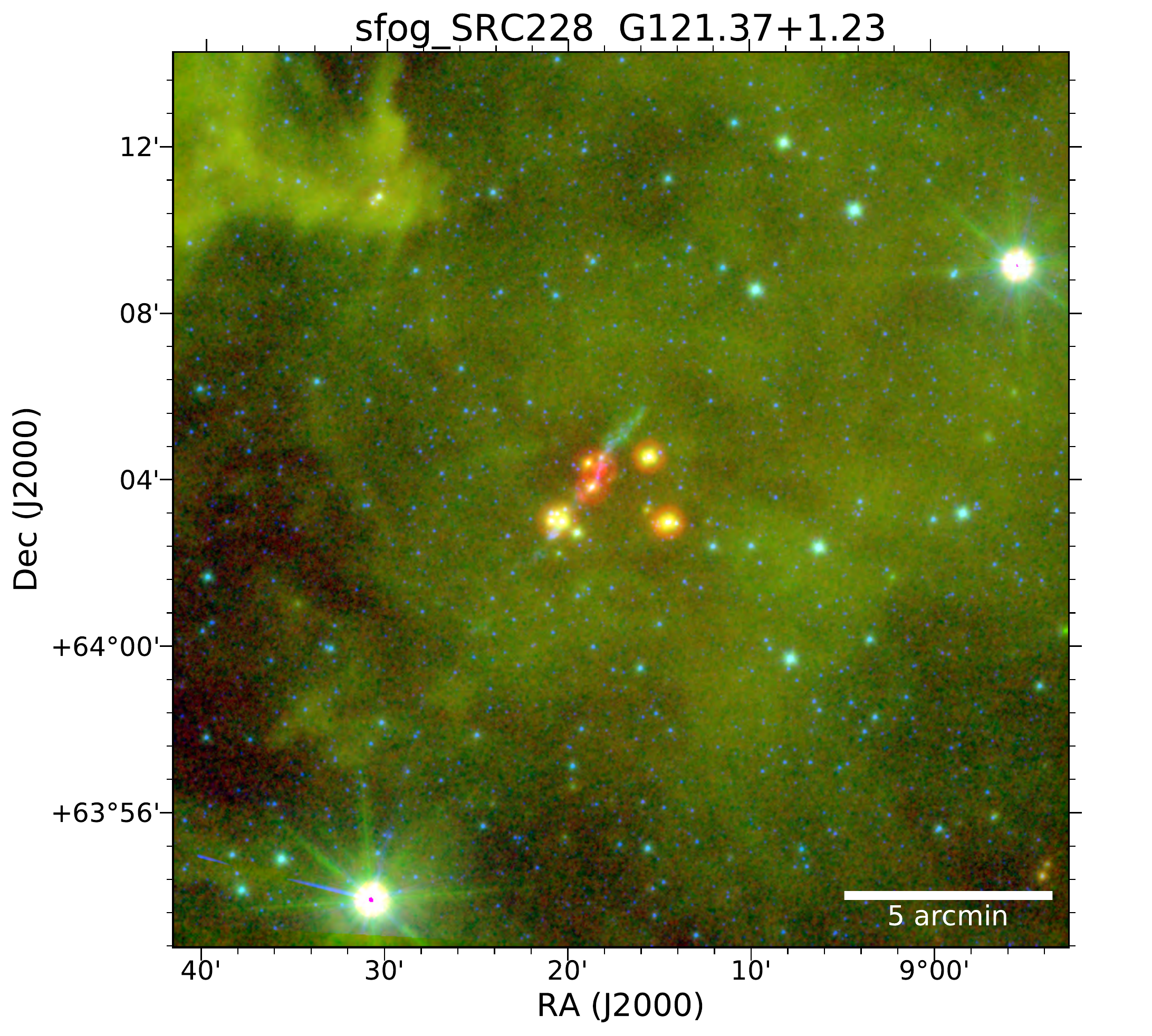}\includegraphics[width=.33\linewidth]{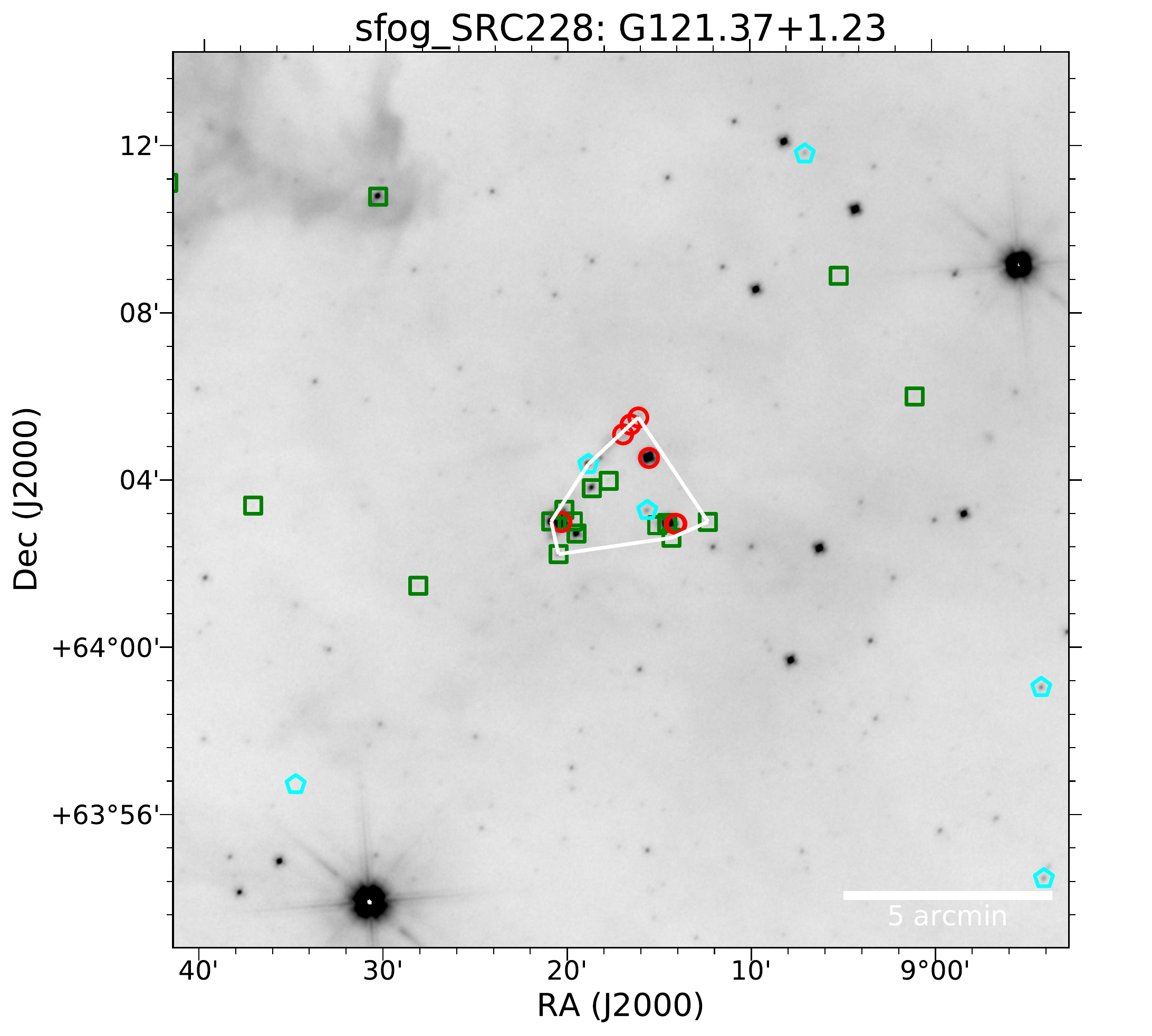}
\includegraphics[width=.33\linewidth]{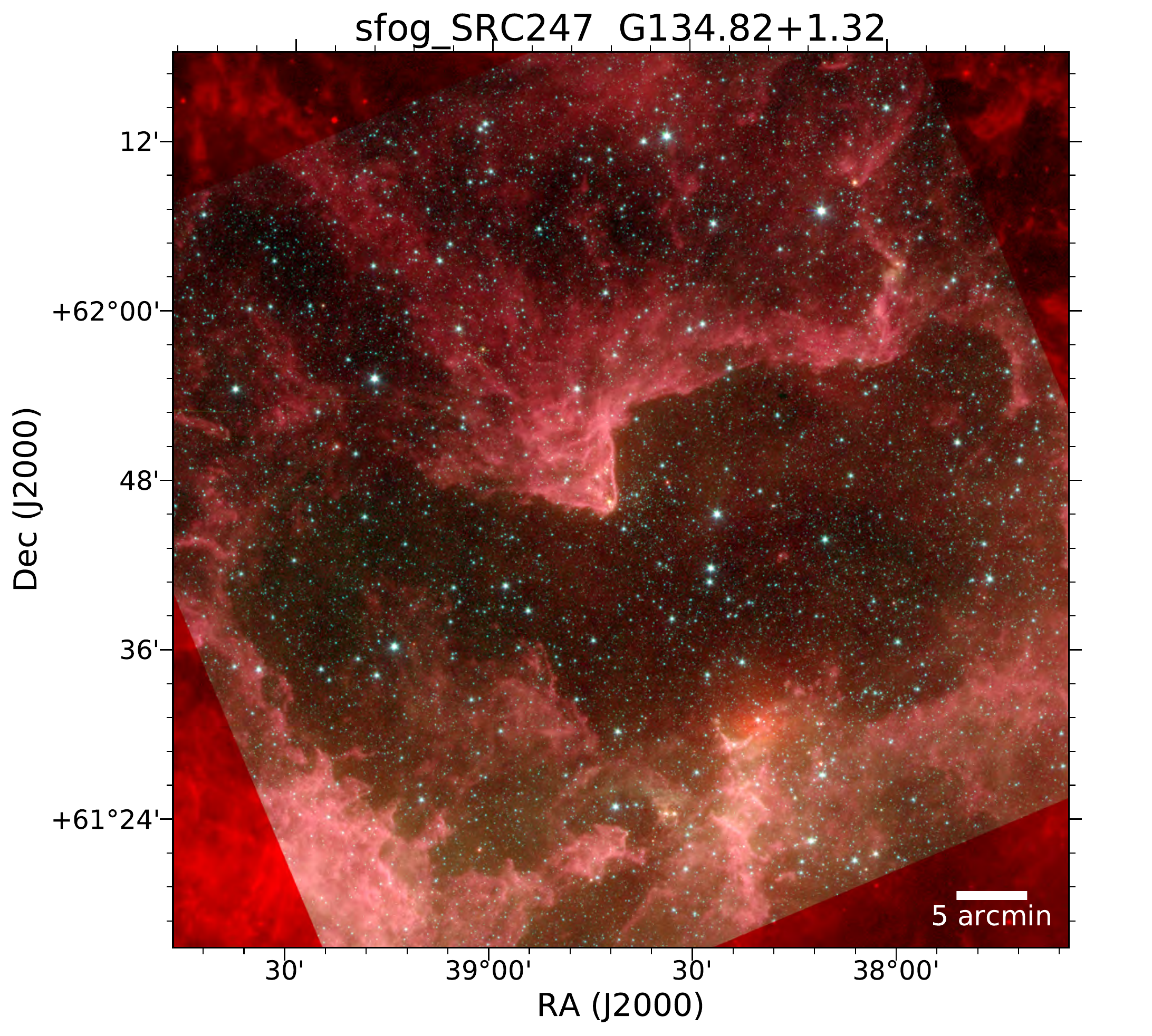}\includegraphics[width=.33\linewidth]{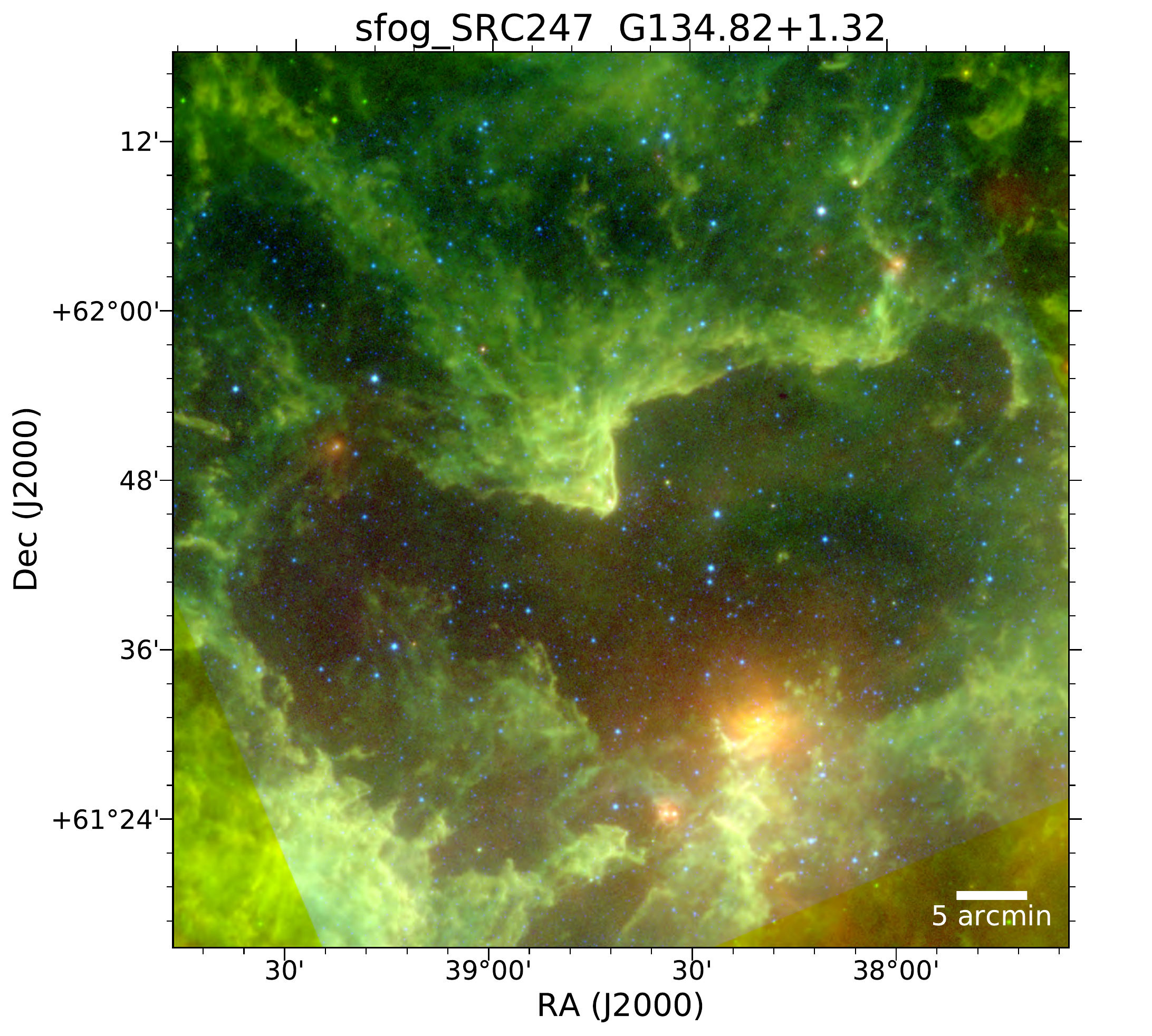}\includegraphics[width=.33\linewidth]{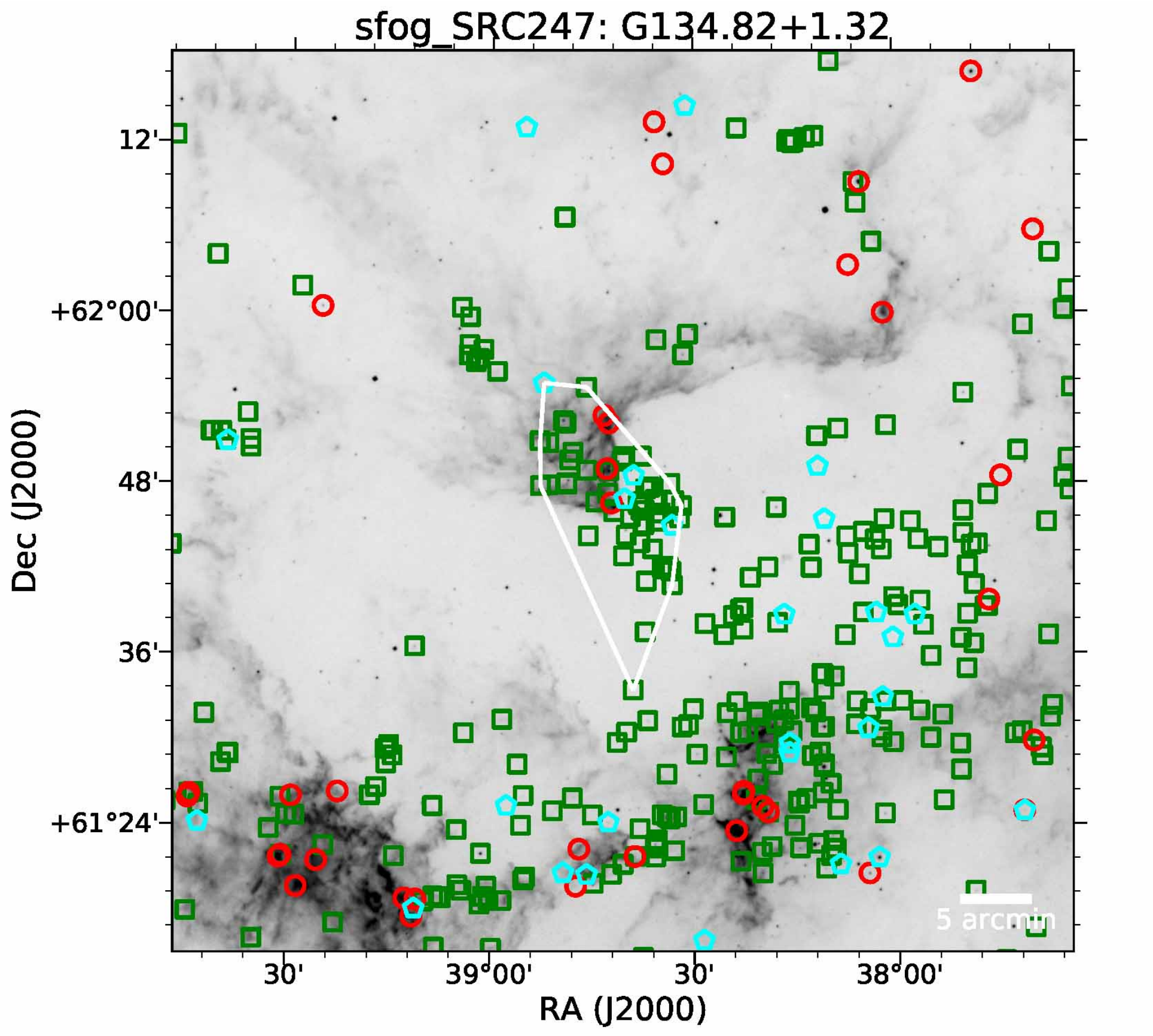}
\caption{ Four examples of clusters identified in the SFOG field: \#147, \#217, \#228, and \#247 from top to bottom. In the left column are 3-color images in {\it WISE} 12~\micron\ (red), IRAC 4.5~\micron\ (green), and IRAC 3.6~\micron\ (blue); the center column contains 3-color images with {\it WISE} 22~\micron\ (red), {\it WISE} 12~\micron\ (green), and IRAC 4.5~\micron\ (blue); and the right column shows the {\it WISE} 12~$\mu$m in reverse grayscale with the identified YSOs and the calculated convex hulls for each cluster overlaid. The symbols show the positions of the Class I (red circles), Class II (green squares), and  Class IIa/III (cyan pentagons) YSOs. All of the cluster images and associated FITS files are available from \citet{winston2020}.    }
\protect\label{fig_clusters1}
\end{figure*}
\end{center}

\begin{center}
\begin{figure*}
\includegraphics[width=.33\linewidth]{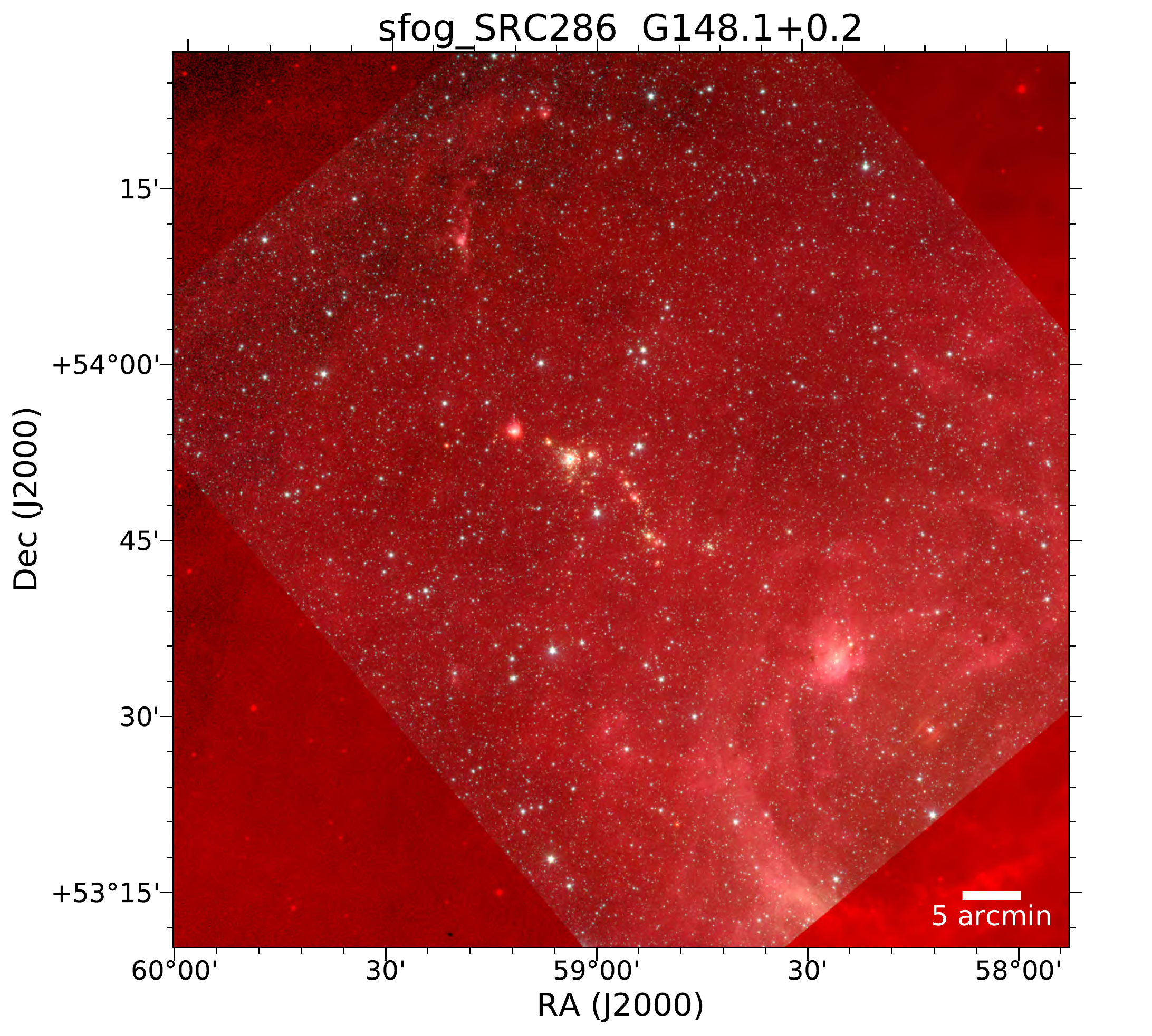}\includegraphics[width=.33\linewidth]{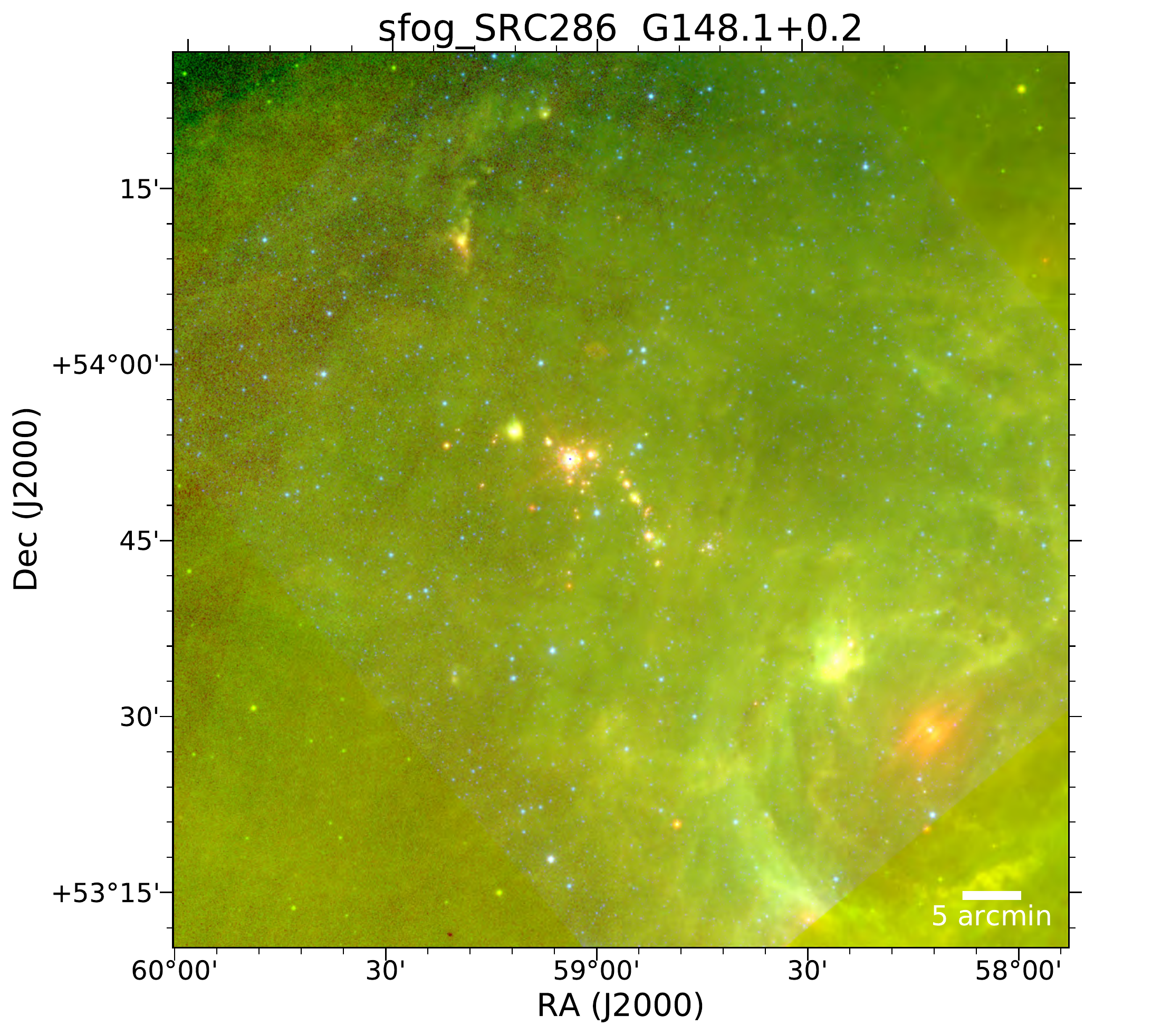}\includegraphics[width=.33\linewidth]{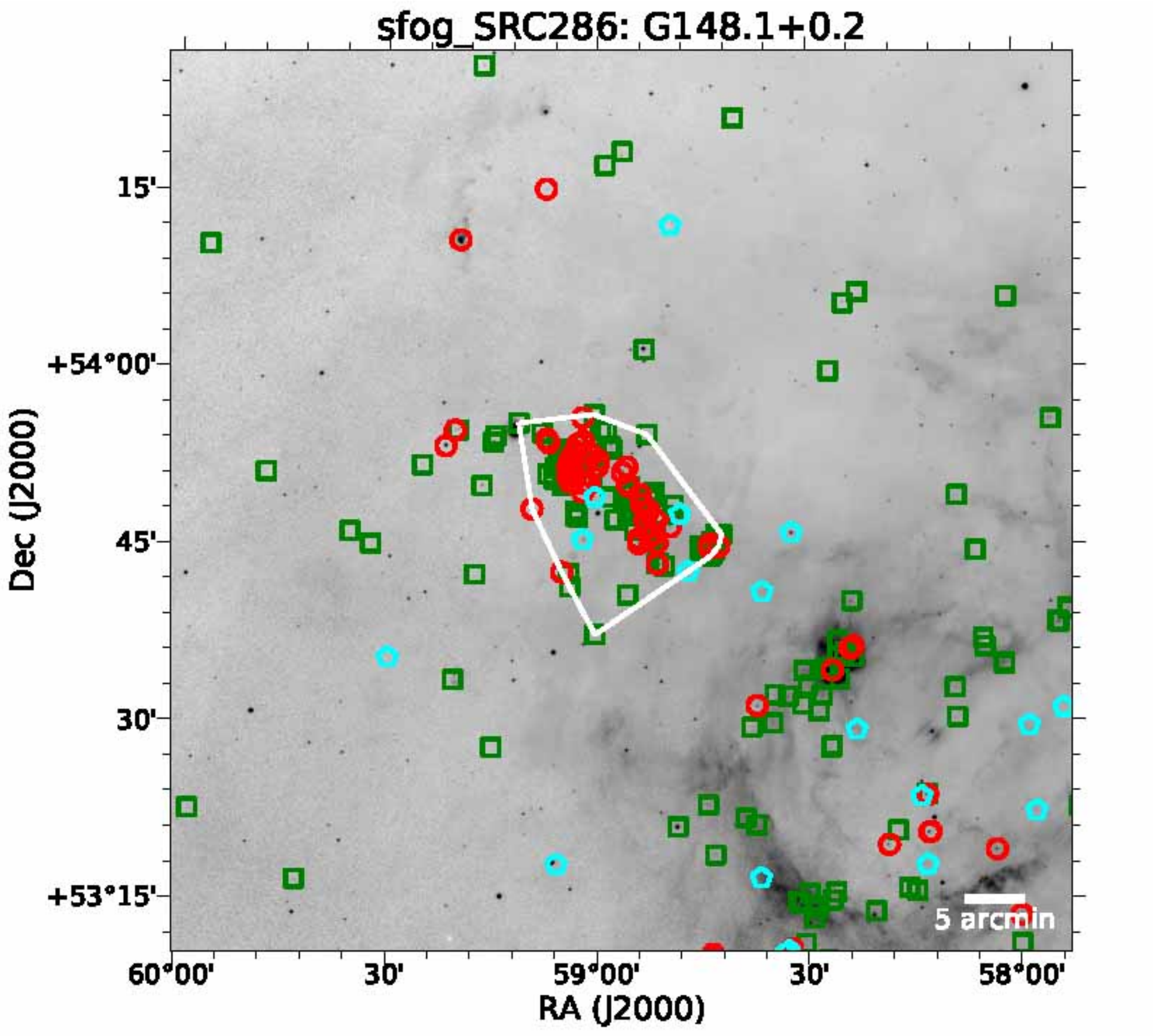}
\includegraphics[width=.33\linewidth]{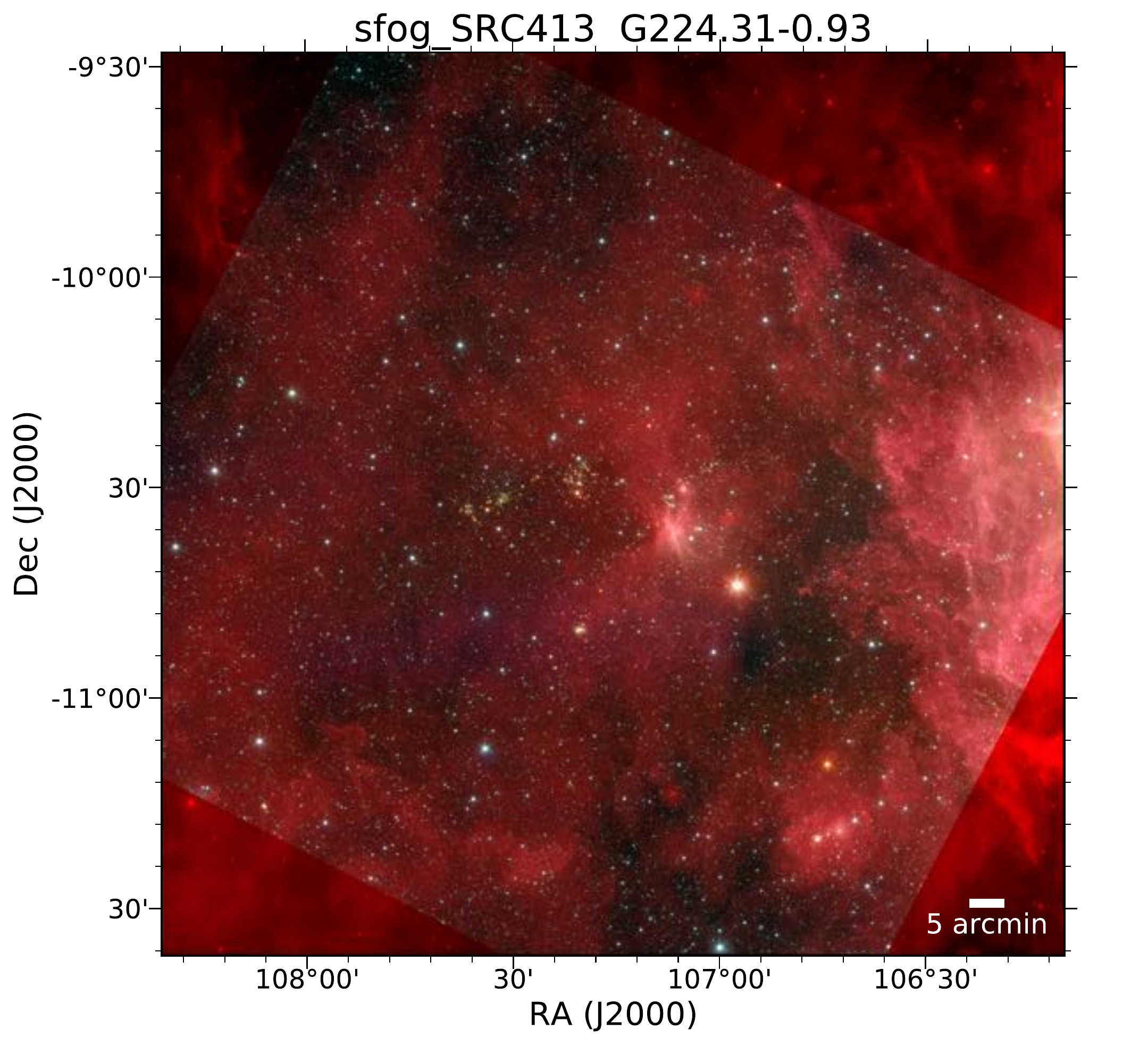}\includegraphics[width=.33\linewidth]{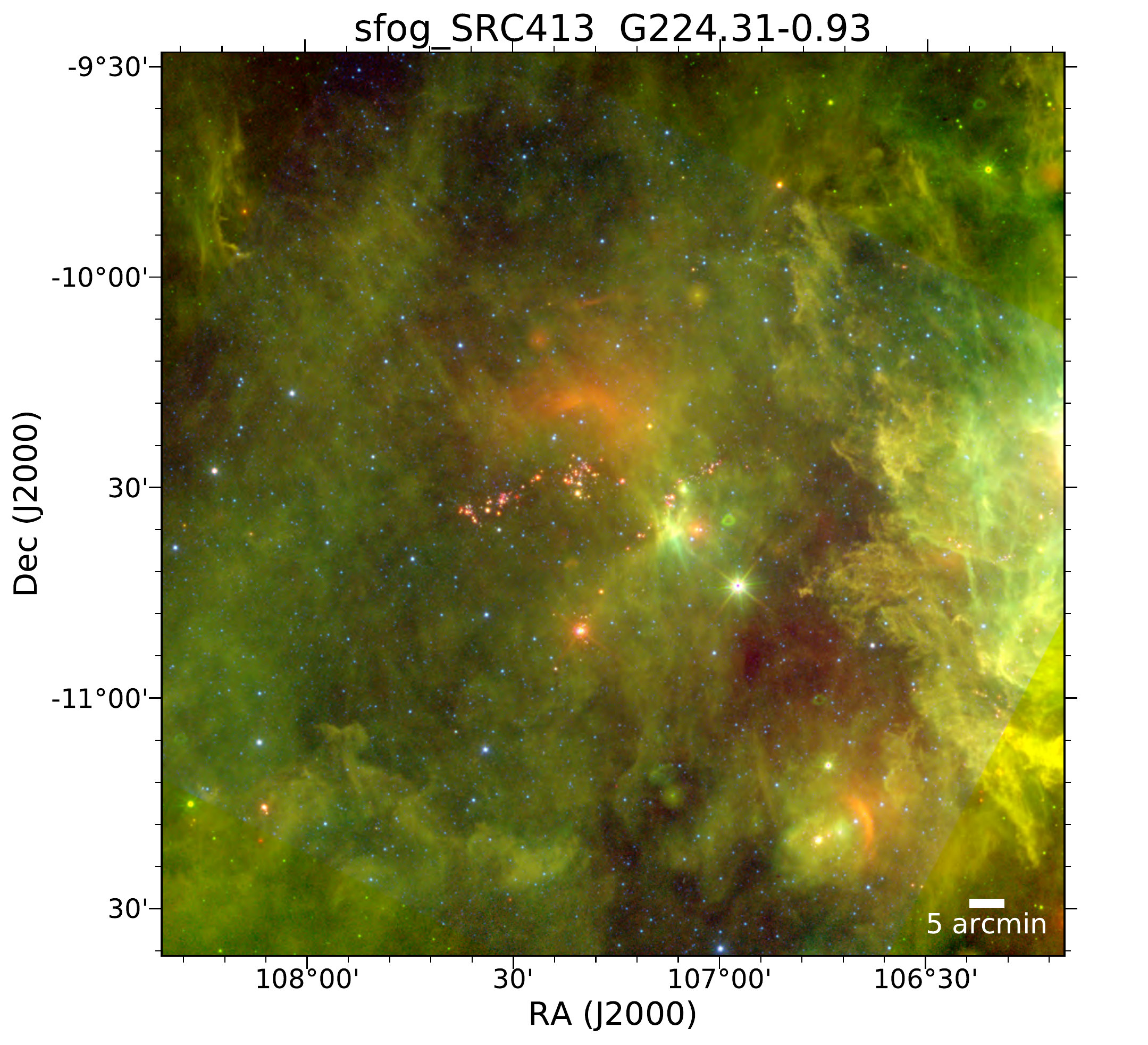}\includegraphics[width=.33\linewidth]{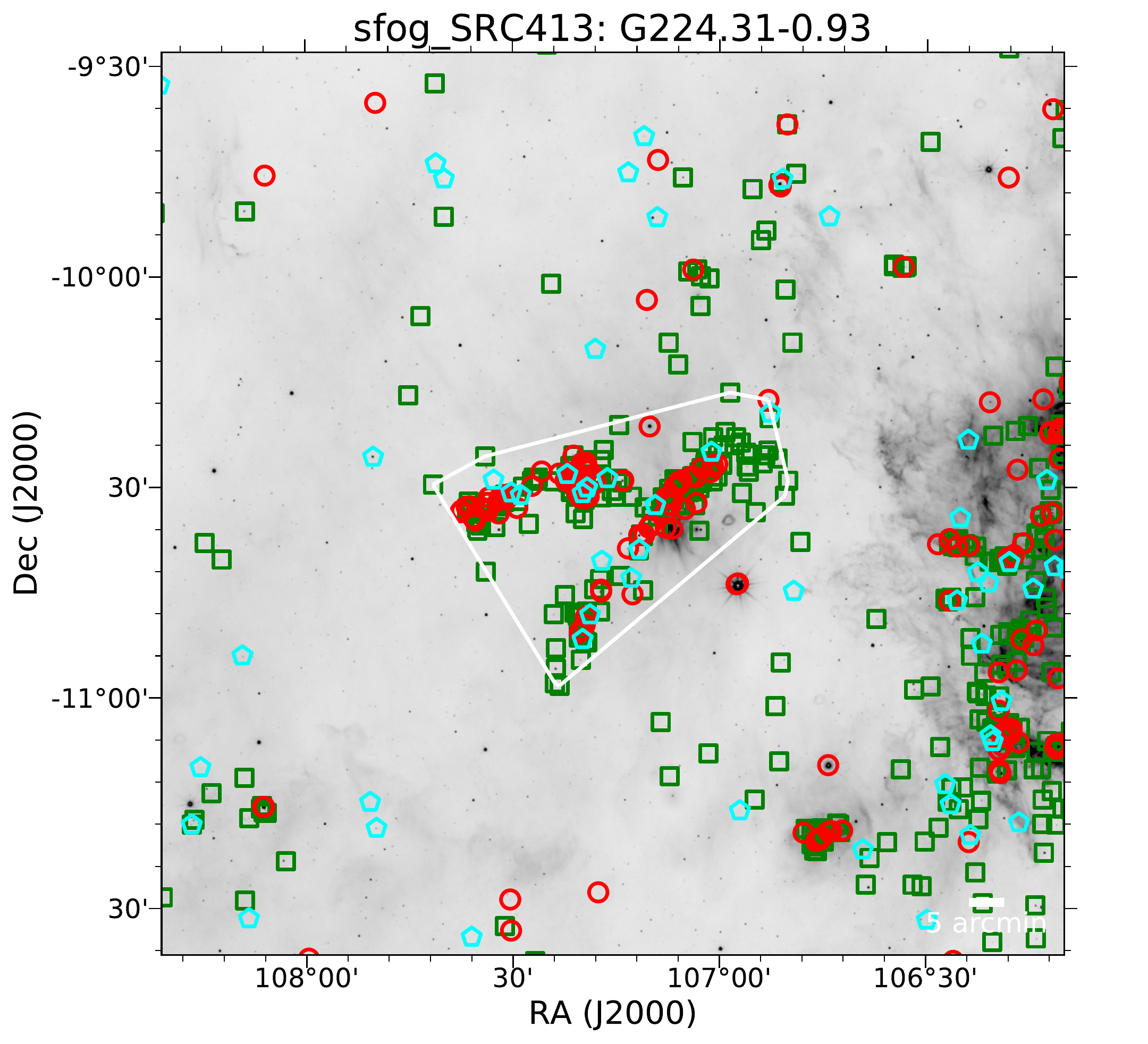}
\includegraphics[width=.33\linewidth]{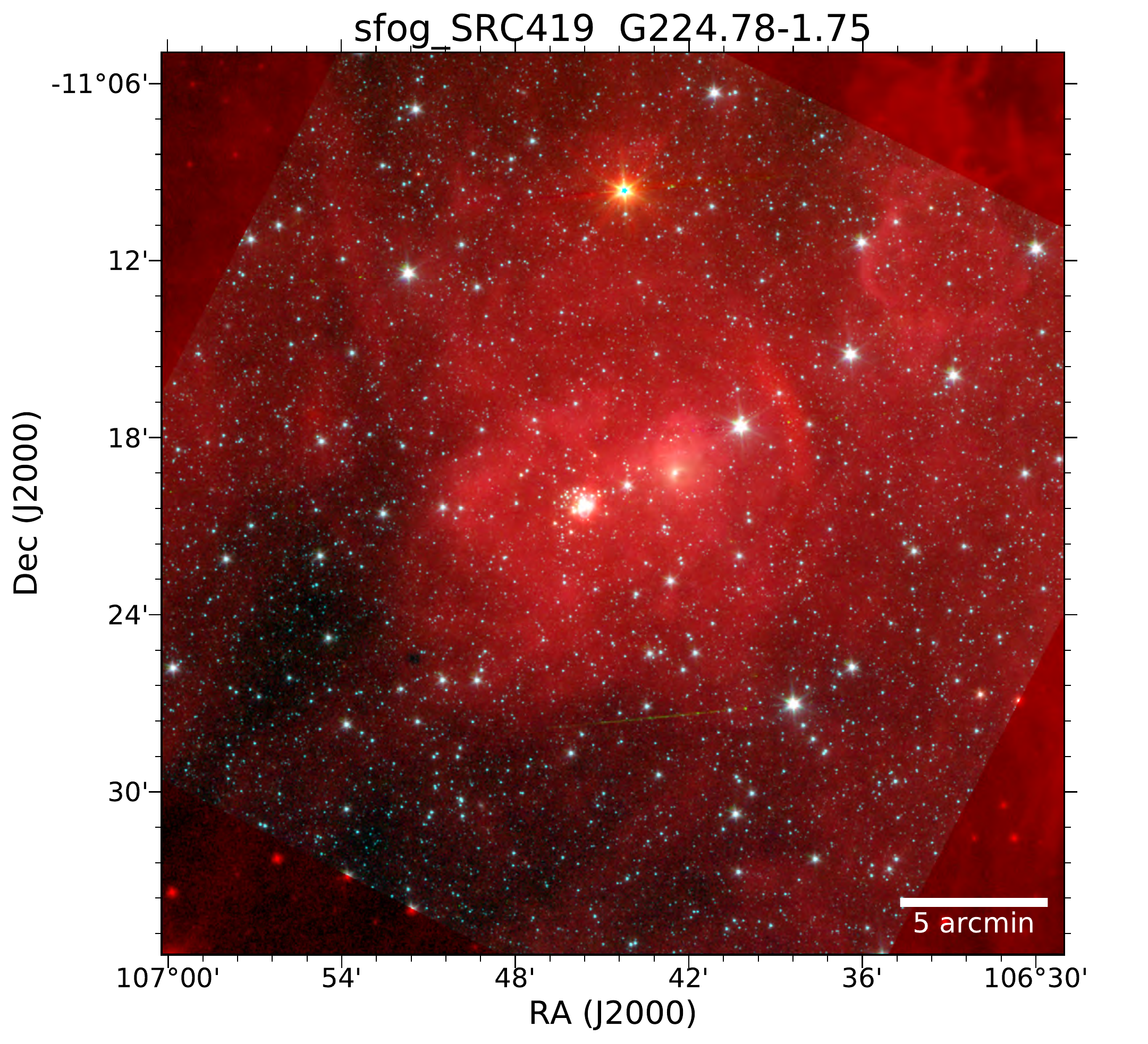}\includegraphics[width=.33\linewidth]{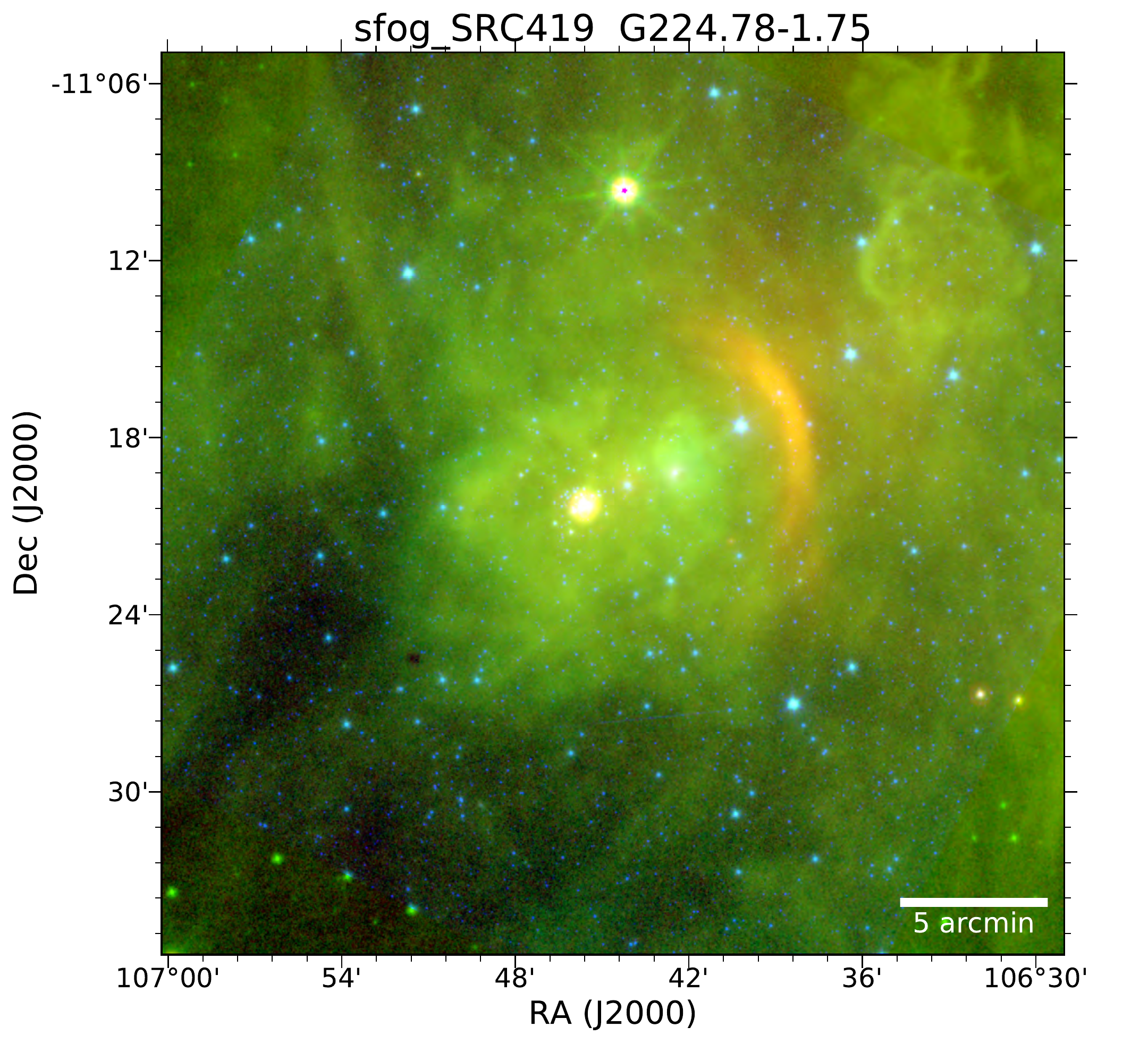}\includegraphics[width=.33\linewidth]{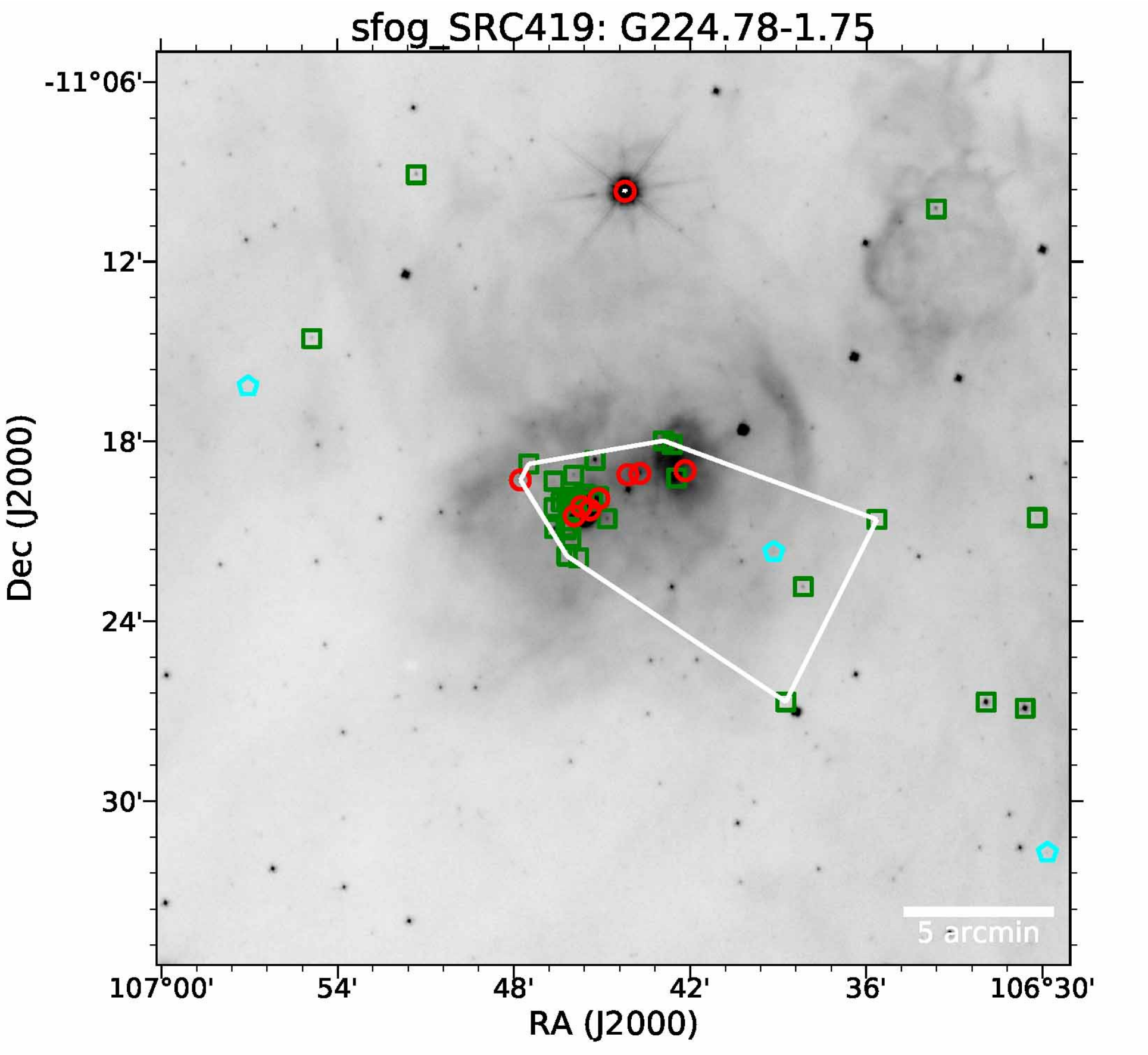}
\includegraphics[width=.33\linewidth]{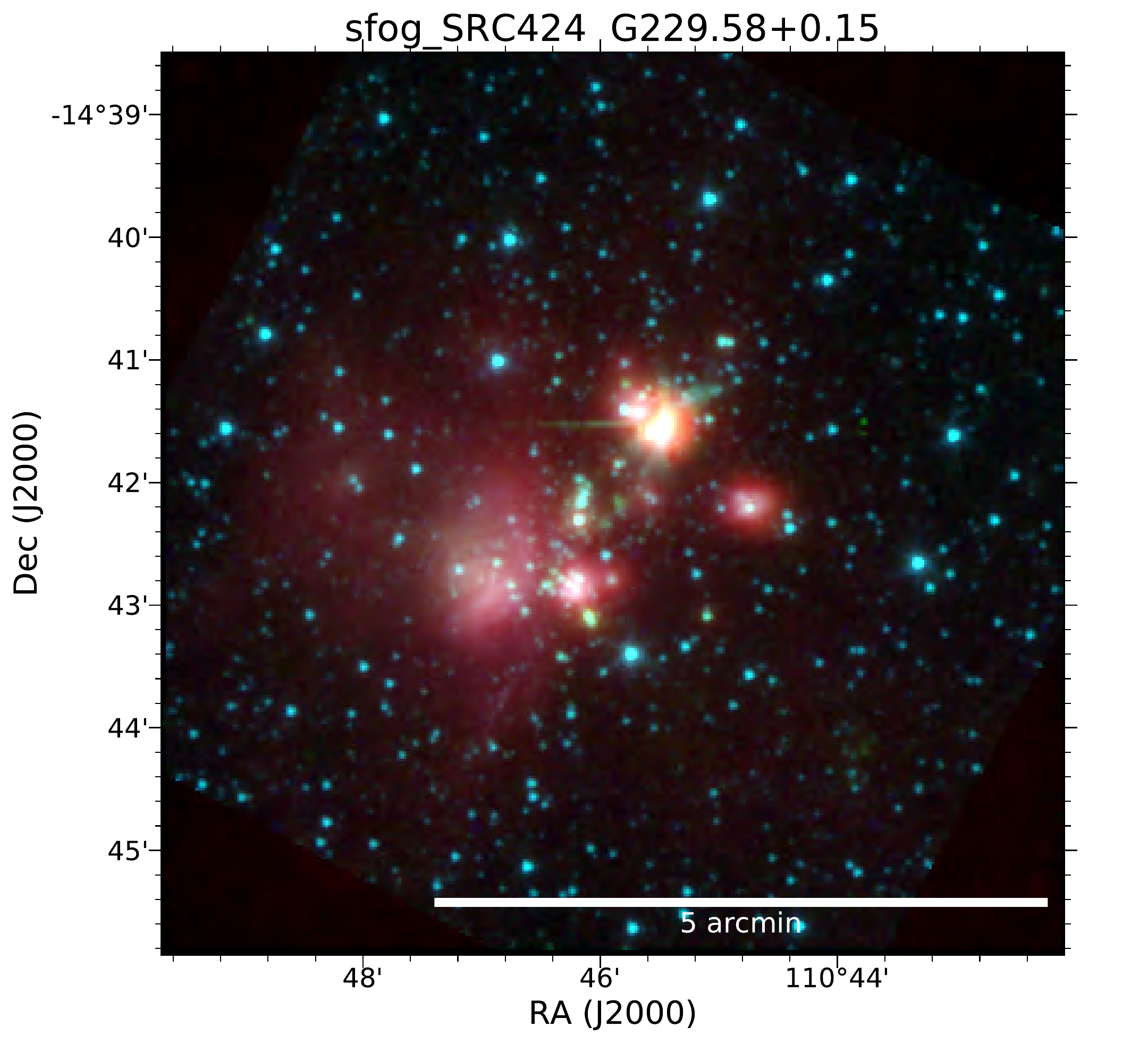}\includegraphics[width=.33\linewidth]{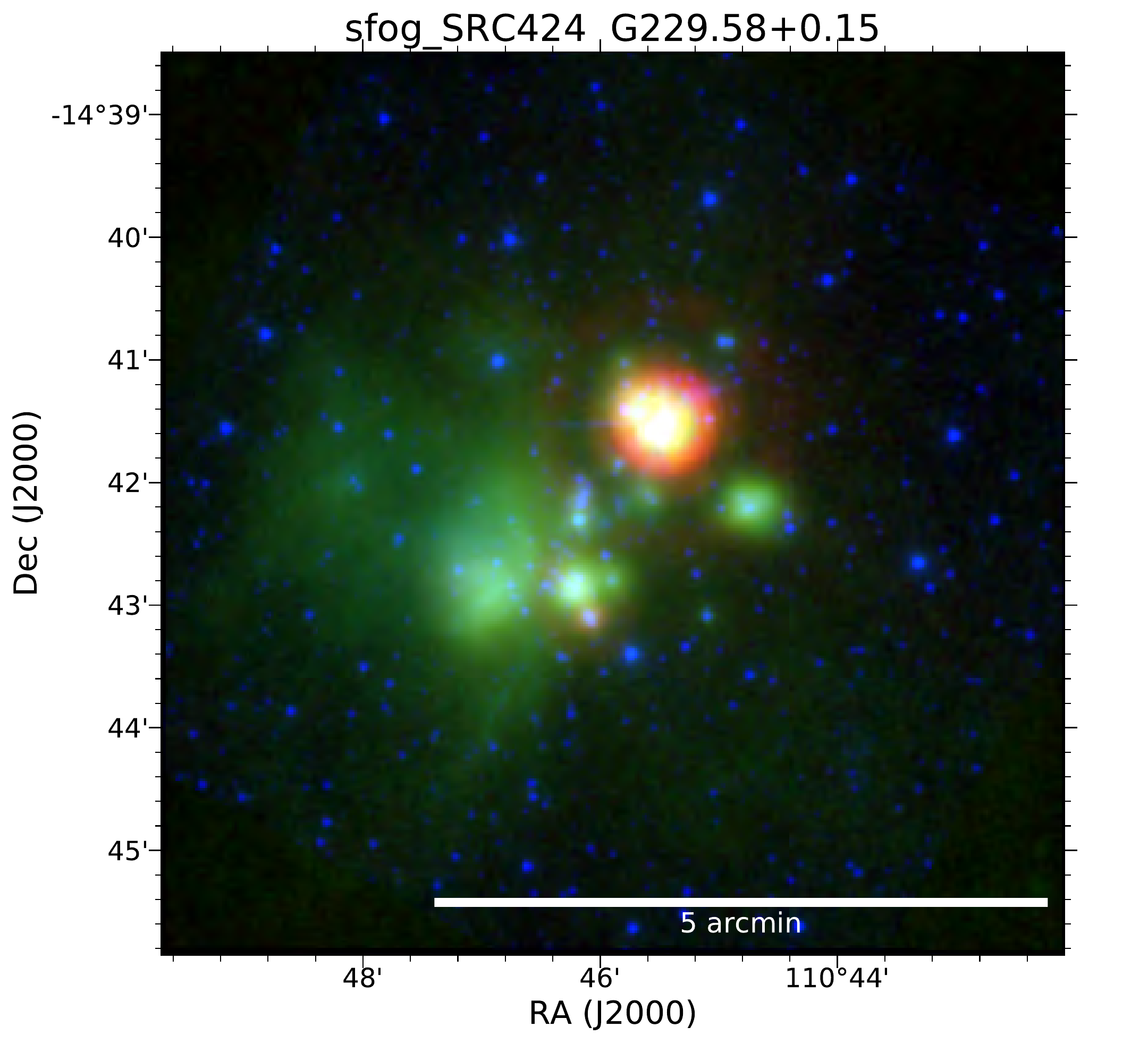}\includegraphics[width=.33\linewidth]{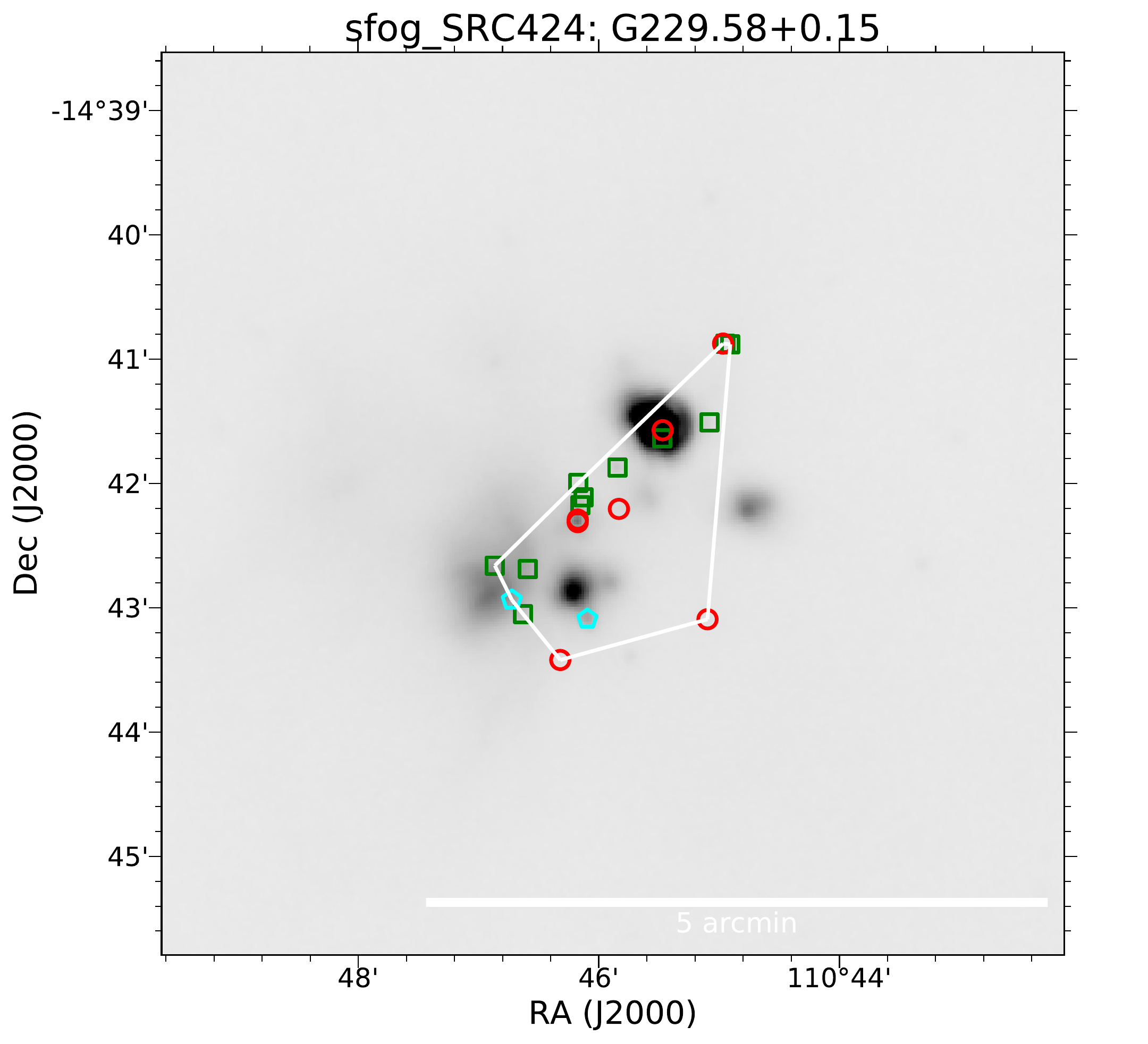}
\caption{Four more examples of clusters identified in the SFOG field: \#286, \#413, \#419, and \#424 from top to bottom. In the left column are 3-color images in {\it WISE} 12~\micron\ (red), IRAC 4.5~\micron\ (green), and IRAC 3.6~\micron\ (blue); the center column contains 3-color images with {\it WISE} 22~\micron\ (red), {\it WISE} 12~\micron\ (green), and IRAC 4.5~\micron\ (blue); and the right column shows the {\it WISE} 12~$\mu$m in reverse grayscale with the identified YSOs and the calculated convex hulls for each cluster overlaid.  The symbols show the positions of the Class I (red circles), Class II (green squares), and Class IIa/III (cyan pentagons) YSOs. All of the cluster images and associated FITS files are available from \citet{winston2020}.    }
\protect\label{fig_clusters2}
\end{figure*}
\end{center}

The approximate area of each cluster was quantified by measuring the convex hull of the associated cluster members. The convex hull is the set of points whose vertices 
include all of the points in the set.   
Figures~\ref{fig_clusters1} and \ref{fig_clusters2} show eight examples of clusters identified across the SFOG field, showcasing the range of sizes and environments over 
which clusters were identified.  
In many cases it is clear that the clusters form part of a larger association or star forming complex.  They are often surrounded by non-clustered YSOs that are 
likely to be associated with the cluster but did not satisfy the DBSCAN criteria.  We present these clusters as starting points for future studies.

The cluster identification of each YSO is listed in Table~\ref{tab:photdescription}.   
Table~\ref{tab:clusters} lists the properties of the largest clusters (number of YSOs $>$ 100) found in the SFOG catalog, sorted by decreasing number of YSOs in the cluster. The table gives the cluster number and name,  the number of YSOs, coordinates of the cluster central point, the circular radius based on the separation of the most distant YSOs, and association with previously identified star forming regions.  The complete version of the table is available in the electronic version of this paper.  The electronic version 
also includes a list of {\it WISE} \ion{H}{2} counterparts, and a full listing of the SIMBAD objects located within the convex hull of each cluster.  These sources are not assumed to be physically associated with the cluster, and no 
attempt has been made to filter the lists or to match them to the YSOs. They are provided as a reference for more in-depth studies.  

Images of each of the individual clusters were constructed using mosaics with a radius of 5 times the estimated convex hull radius of each cluster. The IRAC mosaics were downloaded from the 
IPAC servers, and the {\it WISE} Coadder\footnote{\url{https://irsa.ipac.caltech.edu/applications/ICORE}} was used to generate mosaics of the fields in the {\it WISE} bands. For cases where a cluster was near the edge of the IPAC mosaics of the IRAC data, adjoining mosaics were combined to form images that could be used to display the full cluster.  These mosaics were then used to create a selection of 3-color images and a {\it WISE} 12~\micron\ band grayscale image overlaid with the 
cluster convex hull and the locations of the YSOs by evolutionary class.  
The full set of mosaics for every cluster are available on the Harvard Dataverse SFOG page \citep{winston2020}.

\begin{figure}
\epsscale{1.15}
\plotone{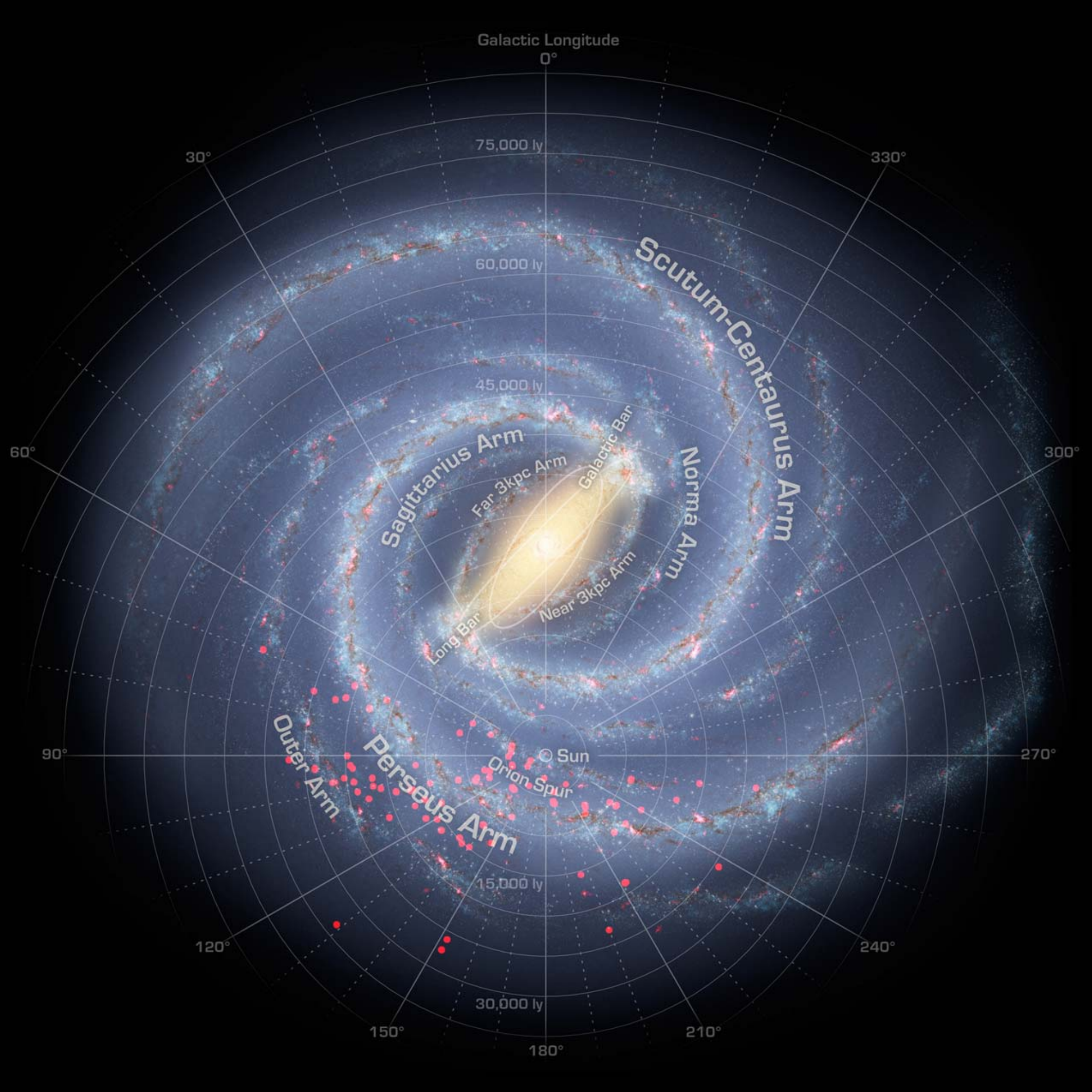}
\caption{ Schematic representation of the Milky Way Galaxy \citep{hurt2008} with red points overlayed showing the locations of the identified clusters with known distance measurements. }
\protect\label{fig_galactic}
\end{figure}

Figure~\ref{fig_galactic} shows the spatial distribution of the identified clusters overlaid on a schematic view of the Milky Way Galaxy \citep{hurt2008}.

As can be seen from Table \ref{tab:clusters}, all of the larger clusters are associated with previously known star forming regions, identified for example through surveys of \ion{H}{2} regions, CO emission, radio continuum emission, detection of dark clouds, or the infrared emission from compact sources or the surrounding nebula. However, we also find 6 of the smaller clusters do not have any SIMBAD sources within their convex hulls and may be newly identified.

\subsection{Catalog contamination}

The true level of contaminants remaining in the catalog used to select YSOs is difficult to estimate precisely given the broad range in spatial coverage and distances covered 
by the SFOG field.  Spectral typing of the YSOs would provide confirmation of their nature, and a program to obtain spectra and X-ray observations of a selection of 
the clusters is currently underway.  We discuss in the following sections possible sources of contamination in the YSO catalog and their effects on the clustering analysis.

\subsubsection{IRAC sample contamination}

To examine the effectiveness of the contamination cuts used in the GLIMPSE360 analysis, a comparison was made to the SMOG field of Paper I. 
By applying the same contamination cuts and selection criteria used in this paper, 1,512 YSOs were selected in the SMOG field.  
Of these, 1,102 objects were matched at 1$\arcsec$ between the two catalogs.  
A majority of the non-matching sources were included in the SFOG catalog due to the differences between the selection criteria required because of the lack of IRAC 5.8 and 8~\micron\ and MIPS 24~\micron\ coverage in the GLIMPSE360 field.  Plotting these sources on an IRAC four-color CCD, 22 objects exhibited colors similar to those of galaxies.  This would imply that 22/1,512 or $<$2\% of the candidate YSOs are likely to be extragalactic contaminants.  

By applying this percentage to the 28,837 objects in the GLIMPSE360 component of our catalog, we would estimate that 420 of these may possibly be contaminants. 
These would likely be found predominantly in the non-clustered population since both foreground and background contaminants tend to be randomly distributed across the field.

\subsubsection{Contamination in the WISE sample} 

In order to assess the validity of the clustering algorithm, each of the regions was visually checked to look for interesting or spurious clusters. 
From this assessment, a total of three of the identified clusters,  \#515, \#516, and \#564, were found to be invalid due to spurious {\it WISE} sources. 
The {\it WISE} photometry contaminant removal process removes confirmed diffraction spike objects, but those objects with tentative 
identification as photometric diffraction spikes are not removed. 
In each of these cases, a visual inspection of the images containing the `cluster'  found them to contain a number of diffraction spike objects surrounding a saturated bright foreground star.  

In removing these three spurious clusters, the number of clusters is reduced to 618 in total.  
The number of YSOs is reduced from 47,405 to  47,338 YSOs.  
The breakdown by evolutionary class becomes:  10,461 Class I, 29,552 Class II, and 7,325 Class IIa/III YSOs.

\subsubsection{AGB Contamination}\label{agbcontam}

The Besan\c con Galactic population synthesis models were used to estimate the AGB population at four points along the Galactic plane \citep{robin03,robin14}.  
The points chosen were at $ l = 70^\circ, 105^\circ, 180^\circ, 250^\circ$  and $b = 1^\circ, 1^\circ, 0^\circ, -1^\circ$, respectively, with a 1 deg$^2$ field at each location.  
The models predicted that these fields contained either one or two AGB stars each.  By scaling these numbers up to the approximate 600 sq. deg size of the SFOG field, we estimate that 
there may be between 600-1,200 AGB stars in the SFOG field.   
This would imply that $\sim$1,200/47,338 or $\sim$2.5\% of the candidate YSOs may be AGB stars.  
The AGB stars would tend to be randomly distributed over the field, and thus less likely to be included in a cluster, so these contaminants are not likely to affect our clustering analysis significantly.

\section{YSO and Cluster Physical Characteristics}

\subsection{SED model fitting}\label{sedfitter}

The Python SEDFitter\footnote{\url{https://sedfitter.readthedocs.io}} package of \citet{robitaille} was used to provide an estimate of the mass, age, disk and accretion properties of the YSOs in the SFOG field. 
The code uses a sample grid of YSO model SEDs with varying age, mass, inclination etc., to compare to the input photometry, with the scale factor $S$ 
(dependent on source distance and luminosity) and the extinction $A_V$ as free parameters.  The code returns a sample of best fit models and their 
associated parameters.  
From our previous SMOG study, it was determined that within each cluster the range in distances for the best fit models covered the entire provided 
distance range, and that therefore the results could not be used in any meaningful way when the distance to the YSO is unknown.  

\begin{figure*}
\includegraphics[width=.333\linewidth]{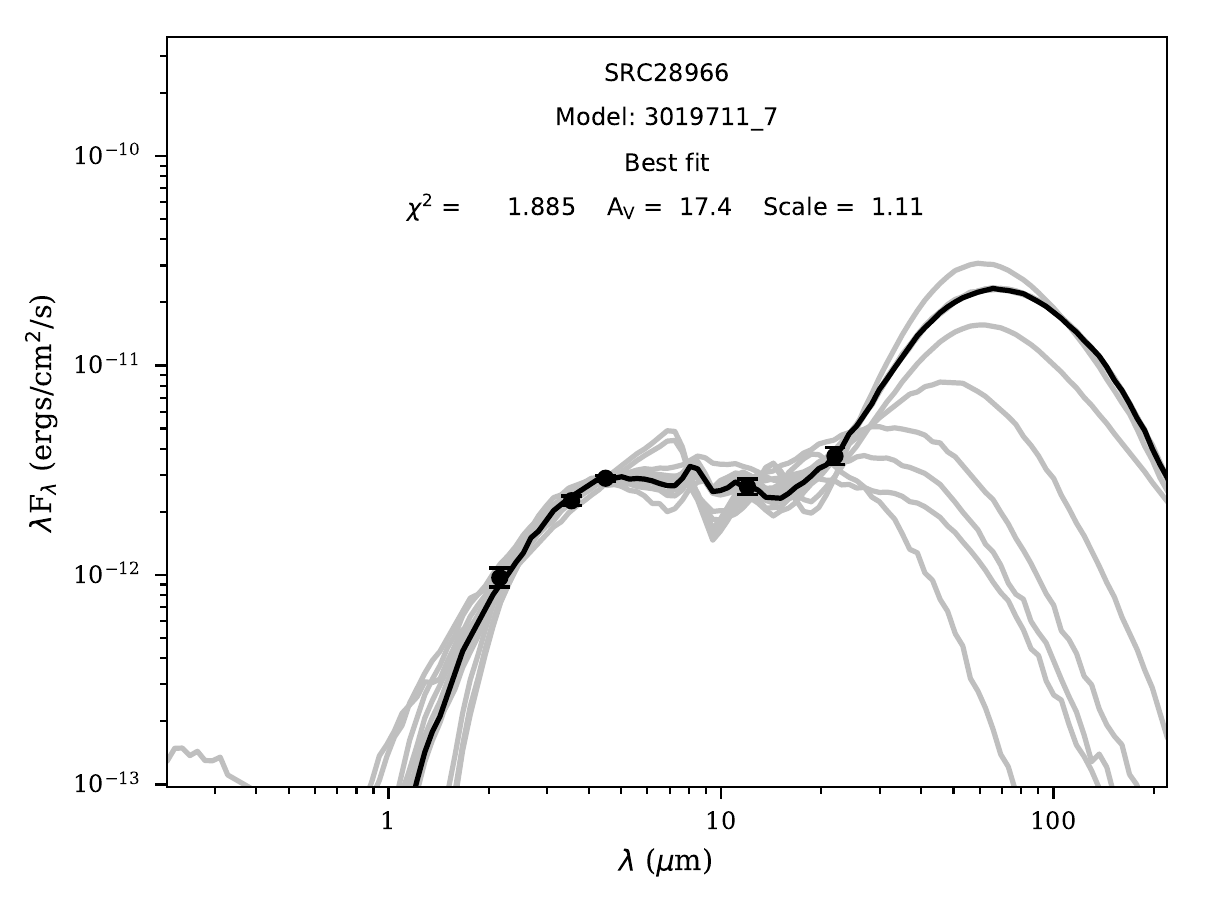}\includegraphics[width=.333\linewidth]{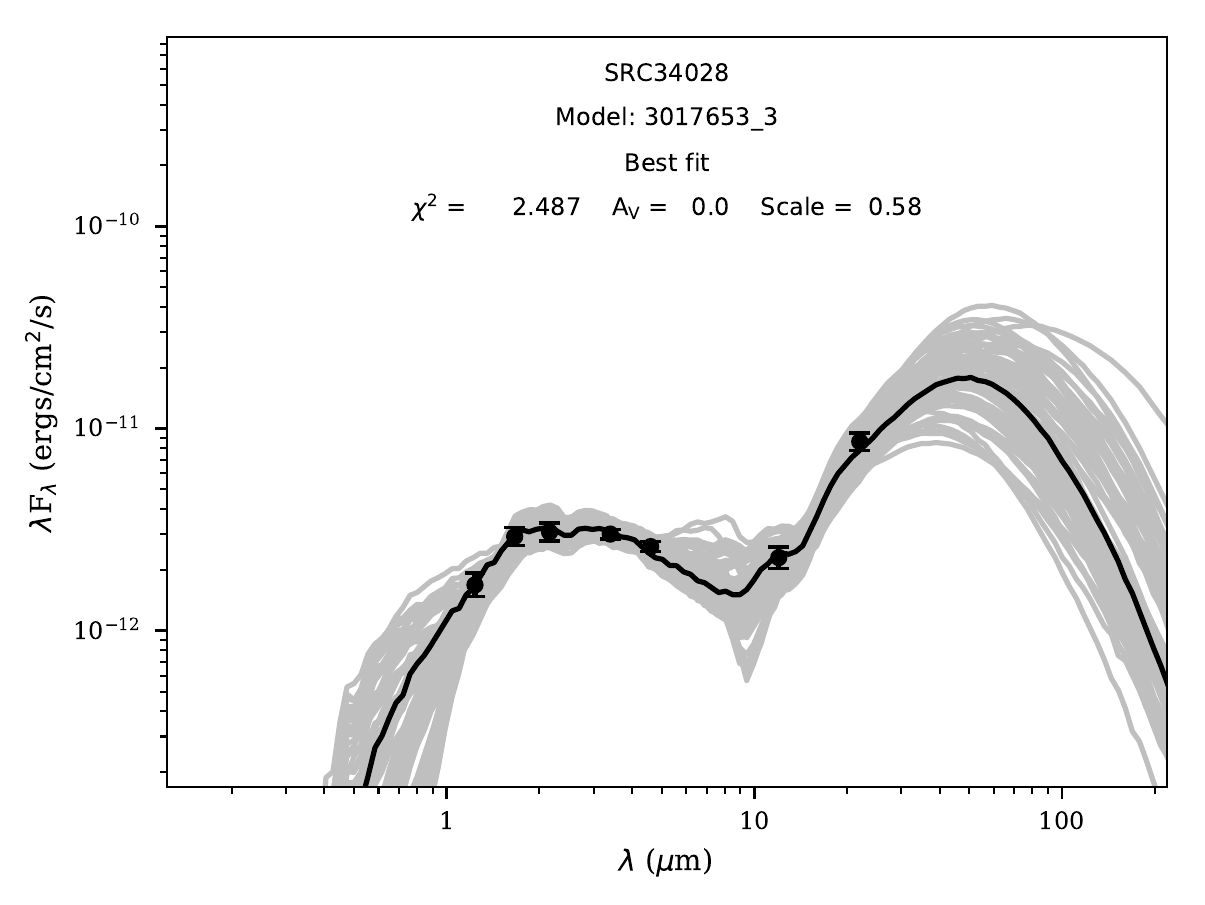}\includegraphics[width=.333\linewidth]{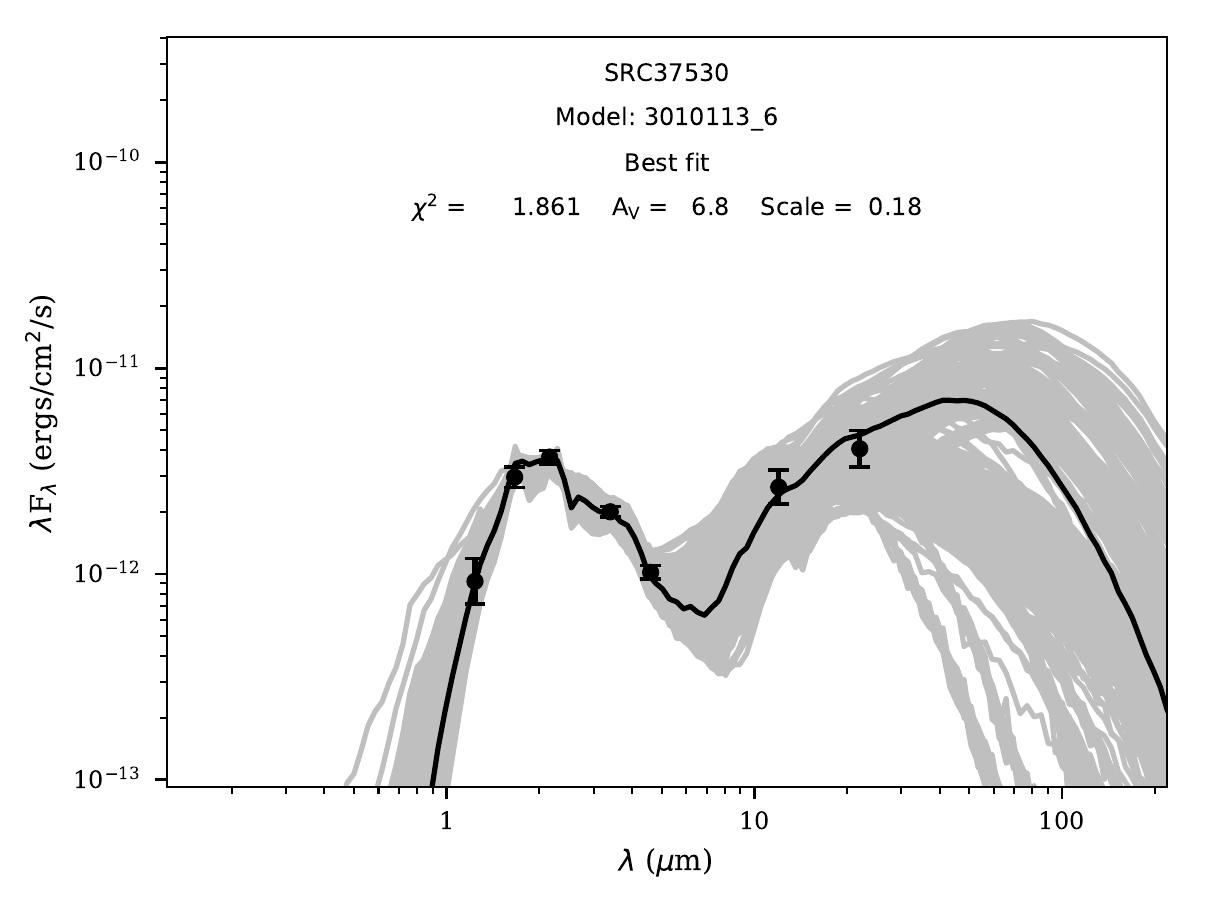}
\caption{ Examples of three SEDFitter model fits to YSOs in the SFOG catalog with detections in 2MASS, IRAC, and WISE: SRC28966, SRC34028, and SRC37530 in clusters 15, 85, and 106, respectively.   }
\protect\label{fig_3seds}
\end{figure*}

Therefore only those clusters with reliable distance estimates from {\it WISE} \ion{H}{2} regions \citep{anderson} located within the cluster's convex hull were used in the SEDFitter modeling.  
Reliable distances were found for 100 of the 618 clusters identified in the SFOG field.   
The \ion{H}{2} regions associated with each cluster and their distances are listed in the full online version of Table~\ref{tab:clusters}.   
The range of Galactocentric distances for these 100 clusters ranged from 7.8~kpc to 18.1~kpc. 
 
\begin{deluxetable*}{cccccccccccccc}
\tabletypesize{\scriptsize}
\tablecaption{SFOG Field Young Stellar Objects: SED Fitter Results }
\tablehead{ 
\colhead{SF} & \colhead{Cl}  & \colhead{Dist} & 
\colhead{R$_{Gal}$}  & 
\colhead{N$_{data}$} & 
\colhead{$\chi^{2}$}   & 
\colhead{A$_V$}  &
\colhead{M$_C$} & 
\colhead{Age} &
\colhead{M$_{disk}$} &  
\colhead{M$_{env}$} & 
\colhead{$\dot{M}$}  & 
\colhead{$T_{*}$} & 
\colhead{$L_{*}$}      \\[-0.3cm]
\colhead{ID}  & \colhead{\#}  & \colhead{kpc}  & 
\colhead{kpc}  & \colhead{} & 
\colhead{}  & 
\colhead{mag} &
\colhead{$M_\odot$} &  
\colhead{yr} &
\colhead{$M_\odot$} &  
\colhead{$M_\odot$} &  
\colhead{$M_\odot$} &  
\colhead{K} &  
\colhead{$L{_\odot}$} 
}
\startdata
SRC588 & 5 & 3.4 & 7.8 & 5 & 2.475 & 0.213 & 2.492 & 3.900e+06 & 1.008e-05 & 5.367e-04 & ... & 7.985e+03 & 4.251e+01 \\
SRC591 & 5 & 3.4 & 7.8 & 5 & 1.856 & 0.000 & 1.068 & 3.815e+05 & 7.670e-05 & 3.610e-03 & 7.183e-07 & 4.264e+03 & 5.582e+00 \\
SRC597 & 5 & 3.4 & 7.8 & 5 & 0.322 & 5.213 & 1.981 & 8.221e+06 & 3.102e-05 & 2.993e-06 & ... & 7.854e+03 & 1.186e+01 \\
SRC600 & 5 & 3.4 & 7.8 & 5 & 0.210 & 5.039 & 1.470 & 1.192e+06 & 1.645e-02 & 1.321e-04 & ... & 4.537e+03 & 3.165e+00 \\
SRC601 & 5 & 3.4 & 7.8 & 5 & 0.829 & 3.181 & 1.638 & 7.610e+06 & 8.970e-03 & 4.193e-08 & ... & 5.322e+03 & 3.185e+00 \\
SRC608 & 5 & 3.4 & 7.8 & 4 & 0.007 & 11.30 & 0.540 & 3.042e+05 & 9.975e-04 & 3.223e-03 & 2.440e-07 & 3.822e+03 & 2.954e+00 \\
SRC613 & 5 & 3.4 & 7.8 & 5 & 0.776 & 0.345 & 1.915 & 6.945e+06 & 3.301e-05 & 2.837e-08 & ... & 6.449e+03 & 1.303e+01 \\
SRC616 & 5 & 3.4 & 7.8 & 4 & 0.393 & 8.739 & 2.009 & 6.060e+04 & 3.965e-03 & 1.839e-01 & 9.353e-06 & 4.316e+03 & 3.828e+01 \\
SRC617 & 5 & 3.4 & 7.8 & 4 & 0.491 & 7.324 & 8.147 & 6.542e+03 & 4.394e-03 & 3.207e+01 & 1.131e-03 & 4.381e+03 & 9.556e+02 \\
SRC622 & 5 & 3.4 & 7.8 & 5 & 0.971 & 0.157 & 3.736 & 1.364e+04 & 1.649e-02 & 6.810e+00 & 1.486e-04 & 4.327e+03 & 1.446e+02 \\
\enddata
\label{tab:SEDfit}
\tablecomments{Table \ref{tab:SEDfit} is published 
in its entirety in the machine readable format.  A portion is
shown here for guidance regarding its form and content.}
\end{deluxetable*}

The SEDFitter routine was run using a fixed distance range based on the distance to the cluster, allowing the $A_V$ to vary from 0-40~$A_V$, for each cluster separately.   
YSOs lacking photometry across a sufficient number of bands were not fit successfully.  
Further, of those YSOs fit, not all had fits with low $\chi^2$ values.  Of the 100 clusters run, 96 contained some YSOs with `good' fits ($\chi^2 < 3.0$). 
In general, four photometric points were required for a 'good' fit to the SED model.
In total, 6,234 YSOs had reliable model fits across the 96 clusters. 
Figure~\ref{fig_3seds} shows three examples of SEDFitter model fits to YSOs in the SFOG catalog.

Table~\ref{tab:SEDfit} lists a sample of the results of the SEDFitter routine for the YSOs in the 96 clusters for which reliable fits were obtained.  
A number of parameters are presented for each model fit, including the best fit to the  object mass, disk mass, age, $A_V$, central temperature, disk accretion rate, etc., and the $\chi^2$ value.  The weighted average values of all parameters for fits with $\chi^2 < 3$ were calculated and are presented for each source.  
The upper and lower limit model fit parameters are also supplied in each case.  The full data table, including all columns, is available online in electronic format.

\begin{figure*}
\epsscale{1.15}
\plotone{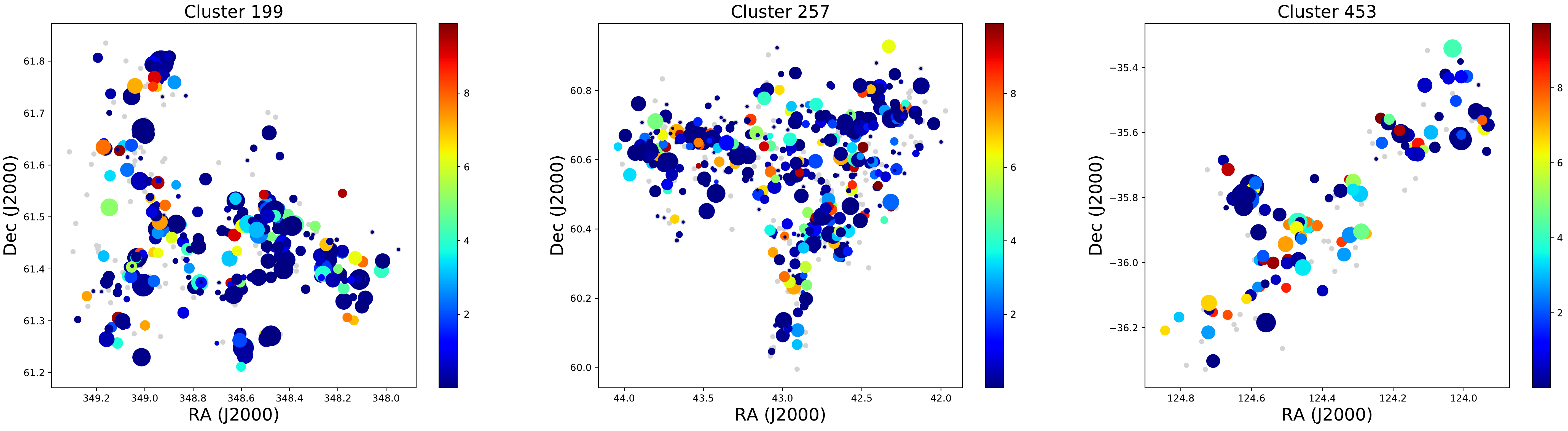}
\caption{ Examples of the SEDFitter results for spatial distributions of ages and masses within each cluster. 
The relative sizes of each symbol indicate how massive the YSO is with respect to the most massive object in that cluster.   
The color bar shows how the color scaling relates to the calculated age of the YSO (in Myr), with the bluer objects being younger, and the redder objects being older.  }
\protect\label{fig_3sedam}
\end{figure*}

Given the uncertainties in the distances and the sparse photometry for each source, we do not consider the individual ages and masses derived from the model 
fits to be entirely reliable and will not discuss them further here. 
However, cumulatively for all clusters, they can provide an insight into the relative ages and masses of the YSOs in different regions 
of the outer Galaxy.  
Figure~\ref{fig_3sedam} shows the age and relative mass of the YSOs in three clusters by their spatial distribution.  
The size of the circles indicates the mass of the YSO relative to the most massive object that we identified in that cluster.  
The color indicates the age, the range in ages in Myr is shown by the colorbar for each plot.  
There were no strong trends in the distribution of age or mass across the 618 clusters.  

\subsection{Initial Mass Function (IMF)}
\begin{figure*}
\epsscale{1}
\plottwo{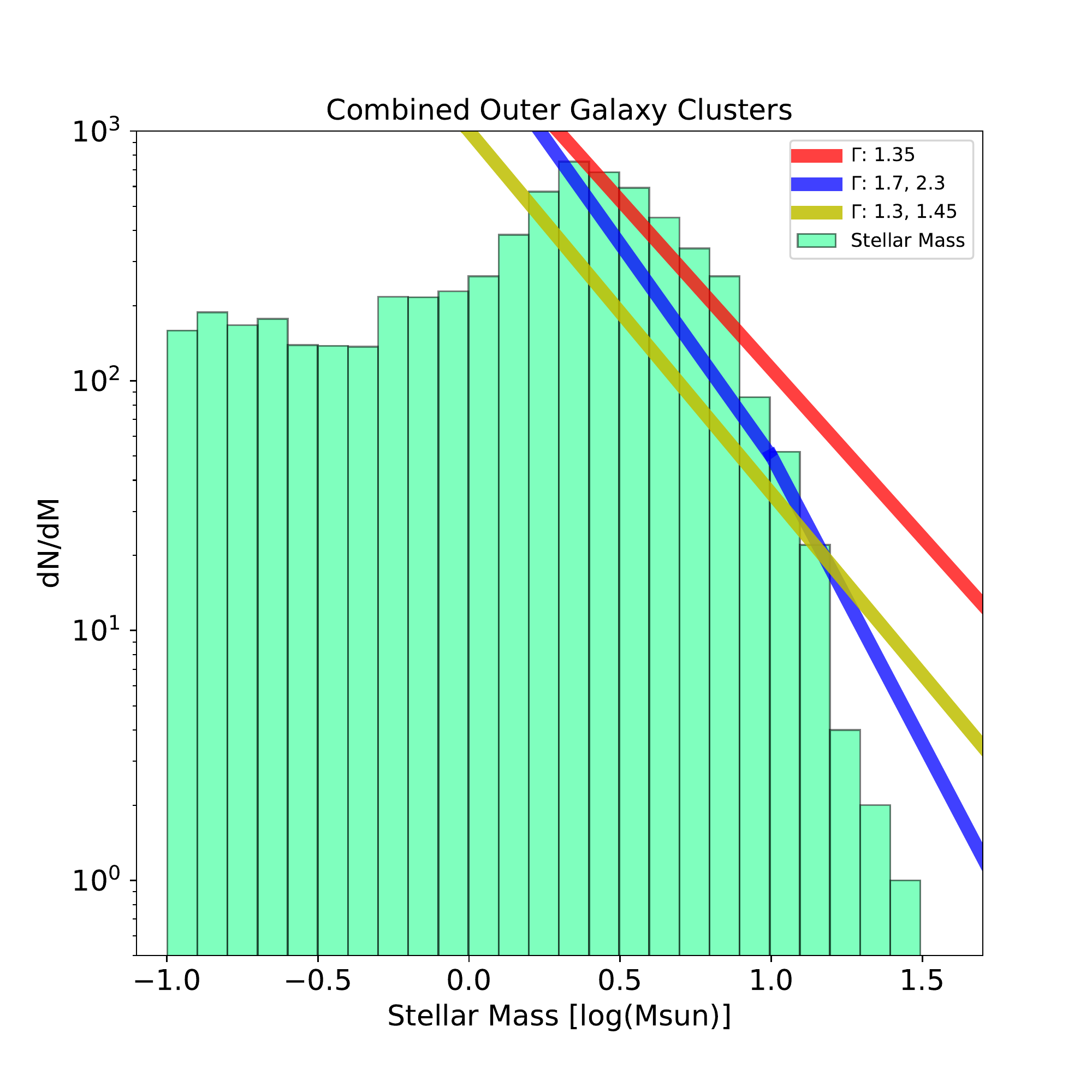}{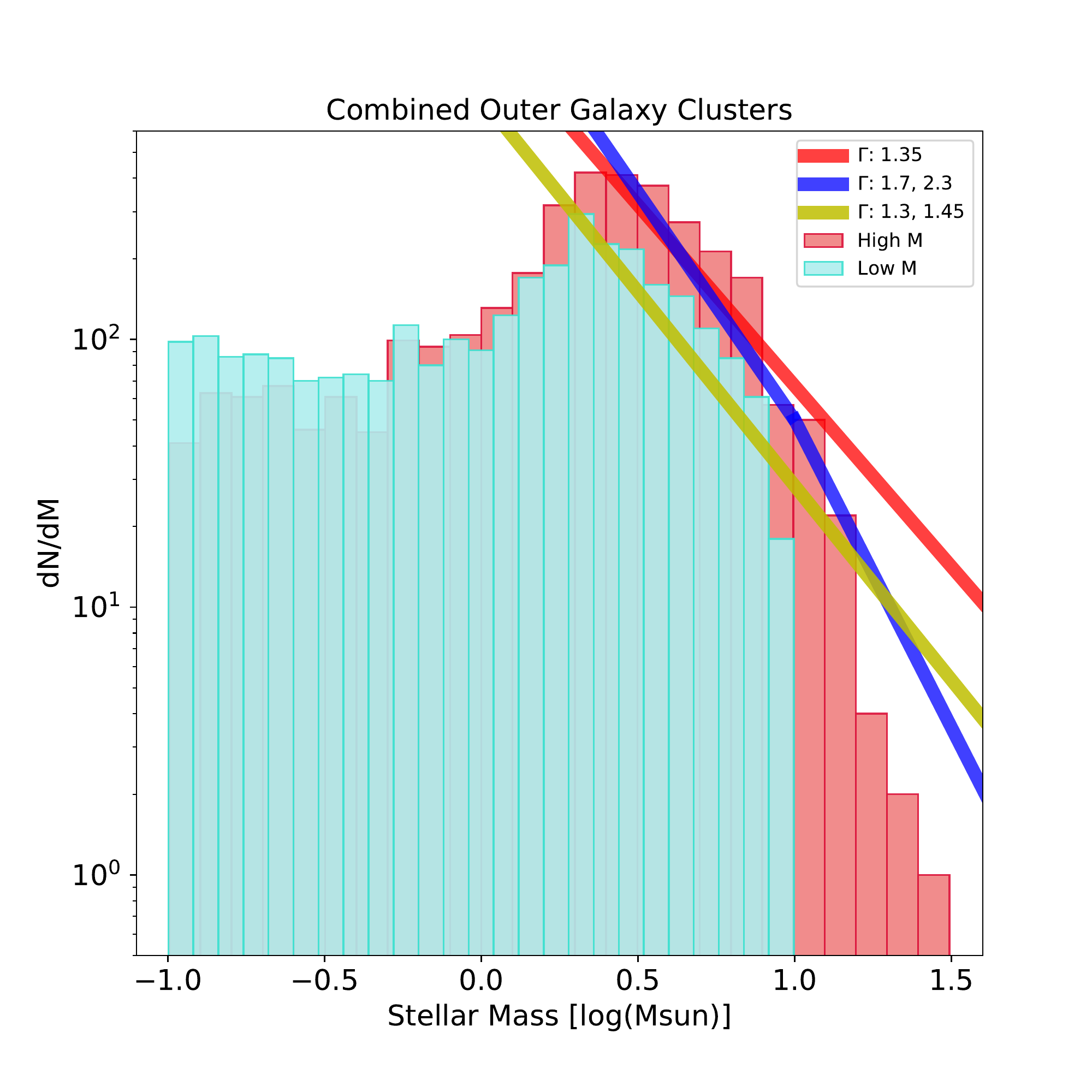}
\caption{ Examination of the Initial Mass Function for the outer Galactic regions identified in the SFOG field.  
{\it Left:} The IMF as determined by combining the SEDFitter calculated masses for the 96 clusters with distance estimates and 
with members with $\chi^2<3$ fits.  The green histogram shows the best fit model masses for all the YSO. 
 The data show reasonable correlation to both the Salpeter slope of the IMF (m$\sim$-1.35) and more recent estimates 
 (m$\sim$-2.7, m$\sim$-2.3) for a broken power law fit.  
{\it Right:}  The IMF split into the 58 clusters with a highest mass member lower than 10~$M_{\odot}$ and the 38 clusters 
with a highest mass member greater than 10~$M_{\odot}$.   
}
\protect\label{fig_imf}
\end{figure*}

The initial mass function (IMF) of the clusters in the outer Galaxy is of great importance to determine whether the environment 
of the outer Galaxy has had an affect on the star formation efficiency and rate.  
Because for each cluster we have the SED-derived masses for only a small sample of YSOs, we examined the IMF of all 96 clusters combined, as shown in Figure~\ref{fig_imf}~(left). 
We then split the clusters into two groups based on the mass of the most massive identified member, with the cut at 10$M_{\odot}$, 
as shown in Figure~\ref{fig_imf}~(right).  There were 58 low mass clusters (2,929 YSOs), and 38 high mass clusters (3,305 YSOs). 
The clusters were then split into two groups based on their Galactocentric radius, with the division between inner and outer Galaxy placed at 11.5~kpc, based on \citet{huang}. 
There were 46 clusters with 4,533 YSOs with $R_{Gal} < 11.5$~kpc, and 50 clusters with 1,701 YSOs with $R_{Gal} > 11.5$~kpc.

\begin{figure*}
\epsscale{1.}
\plotone{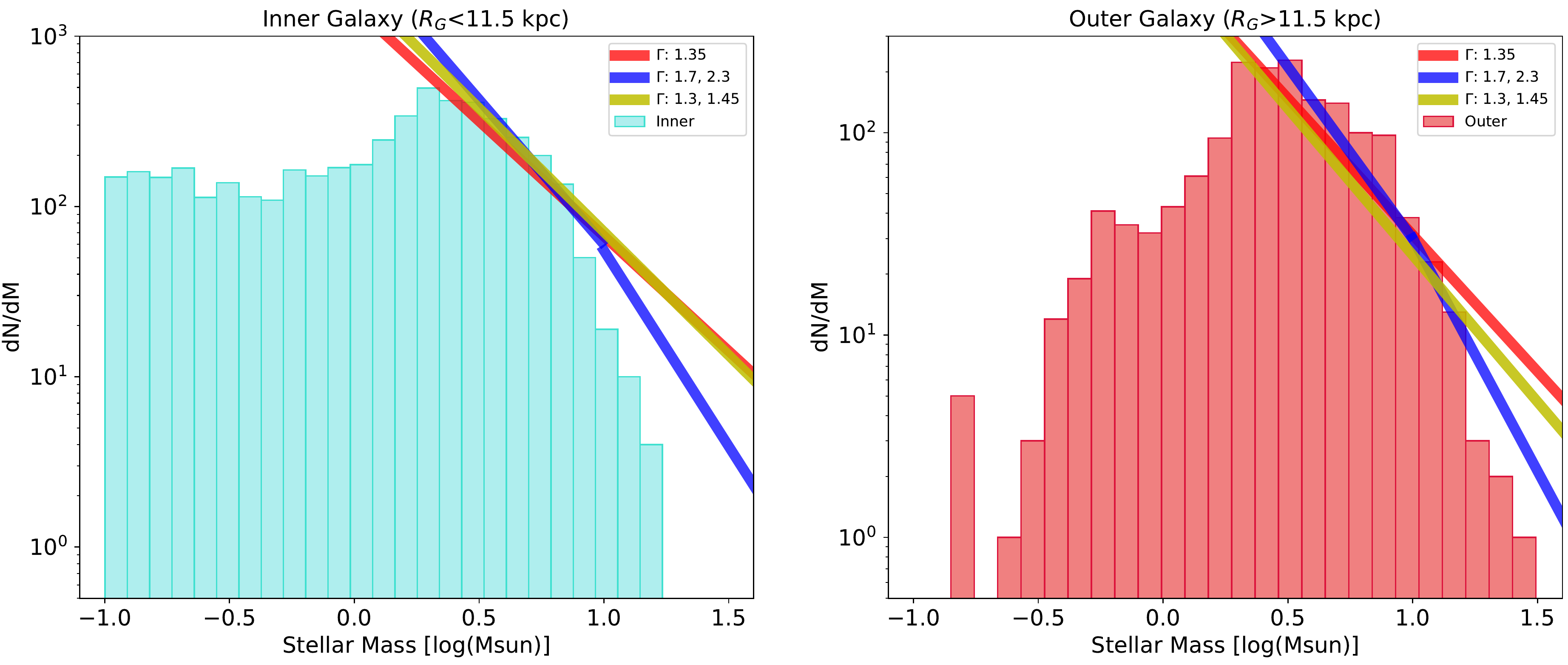}
\caption{ Examination of the Initial Mass Function for the outer Galactic regions identified in the SFOG field.  
{\it Left:} The IMF for the 46 clusters within a Galactocentric radius of 11.5~kpc.
{\it Right:} The IMF for the 50 clusters at Galactocentric radius greater than 11.5~kpc.
}
\protect\label{fig_imfnf}
\end{figure*}

In the figures, the red line plots the power law slope with $\Gamma = 1.35$, roughly the value of the Salpeter slope \citep{salpeter}.  
The blue line shows the \citet{miller} broken power law fit with slopes of $\Gamma_1 = 1.7$ and $\Gamma_2 = 2.3$.   
The \citet{kroupa} IMF with three distinct values of $\Gamma$ for the low, solar, and high mass regions  is shown in yellow, 
as presented by \citet{weisz} who used a slope of $\Gamma = 1.45$ above two solar masses.  

Linear regression fits were made to the high-mass end of the resultant histograms in the range 3~$M_{\odot}< M_*<10~M_{\odot}$.  
The slopes of the fits are listed in Table~\ref{tab:imfslopes}. 
\begin{deluxetable}{lrr}
\tablewidth{0pt}
\tabletypesize{\small}
\tablecaption{ IMF slopes and uncertainties\label{tab:imfslopes}}
\tablehead{
\colhead{Clusters} & \colhead{$\Gamma$}  & \colhead{$\epsilon$} 
}
\startdata
 All Clusters          & 1.918 & 0.419  \\ 
 $Max M_{\odot} < 10$  & 1.929 & 0.355  \\ 
 $Max M_{\odot} > 10$  & 1.275 & 0.345  \\ 
 $R_{Gal} < $ 11.5~kpc & 1.873 & 0.308  \\ 
 $R_{Gal} > $ 11.5~kpc & 1.146 & 0.241  \\ 
\enddata
\end{deluxetable}
The slopes for the complete sample of YSOs, and the nearby and low mass clusters, are broadly consistent with the values of $\Gamma$ presented 
in the literature by \citet{parravano} with $\Gamma = 1.7-2.1$, as discussed in detail in Paper I.  
The high mass and more distant clusters have shallower slopes than the others, though still consistent with the Salpeter value.  
The 1-$\sigma$ uncertainties overlap for the high/low mass cluster slopes, though they do not quite overlap for the near/far cluster slopes.  
The near/far distance appears to be due to the difference in the modeled populations of the high and low mass clusters, and not necessarily to a 
difference in the environment of the outer Galaxy.  

To further investigate the comparison of the SEDfitter generated IMFs of the 
inner and outer galaxies, we undertook a comparison of the Cygnus-X Legacy Survey catalog of IRAC and MIPS identified YSOs (Winston et al., in preparation).  This catalog contains 
a total of 30,646 YSOs - 2,029 Class I, 27,672 Class II, and 945  Class IIa/III young stars - across the Cygnus X North and South fields.  These YSOs were run through the SEDFitter routine in the same way as the SFOG data, with a fixed distance of 1.4~kpc, and the IMF constructed.  The linear regression fit to the data between 
$3~M_{\odot}<M_*<10~M_{\odot}$ was $\Gamma = 2.315 \pm 0.298$. 
This value is consistent with the literature values and with the slope measured for all the SFOG clusters and those with $Max M_{\odot}<10$ or $R_{Gal}<11.5$~kpc.

The IMF is generally presumed to be universal, however the precise origins of the IMF, its relation to the Core Mass Function (CMF), and the effects of galactic and local star-forming environment on its evolution are not fully understood.  \citet{2019BAAS...51c.335P} provide a useful overview of our current understanding in their recent white paper.  An accurate and in-depth study of the IMF in the outer Galaxy is beyond the scope of this paper.  The results presented here should be taken as an indication that the IMF in the Outer Galaxy is similar to that of the Inner Galaxy, and that this may indicate that metallicity does not greatly impact the IMF.  
However, there are a number of caveats, both astrophysical and analytical, which affect the SEDFitter derived IMF that we will now outline.  

Variations in the IMF of an astrophysical origin could be due to e.g. metallicity \citep[e.g.][]{2020bugm.conf...15K}, feedback from massive stars in the cluster either triggering or impeding star formation \citep[e.g.][]{2013MNRAS.435..917W, 2016MNRAS.460.3272K}, jets and outflows \citep[e.g.][]{2018MNRAS.476..771C}, disk evolution \citep[e.g.][]{2016ApJ...825..125P}, and clustered vs. distributed star formation \citep[e.g.][]{2011MNRAS.410.2339B}.  Recent theory suggests that massive stars may form a few Myrs after initial low mass star formation begins, which would also lead to a variation in IMF with cluster age \citep{2019MNRAS.490.3061V}.  Faster disk evolution in intermediate and massive stars has also been reported by \citet{2016ApJ...825..125P} in the M17 SWex star forming region that would also bias the slope of the IMF in this mass regime.  

The limitations of the data can impact the accuracy of the SEDFitter results.    
The distance estimates to the clusters are derived from kinematic distances, which can have uncertainties of at least 10-20\% and are subject to systematic uncertainties \citep{anderson}.  \citet{2020A&A...633A..99C} find in their {\it Gaia} DR2 study of open clusters that not all previously identified members are physically associated with the regions, this would also affect the SEDFitter results, which assume a single distance to each cluster.    

Further, each cluster also has a spatial 'depth' and so a YSO at the near edge of the region may lie a few hundred parsecs nearer than one at the far edge. The clusters are all of different ages, and within each cluster the subclusters may also have a range in age.  Thus the disk fraction, and hence percentage of the population detected with Spitzer, will vary - this should be accounted for when stacking the clusters in the IMF, but cannot be done here. 

There are also the issues of varying levels of diffuse mid-IR emission and source crowding in both \Sp\ and to a greater extent the {\it WISE} data.  Diffuse emission reduces sensitivity in the IRAC bands, meaning fainter sources will not be detected.  Source crowding is primarily a spatial resolution issue and is most prevalent in the central cores of clusters. This leads to an underestimation of the disk bearing population, particularly of the lower mass YSOs. 

The mid-IR data also does not provide a measure of the diskless population of the clusters.  These populations are generally identified using spectroscopy or X-ray observations.  The coverage of X-ray studies is highly constrained in comparison to IR surveys and is generally focused on the centers of known star-forming regions \citep{win09, win11, 2013ApJS..209...32B}. These studies show large populations of YSOs without disks, and trends in disk fractions with proximity to OB stars.  

Our selection method is not very sensitive to massive stars, and may not detect all the of known massive stars in a region. This would lead to a steeper IMF slope at higher masses.  A comparison with \citet{2014yCat....102023S} shows that of the 3302 objects in their catalog with spectral types 'O' or 'B' that were located within the convex hulls of the clusters, 591 were matched to a YSO in the SFOG catalog within 2\farcs0, or 18\% of the known massive population.  

The accuracy of the SEDFitter analysis is limited by the small number of near- and mid-IR bands available for model fitting and the reddening across those bands. \citet{2019ApJ...881...37P} find a degeneracy between stellar effective temperature and extinction that can lead to inaccurately modeled masses.  The presence of disk material adds to this issue, especially as disk/envelope masses are poorly constrained without far-IR and longer photometric data.

\section{Comparison to other Catalogs}\label{compare}

The SFOG catalog of YSOs and identified clusters was compared to a number of other published surveys and databases covering the outer Milky Way Galaxy.

\subsection{Other Databases of YSOs and Clusters}
The individual YSO candidates were matched to the online SIMBAD database to within a 2$\arcsec$ radius for all 47,338 YSOs, with 7,343 matches. 
Of these, 4,428 had been previously identified as a type of young stellar object, leaving 42,910 possibly new YSO candidates (some previous studies may not be listed on the SIMBAD database).  
Table~\ref{tab:Simbad} provides a sample listing of the YSOs with matches within 2$\arcsec$ of a SIMBAD source, 
giving the source identification and the object type. 
The full table is available online in electronic format, and includes further selected information pertaining to the SIMBAD objects. 

In their recent paper, \citet{armentrout} list 166 \ion{H}{2} regions previously thought to be radio quiet, that they confirm to be weak 
radio sources.  Of the 166, 37 were found to lie within the convex hull of one of the 618 clusters in the SFOG field. Twenty of these matches were found between 90\degr\ and 115\degr\ Galactic longitude.  There was no significant difference in the median number of YSOs or the median effective hull radius between the clusters in the radio quiet Armentrout sample or the {\it WISE} \ion{H}{2} sample.  
Further, when matching the \ion{H}{2} regions to the catalog of YSOs, it was found that 58 had a YSO located within 10$\arcsec$ and that 
107 had a YSO located within 1$\arcmin$.  

The positions of the clusters were also compared to those of \citet{avedisova}, who reported 66,887 clusters over the whole Galactic plane. 
Of these, 24,101 were within the boundaries of the SFOG field. Of the 618 clusters we found in the SFOG field, 260 were matched to one or more of the \citet{avedisova} 
clusters, while 358 were new to this survey.  

\subsection{{\it Gaia} DR2}
The {\it Gaia} DR2 data release was also searched for matches to the SFOG catalog. Of the 47,338 YSOs identified by SFOG, 25,919 had a {\it Gaia}  counterpart within 0\farcs5.  These matched sources were then compared to the \citet{2018AJ....156...58B} catalog of calculated parallax distances for {\it Gaia}  DR2 to determine the individual distances to the SFOG YSOs.  Of the 25,919 matches, 23,424 had a distance listed in \citet{2018AJ....156...58B}.  A comparison was then made to the 100 identified clusters with known distances to determine if the distances of the YSOs in those clusters corresponded to the {\it WISE} \ion{H}{2} catalog distances.  The clustered YSOs showed both a large range in individual uncertainty of a few kiloparsecs and a spread within the cluster YSOs of a few kiloparsecs.  While some individual YSOs had distances corresponding to that from {\it WISE}, the median and average values for the cluster did not.  It can be noted that the {\it Gaia} distance uncertainties increase greatly beyond $\sim$0.5~kpc, and that this may account for the lack of consistency in cluster distance estimates.  
Given the lack of consistency in the cluster distances, it was decided not to use these in the SEDFitter routine or to report them in the catalog.  

\citet{2020A&A...633A..99C} calculated {\it Gaia} DR2 distances to previously identified open clusters, using the mode of distance likelihoods of known members. Of these, 27 matched to an SFOG cluster and 12 had distances in the {\it WISE} HII catalog.  The differences in the measured distances varied widely (from 250-7300~pc).  Given that young clusters are embedded and {\it Gaia} is more likely to detect foreground stars or the front cluster population, we decided to use the kinematic distances in this paper.

\subsection{Other IR Surveys}

The first comparison was made to the SMOG field of Paper I, to assess the effect of the loss of the 5.8 and 8~$\mu$m IRAC bands on the 
detection rate of YSOs.   The SMOG field 2MASS+IRAC selection technique identified 3835 YSOs.  
The same GLIMPSE360 criteria used in this paper identified 1,512 YSOs.  
Of these, 1,102 objects matched between the two catalogs.  This suggests that we are detecting 29\% of the original catalog with the GLIMPSE360 criteria.  Much of this difference in the number of YSOs detected is attributable to the more stringent 3.6~$\mu$m cut around 14~mag in the SFOG selection criteria; many of the undetected SMOG YSOs are fainter and redder than those selected by the SFOG criteria.  This would imply that the total YSO population may be at least as high as 73,247 across the field to the depth of the SMOG observations.  
Of the remaining 410 unmatched YSOs that were identified with the new method and not in the original catalog, about seventy fall on the edges of the field where there was no overlap between the two IRAC FOVs, and thus were not in the original `cleaned' source catalog after contaminant removal.  A further $\sim$30 SFOG objects fall in the galactic contaminants color-space of the IRAC four-band CCD.  The rest show weak [3.6-4.5] color excess that was not deemed sufficient for selection in the original SMOG criteria.  The spatial distribution of the unmatched sources shows that the majority ($>$75\%) are associated with matched YSOs and/or identified clusters.  

The catalog of \citet{toth} {\it AKARI} YSOs was compared to the SFOG YSOs. The {\it AKARI} catalog was based on the Far-Infrared Surveyor All-Sky Survey catalog composed of photometry at four IR wavelength bands centered at 65, 90, 140 and 160~\micron. Of the 44,001 {\it AKARI} YSOs, 14,986 were located in the SFOG field. 
Of these, 49 matched to the SFOG catalog within 2$\arcsec$ and 261 to within 5$\arcsec$.  However, given the {\it AKARI} spatial resolution of 1\arcmin\ to 1\farcm5 at these wavelengths, the {\it AKARI} sources contain many individual YSOs and likely significant emission from the clouds surrounding these clusters, as well as deeply embedded YSOs, and are therefore not likely true or unique matches to the IRAC-identified YSOs in the SFOG survey.

We compared the {\it WISE}\ identified YSOs in SFOG  to the \citet{marton} {\it WISE}\ Single Vector Machine (SVM) selected YSO candidates.  
They identify 133,980 Class I/II YSOs across the whole sky, with about 16,945 candidate YSOs within the SFOG field.  
Within a radius of 1\farcs5, we match 5,537 of 20,892 of our {\it WISE} detected YSOs, or 26\%, to the \citet{marton} sample.  
Including the full SFOG catalog, we match 6,425 of 47,338, or 14\%, of our YSOs to the Marton sample.  
A comparison of the CCDs of the two samples of YSOs, shows that ours has a more conservative color cut, with the Marton sample selecting YSOs with {\it WISE1$-$WISE2} $< 0$, which accounts for most of the difference in the catalogs.  

We compared our detection of YSOs in the W5 star-forming region to that reported in \citet{2008ApJ...688.1142K}, who published a {\it Spitzer} IRAC and MIPS  survey of the region conducted during the cryogenic mission phase.  They identify 3 deeply embedded objects, 171 Class I, 1,809 Class II, 79 transition disks, (for a total of 2,062 IR-excess YSOs) and 15,709 Class IIa/III stars.  
Of these, we match 1,091 objects within a 1\arcsec\ radius; 
none of the deeply embedded objects, 88 Class I (52\%), 947 Class II (52\%), 18 transition disks (23\%), and 38 Class IIa/III (0.2\%).  The \citet{2008ApJ...688.1142K} Class IIa/III objects have weak excess and are positionally associated with the region, hence the very low detection rate in the SFOG sample.  The 50\% detection rate for the IR-excess YSOs roughly corresponds to the brighter YSOs in the two shorter IRAC bands.  The spatial distribution of the YSOs and the identified clusters match closely to those found in this paper. 

The SFOG catalog was also compared to the \citet{2011ApJ...743...39R} four-band IRAC and MIPS 24~\micron\ survey of the W3 star-formation region. They identified 1,566 YSOs, of which we match 446 SFOG YSOs to within 1\arcsec\ radius.  
The \citet{2011ApJ...743...39R} evolutionary classes are:
184 Class 0/I, 560 deeply embedded Class 0/I, 549 Class II, and 273 embedded Class II.  We match 
65 (35\%), 35 (2.2\%), 343 (63\%), 3 (1.1\%) of these classes, respectively.  The SFOG catalog does not identify the deeply embedded YSOs; some of these were selected using {\it Spitzer} MIPS photometry and so may not show excess emission in the shorter wavelengths, the remainder were below 14th magnitude in the IRAC 3.6~\micron\ channel and so were not selected in SFOG.  
However, the spatial distribution of the W3 YSOs in the SFOG catalog traces a similar cluster distribution to that of \citet{2011ApJ...743...39R}.  

We made a further comparison of the SFOG catalog to the \citet{2011ApJS..193...25R} {\it Spitzer} IRAC and MIPS survey of the North American \& Pelican nebula star-forming regions.  They report a total of 2,196 YSOs in the field, with 262 that lie in the overlap region with the SFOG catalog field. Of these, 132 are matched to within a 1\arcsec\ radius of an SFOG YSO, being again the brighter 3.6 and 4.5~$\mu$m sources.  

We thus draw two conclusions from the comparisons to these  well-studied regions and other IR-based YSO catalogs. First, we are finding roughly half of the previously identified YSOs; those that are less embedded and brighter at 3.6~$\mu$m. We also used more conservative color cuts and brighter magnitude limits to minimize the number of spurious YSO identifications. Secondly, from the YSOs we do detect, we find that we are reliably identifying the overall spatial structure and main clusters in these regions that were found in the other surveys that used deeper integrations or had coverage in the IRAC 5.8 and 8~\micron\  and MIPS 24~\micron\ bands.

\section{Summary}\label{summ}

We have undertaken a  study of the 600 deg$^2$ SFOG field comprising the GLIMPSE360 and SMOG survey regions.  
We combined the {\it Spitzer} data with 2MASS near-IR photometry, and used the {\it WISE} catalog of the field to identify more embedded YSOs.  

\begin{itemize}

\item We identify 42,757 YSOs with IR-excess emission in the GLIMPSE360 and {\it WISE} data. When combined with the SMOG field and after removing the spurious {\it WISE} sources, we find a total of 47,338 YSOs. 

\item The evolutionary class of the YSOs was determined from the SED slope: 
10,461 Class I, 29,552 Class II, and 7,325 Class IIa/III.  

\item We identify 618 reliable clusters in the  SFOG field.   
The ratio of YSOs identified as members of clusters was 25,528/47,338, or 54\%.  
The smallest cluster has 5 members, and the largest has 1,177 members, with a median size of 17 YSOs. 
Of the 618 clusters, 47 have more than 100 members, and 22 have more than 200 members.

\item One hundred clusters had a distance estimate from \ion{H}{2} regions within their convex hulls. The SEDs of the YSOs in these clusters were fitted using the SEDFitter routine, of these 96 had reliable fits. 

\item From the modeled masses, the IMF was constructed for the clusters across the SFOG field.  
The slope of the combined IMF was found to be $\Gamma = 2.38 \pm 0.20$ above 3~M$_{\odot}$. 
Dividing the clusters by Galactocentric distances, the slopes were  $\Gamma = 1.87 \pm 0.31$ above 3~M$_{\odot}$ for $R_{Gal} < 11.5$~kpc  
and $\Gamma = 1.15 \pm 0.24$ above 3~M$_{\odot}$ for $R_{Gal} > 11.5$~kpc. 
These values are consistent with each other within the uncertainties, and with those obtained in the inner Galaxy high-mass SFRs. 
The slopes are likely also consistent with a universal Salpeter IMF.  

\end{itemize}
\hfill

\acknowledgments
This work is based on observations made with the {\it Spitzer} Space Telescope, which is operated by the Jet Propulsion Laboratory, California Institute of Technology 
under NASA contract 1407.   We gratefully acknowledge funding support for this work from NASA ADAP grant NNX16AF37G.
This publication makes use of data products from the Two Micron All Sky Survey, which is a joint project of the University of Massachusetts and the Infrared Processing and 
Analysis Center/California Institute of Technology, funded by the National Aeronautics and Space Administration and the National Science Foundation. 
Support for the IRAC instrument was provided by NASA through contract 960541 issued by JPL. 
This publication makes use of data products from the Wide-field Infrared Survey Explorer, which is a joint project of the University of California, Los Angeles, and the Jet Propulsion 
Laboratory/California Institute of Technology, funded by the National Aeronautics and Space Administration.
This research made use of Montage. It is funded by the National Science Foundation under Grant Number ACI-1440620, and was previously funded by the National Aeronautics 
and Space Administration's Earth Science Technology Office, Computation Technologies Project, under Cooperative Agreement Number NCC5-626 between NASA and the 
California Institute of Technology.
This research has made use of NASA’s Astrophysics Data System.

\facility{Spitzer (IRAC)}
\software{Astropy \citep{astropy}, SAOimageDS9 \citep{Joye2003}, Montage \citep{Berriman2008}, SEDFitter \citep{robitaille}}


\appendix

\section{IRAC \& 2MASS Source Selection}\label{irysos_appen}

The removal of contaminating sources and the selection of YSOs in the GLIMPSE360 field were undertaken in a different manner to those implemented 
for the SMOG field, with IRAC four band coverage which was based on the methods of \citet{gut08} and \citet{gut09}.  The location of the SMOG field 
contaminants in color-space were used as the basis for identifying the locations of the contaminants in the two band IRAC data of GLIMPSE360.    

Background galaxy contaminants, including candidate AGNs and PAH galaxies, and saturated sources were removed using a cut in color-magnitude space. 
\begin{equation}
\begin{split}
[3.6]  <  6.0 \mbox{  and  }  \\   
[4.5]  <  5.5 \mbox{  and  }  \\   
[3.6]  >  16.0 \mbox{  and  }  \\   
[3.6]  >  14.0  \mbox{  and }  [3.6 - 4.5] < 0.5  \mbox{  and  }  \\ 
[3.6]  >  2 ( [3.6 - 4.5]  +  0.5 ) + 14  
\end{split}
\end{equation}
The full GLIMPSE360 Archive contained 49,378,049 sources.  The catalog contained 7,527,352 objects after contaminants were removed.  
The SFOG fields lie in the direction of the outer Galaxy where the level of shielding from the Galactic center from both the stellar 
population and dust component is reduced and thus the extragalactic background is expected to be higher.  

Three combinations of 2MASS and IRAC bands were used to select for less extincted objects. The photometry was first dereddened, and only sources 
with good values of extinction were included for selection.  
\begin{equation}
\begin{split}
(A_H - A_{4.5}/A_J - A_H) ( [J - H] - 0.6 + \sigma_{JH})  + 1.0  + \sigma_{H4.5}  < [H - 4.5] \mbox{  and } \\
[J - H] > 0 
\end{split}
\end{equation}

\begin{equation}
\begin{split}
(A_K - A_{4.5}/A_H - A_K) ([H-K] + \sigma_{HK}) + 0.4 + \sigma_{K4.5} < [K - 4.5]    \mbox{  and } \\
[H - K] > 0       \mbox{  and } \\
[K - 4.5] > 0.2 + \sigma_{K4.5}
\end{split}
\end{equation}

\begin{equation}
\begin{split}
[3.6 - 4.5] -  \sigma_{3.6,4.5}  >  0   \mbox{  and  }  \\   
[K - 3.6]  -  \sigma_{K3.6}  >  0.2*[3.6 - 4.5] + 0.3     \mbox{  and  }  \\   
[K - 3.6]  - \sigma_{K3.6}  >  -1.0 ( [3.6 - 4.5] - \sigma_{12} )  +  0.8  
\end{split}
\end{equation}

For these three selection criteria:  20,339, 24,560, and 16,941 candidate YSOs were selected respectively, for a combined total of 28,837 YSOs identified using the GLIMPSE360 photometry.

\section{WISE Source Selection}\label{wise_ysos_appen}

Following the process laid out by \citet{fischer}, spurious detections were cleaned from the catalog.  The first step was to remove those sources with 
uppercase flags in bands W1, W2, and W3.  Upper limits in bands W1, W2, W3 are then removed, and a saturation cut-off of $W1 > 5$ applied to the data.  
The initial catalog contained 14,483,596 sources.  The remaining catalog contained 543,457 sources, or $\approx 4$\% of the original catalog.  
The contaminating background galaxies and source selection criteria for the {\it WISE} data were taken from \citet{koenig, fischer} and adapted to the 
requirements of the GLIMPSE360 field.  
We adjusted the criteria for removal of AGN and star-forming galaxy contaminants slightly from those of \citet{koenig} as follows:
\begin{equation}
\begin{split}
SFG =  & [W2 - W3] > 2.3   \mbox{  and  }  \\
            & [W1 - W2]  < 1.0   \mbox{  and  }  \\
            & [W1 - W2] < 0.46([W2 - W3] - 0.78)   \mbox{  and  }  \\
            & [W1] > 14 
\end{split}
\end{equation}
\begin{equation}
\begin{split}
AGN = & [W1] > 1.8( [W1 - W3] + 4.1 ) \mbox{  and  } \\
            & [W1] > 14  \mbox{  or  }   \\
            & [W1]  >  [W1 - W3] + 10.0  
\end{split}
\end{equation}
Of those 14,483,596 sources,  13,940,139 were identified as contaminants, leaving a cleaned catalog of 543,457 sources.  

The clean catalog was then searched for YSOs according to criteria taken from the \citet{fischer} and \citet{koenig} papers. 
From these there were 2,245 transition disk candidates and 7,094 YSOs identified from the four-band 
{\it WISE} diagram, and 7,244 Class I and 10,447 Class II sources identified from the {\it WISE} 3-band diagram.  
There was a total of 20,892 candidate YSOs identified with {\it WISE}.

\section{IRAC+WISE Source Selection}\label{irws_ysos_appen}

The IRAC+{\it WISE} YSOs were selected following the same selection cut-offs as were applied to the {\it WISE} sources. 
The cleaned {\it WISE} catalog was matched to the contaminant-removed IRAC catalog, and sources with a 1\farcs5 or closer 
spatial coincidence were considered to be the same object, giving 318,588 objects in the joined catalog.  
The YSOs were then selected by replacing the {\it WISE} bands 1 \& 2 with IRAC bands 1 \& 2, and replacing this 
photometry in the {\it WISE} color-color diagram YSO selections.  
The 2MASS selection criteria were not applied here as they replicate the IRAC+2MASS selection from the GLIMPSE360 data.
A total of 11,196 candidate YSOs were selected using this method.



\end{document}